\documentclass[article]{JHEP3}
\pdfoutput=1

\usepackage{amsmath,amssymb,amsthm,amscd}
\usepackage{epsfig}
\usepackage{psfrag,comment}
\usepackage{graphicx}

\newcommand{\CC}{{\cal C}}

\newcommand{\CF}{{\cal F}}

\newcommand{\CN}{{\cal N}}

\newcommand{\CP}{{\cal P}}
\newcommand{\CQ}{{\cal Q}}
\newcommand{\CR}{{\cal R}}
\newcommand{\CS}{{\cal S}}

\def\BZ{{\mathbb Z}}
\def\BR{{\mathbb R}}
\def\BC{{\mathbb C}}

\newcommand{\be}{\begin{equation}}
\newcommand{\ee}{\end{equation}}
\newcommand{\ba}{\begin{aligned}}
\newcommand{\ea}{\end{aligned}}
\newcommand{\bea}{\begin{eqnarray}}
\newcommand{\eea}{\end{eqnarray}}
\newcommand{\bean}{\begin{eqnarray*}}
\newcommand{\eean}{\end{eqnarray*}}

\def\r{\right\rangle}

\def\1{\boldsymbol{1}}
\def\0{|\1\r}

\def\re{{\mathbb{R}}{\mathrm{e}}}

\def\disc{{\mathrm{Disc\,}}}

\newcommand{\rme}{{\rm e}}
\newcommand{\rmi}{{\rm i}}
\newcommand{\rmd}{{\rm d}}

\def\XXint#1#2#3{{\setbox0=\hbox{$#1{#2#3}{\int}$}
     \vcenter{\hbox{$#2#3$}}\kern-.5\wd0}}

\newdimen\tableauside\tableauside=1.0ex
\newdimen\tableaurule\tableaurule=0.4pt
\newdimen\tableaustep
\def\phantomhrule#1{\hbox{\vbox to0pt{\hrule height\tableaurule width#1\vss}}}
\def\phantomvrule#1{\vbox{\hbox to0pt{\vrule width\tableaurule height#1\hss}}}
\def\sqr{\vbox{%
  \phantomhrule\tableaustep
  \hbox{\phantomvrule\tableaustep\kern\tableaustep\phantomvrule\tableaustep}%
  \hbox{\vbox{\phantomhrule\tableauside}\kern-\tableaurule}}}
\def\squares#1{\hbox{\count0=#1\noindent\loop\sqr
  \advance\count0 by-1 \ifnum\count0>0\repeat}}
\def\tableau#1{\vcenter{\offinterlineskip
  \tableaustep=\tableauside\advance\tableaustep by-\tableaurule
  \kern\normallineskip\hbox
    {\kern\normallineskip\vbox
      {\gettableau#1 0 }%
     \kern\normallineskip\kern\tableaurule}%
  \kern\normallineskip\kern\tableaurule}}
\def\gettableau#1{\ifnum#1=0\let\next=\null\else
\squares{#1}\let\next=\gettableau\fi\next}
\preprint{
{\small{\textsf{YITP-SB-13-07}}}}

\title{The Resurgence of Instantons: Multi--Cut Stokes Phases and the Painlev\'e II Equation}

\author{Ricardo Schiappa$^1$ and Ricardo Vaz$^2$
\\
$^1$CAMGSD, Departamento de Matem\'atica, Instituto Superior T\'ecnico,\\ 
Av. Rovisco Pais 1, 1049--001 Lisboa, Portugal\\
\\
$^2$C.N.~Yang Institute for Theoretical Physics, Stony Brook University,\\ 
Stony Brook, NY 11794--3840, United States\\
\\
\email{schiappa@math.ist.utl.pt}, \quad
\email{ricardo.vaz@stonybrook.edu}

}

\abstract{
Resurgent transseries have recently been shown to be a very powerful construction in order to completely describe nonperturbative phenomena in both matrix models and topological or minimal strings. These solutions encode the full nonperturbative content of a given gauge or string theory, where resurgence relates every (generalized) multi--instanton sector to each other via large--order analysis. The Stokes phase is the adequate gauge theory phase where an 't~Hooft large $N$ expansion exists and where resurgent transseries are most simply constructed. This paper addresses the nonperturbative study of Stokes phases associated to multi--cut solutions of generic matrix models, constructing nonperturbative solutions for their free energies and exploring the asymptotic large--order behavior around distinct multi--instanton sectors. Explicit formulae are presented for the $\BZ_2$ symmetric two--cut set--up, addressing the cases of the quartic matrix model in its two--cut Stokes phase; the ``triple'' Penner potential which yields four--point correlation functions in the AGT framework; and the Painlev\'e II equation describing minimal superstrings. 
}


\begin{document}



\vfill

\eject

\allowdisplaybreaks

\section{Introduction and Summary}

For almost 40 years the large $N$ limit of gauge theory \cite{th74} has been source to many fascinating results and ideas, with large $N$ duality \cite{m97} playing a definite central role. Given some nonabelian gauge theoretic system, this limit produces an \textit{asymptotic} perturbative expansion\footnote{This is, of course, the very well known topological or genus expansion due to 't~Hooft \cite{th74}.}, in powers of $1/N^2$ (or, from the point--of--view of large $N$ duality, in powers of the closed string coupling $g_s^2$). What this means is that the genus $g$ perturbative contributions to the free energy\footnote{If one has the gauge theory in mind, in here $t = N g_s$ is the 't~Hooft coupling. If one has the dual closed string theory in mind, in here $t$ is a geometric modulus associated to the background geometry.}, $F_g (t)$, will display large--order behavior of the type $F_g \sim (2g)!$ and the perturbative expansion will have zero radius of convergence \cite{s90}. But, more importantly, what this implies is that the perturbative series is not enough to define the free energy and nonperturbative corrections of the type $\sim \exp \left( - N \right)$ are needed in order to properly make sense out of this expansion.

The study of nonperturbative corrections, their relation to the large--order growth of the perturbative expansion and their application within the resummation of perturbation theory has a long history. It is also almost 40 years since these topics were first considered within the quantum mechanical context of the quartic anharmonic oscillator \cite{bw73}. Later, they were extended to the usual perturbative expansion in quantum field theoretic systems; see, \textit{e.g.}, \cite{z81} for a review of early developments. In the present work we are interested in yet another extension of these ideas, namely towards trying to understand, nonperturbatively, the $1/N$ expansion. This topic has a more recent history where we will follow previous work in \cite{m06, msw07, m08, msw08, em08, ps09, mpp09, gikm10, kmr10, asv11} (we refer the reader to the introduction of \cite{asv11} for a quick overview of these developments, or to \cite{m10} for an excellent review of some of the main ideas which were put forward in the aforementioned references). For the moment let us simply mention that most of the original ideas and results in \cite{z81} have their counterparts within the large $N$ context. In particular, the leading growth of the free energies $F_g (t)$ is dictated by a suitable instanton action, $A(t)$, whose physical origin is associated to eigenvalue tunneling, at least within the matrix model context \cite{d91, d92, msw07}. Moreover, the subleading growth is associated to the one--loop amplitude around the one--instanton sector \cite{d91, d92, msw07}. Further corrections to the large--order $(2g)!$ growth arise from higher loop amplitudes around some fixed instanton sector---corrections in $1/g$---and from higher instanton numbers---corrections in $1/n^g$, where $n$ is the instanton number \cite{asv11}.

A very interesting novelty emerges as one addresses the asymptotic perturbative expansion around some fixed multi--instanton sector. Although one might \textit{na\"\i vely} expect that the large--order growth of this asymptotic expansion would be controled by sectors with different instanton numbers (either higher or lower), it turns out that this expectation is \textit{incomplete}: one actually needs to introduce \textit{new} nonperturbative sectors in order to match all large--order results \cite{gikm10, asv11}. In the examples studied so far, these ``generalized'' multi--instantons have instanton actions with opposite sign as compared to the ``physical'' multi--instantons, and they may be assembled all together into a so--called transseries solution to the problem at hand---where resurgent analysis relates every generalized nonperturbative sector to each other via large--order analysis. In this way, by checking that no further corrections exist deep in the large--order asymptotics, one is led to the assumption that these resurgent transseries solutions completely encode the full nonperturbative content of a given gauge or string theory. Transseries solutions were first introduced in the string theoretic context in \cite{m08}, and this approach was later extended in \cite{asv11} (see, \textit{e.g.}, \cite{e0801} for a mathematical overview). The use of resurgent analysis as a tool towards fully understanding generalized nonperturbative sectors of string theoretic systems was first introduced in \cite{gm08, gikm10}, and those approaches were later extended in \cite{asv11} (see, \textit{e.g.}, \cite{cnp93, ss03} for mathematical overviews). In particular, we refer the reader to \cite{asv11} for a complete exposition of the ideas and techniques we shall use in the course of the present work. In fact, it is precisely one of our main goals in this paper to extend the techniques of resurgent analysis and transseries solutions to other examples beyond the ones in \cite{asv11}. Furthermore, let us mention that ideas from resurgence and transseries have also appeared in, \textit{e.g.}, \cite{ddp90, v93, jzj04}, in a quantum mechanical context, and have recently been shown to be very promising in the study of quantum field theory \cite{s02, au12a, au12b, du12a, du12b, fggp13}.

In our endeavor to understand the nonperturbative structure of the large $N$ limit, we have begun with a simpler class of gauge theoretic systems: random matrix models\footnote{Although further motivated by their relation to topological strings, these models are probably the simplest already showing all features of large $N$ resurgence, without the added complication of renormalon physics \cite{au12a, au12b}.}. In this case, it is very well known that, at large $N$, matrix eigenvalues cluster into the cuts of a corresponding spectral geometry \cite{bipz78}. Within the context of large $N$ duality, this was later understood as relating matrix models to B--model topological string theory in local Calabi--Yau geometries \cite{dv02, dv02a, emo07} (see \cite{m04} for an excellent review). Depending on the potential appearing in the matrix partition function, and the phase in which the model is to be found, generically the spectral geometry will correspond to a multi--cut configuration. This is an important point, specially in light of the question: is it always the case that, as one considers the large $N$ limit of some given gauge theory, one will find an expansion of 't~Hooft type with a closed string dual? Within the matrix model context, this question was first raised in \cite{bde00} and answered negatively.

Let us dwell on this point for a moment as it is also at the basis of the class of examples we choose to address in this work. The nature of the large $N$ asymptotic limit depends very much on which gauge theoretic phase one considers \cite{bde00, mpp09} and is analogous to the study of Stokes phenomena in classical analysis. When considering single--cut models, one finds the familiar $1/N^2$ expansion \cite{ackm93, eo07, eo08}, \textit{i.e.}, one finds a topological genus expansion with a closed string dual. However, this is not usually the case when considering multi--cut models, where one finds large $N$ theta--function asymptotics instead \cite{bde00, e08, em08}, \textit{i.e.}, there is no genus expansion and possibly no closed string dual. These two distinct large $N$ asymptotics are associated to what we call Stokes and anti--Stokes phases, respectively, generalizing the usual Stokes and anti--Stokes lines in classical analysis \cite{m10}. In fact, in the Stokes phase the single cut would correspond to the leading saddle, with pinched cuts corresponding to exponentially suppressed saddles \cite{msw07}. On the other hand, in the anti--Stokes phase the many cuts correspond to many different saddles of similar order, where their joint contribution translates into an oscillatory large $N$ behavior \cite{bde00}. Of course one should start the analysis in the opposite direction: having identified Stokes and anti--Stokes phases with particular large $N$ asymptotics, one may then ask what spectral geometry configurations appear in each distinct phase. The point of interest to us in here is that there are regions of moduli space where the Stokes phase is actually realized by a \textit{multi}--cut configuration (essentially, by configurations where all cuts are equal, \textit{i.e.}, they have the precise same eigenvalue filling). It is this type of multi--cut configurations which we investigate and explore in this work, within the framework of resurgent transseries.

This paper is organized as follows. We begin in section \ref{sec:multi} by reviewing background material concerning both matrix models and resurgent transseries. We briefly review saddle--point and orthogonal polynomial approaches to solving (multi--cut) matrix models, and then go through the basics of resurgence and transseries. In section \ref{sec:inst} we address the multi--instanton analysis when we have two cuts with $\BZ_2$ symmetry (ensuring we are in a multi--cut Stokes phase). This is done using methods of spectral geometry which essentially generalize previous work in \cite{msw07, msw08}. Elliptic functions and theta functions which, due to the elliptic nature of the spectral curve, appear during the calculation, end up canceling in the final result thus providing further evidence on the nature of the Stokes phase. This is actually an interesting point of the calculation, as, on what concerns the perturbative sector, it was source to some confusion in early studies of $\BZ_2$ symmetric spectral configurations. In fact, the original saddle--point calculation of the two--point resolvent in a $\BZ_2$ symmetric distribution of eigenvalues \cite{a96, aa96}, with an explicit elliptic function dependence, did not match the corresponding orthogonal polynomial calculation \cite{d97, kf98, bd98}, which saw no trace of these elliptic functions. The reason for this was that \cite{a96, aa96} worked in a fixed canonical ensemble, while in the spectral curve approach to solving some given multi--cut scenario one needs to address the full grand--canonical ensemble as later explained in \cite{bde00}. We shall explicitly see in section \ref{sec:inst} what is the counterpart of those ideas within the multi--instanton context. Section \ref{sec:lo} presents applications of these general multi--instanton results into two different examples. On the one hand we study the two--cut phase of the quartic matrix model. This example further explores the quartic matrix model along the lines of \cite{asv11}, in particular as we construct a two--parameter transseries solution in this phase. Instantons in this example are associated to $B$--cycle eigenvalue tunneling \cite{msw07} and we provide tests of our analytical results by comparing against the large--order behavior of perturbation theory. Earlier results addressing the asymptotics of this model were presented in \cite{bi99} and we extend them in here within the context of transseries and resurgent analysis. On the other hand, we address the example of the ``triple'' Penner potential which is associated to the computation of four--point correlation functions within the AGT framework \cite{agt09}. This is an interesting example as it is actually exactly solvable via generalized Gegenbauer polynomials and where instantons are associated to $A$--cycle eigenvalue tunneling \cite{ps09}. In section \ref{sec:P2} we turn to the asymptotics of multi--instanton sectors, and we do this in the natural double--scaling limit of the two--cut quartic matrix model, which is the Painlev\'e II equation. This equation describes 2d supergravity, or type 0B string theory \cite{dss90, kms03, ss04, ss05}, and we fully construct its two--parameter transseries solution, checking the existence of generalized multi--instanton sectors via resurgent analysis. Earlier results addressing the asymptotics of this model were presented in \cite{dz95} and we extend them in here within the context of transseries and resurgent analysis. In particular, we compute many new Stokes constants for this system (in this way verifying and generalizing the one known Stokes constant \cite{kkm05}), and present the complete nonperturbative free energy of type 0B string theory. We close in section \ref{sec:concl} with a discussion of some ideas which could lead to future research. Let us also stress that, due to the nature of the large--order analysis, we have generated a large amount of data concerning both the two--cut quartic matrix model and the Painlev\'e II equation. Due to space constraints it is impossible to list all such results in the paper, but we do present some of this data in a few appendices.

\section{Revisiting Multi--Cut Matrix Models}\label{sec:multi}

Let us begin by setting our notation concerning both saddle--point and orthogonal polynomial approaches to solving matrix models, with emphasis on multi--cut configurations. We shall also review the required background in order to address the construction of (large $N$) resurgent transseries solutions for these multi--cut configurations, when in their Stokes phases.

\subsection{The Saddle--Point Analysis}\label{sec:sp}

Let us first address the saddle--point approximation to computing the one--matrix model partition function (within the hermitian ensemble, $\beta=1$) in a general multi--cut set--up; see, \textit{e.g.}, \cite{bipz78, fgz93, a96, m04, msw08}. In such configurations the $N$ eigenvalues condense into $s$ different cuts ${\cal C}_1 \cup \dots \cup {\cal C}_s = [x_1, x_2] \cup \dots \cup [x_{2s-1}, x_{2s}]$, and, in diagonal gauge, the partition function is written as
\be\label{Z multi cut}
Z ( N_1, \dots ,N_s ) = \frac{1}{N_1! \cdots N_s!}\, \int_{\lambda^{(1)}_{k_1} \in \, {C}_1} \cdots \int_{\lambda^{(s)}_{k_s} \in \, {C}_s}\, \prod_{i=1}^N \left( \frac{\rmd \lambda_i}{2 \pi} \right) \Delta^2 (\lambda_i)\, \rme^{- \frac{1}{g_s} \sum_{i=1}^N V(\lambda_i)},
\ee
\noindent
with 't~Hooft coupling $t = N g_s$ (fixed in the 't~Hooft limit). In the above expression the $\{ \lambda^{(I)}_{k_I} \}$ are the eigenvalues sitting on the $I$--th cut, with $k_I=1, \dots, N_I$ and $\sum_{I=1}^s N_I = N$, and $\Delta (\lambda_i)$ is the Vandermonde determinant. In this picture it is natural to consider the hyperelliptic Riemann surface which corresponds to a double--sheet covering of the complex plane, $\BC$, with precisely the above cuts. One can then define $A$--cycles as the cycles around each cut, whereas $B$--cycles go from the endpoint of each cut to infinity on one of the two sheets and back again on the other. For shortness, we shall refer to $\CC$ as the contour encircling all the cuts, \textit{i.e.}, $\CC = \bigcup_{I=1}^s A^I$.

The large $N$ saddle--point solution is usefully encoded in the planar resolvent, defined in closed form as
\be\label{multi-genus0resolvent}
\omega_0(z) = \frac{1}{2t} \oint_{\CC} \frac{\rmd w}{2\pi\rmi}\, \frac{V'(w)}{z-w}\, \sqrt{\frac{\sigma_s (z)}{\sigma_s (w)}},
\ee
\noindent
where we have defined 
\be
\sigma_s (z) \equiv \prod_{k=1}^{2s} (z-x_k)
\ee
\noindent
and where one still needs to specify the endpoints of the $s$ cuts, $\{ x_k \}$. One may now describe the large $N$ matrix model geometry via the corresponding spectral curve, $y(z)$, which is given in terms of the resolvent by
\be\label{sp_curve}
y(z) = V'(z) - 2 t\, \omega_0(z) \equiv M(z)\, \sqrt{\sigma_s (z)}.
\ee
\noindent
If the potential $V(z)$ in the matrix model partition function \eqref{Z multi cut} is such that $V'(z)$ is a rational function with simple poles at $z = \beta_i$,  $i = 1, 2, ..., k$ and with residues $\alpha_i$ at each pole, the expression for $M(z)$ in the expression above is simply
\be\label{M multicut} 
M(z) = \oint_{(\infty)} \frac{\rmd w}{2 \pi \rmi}\, \frac{V'(w)}{w-z}\, \frac{1}{\sqrt{\sigma_s (w)}} + \sum_{i=1}^k \frac{\alpha_i}{(\beta_i-z)\, \sqrt{\sigma_s (\beta_i)}}.
\ee
\noindent
At this stage one still needs to specify the endpoints of the cuts. If the eigenvalue distribution across all cuts is properly normalized, the planar resolvent will have the asymptotic behavior $\omega_0(z) \sim \frac{1}{z}$ as $z \to + \infty$. In turn, this asymptotic condition implies the following set of constraints
\be\label{asympcond}
\oint_{\CC} \frac{\rmd w}{2\pi\rmi}\, \frac{w^n\, V'(w)}{\sqrt{\sigma_s (w)}} = 2t\, \delta_{ns}, 
\ee
\noindent
with $n = 0,1,\dots,s$. These are $s+1$ conditions for $2s$ unknowns, where the remaining $s-1$ conditions still need to be specified and they come from the number of eigenvalues $N_I$ one chooses to place at each cut. This distribution of eigenvalues may be equivalently described by the partial 't~Hooft moduli $t_I = g_s N_I$, which may be written directly in terms of the spectral curve:
\be\label{t_I}
t_I = \frac{1}{4\pi\rmi} \oint_{A^I} \rmd z\, y(z), \qquad I = 1, 2, \ldots, s.
\ee
\noindent
Notice that, as expected, these are only $s-1$ conditions as they are not all independent, \textit{i.e.}, $\sum_{I=1}^s t_I = t$. Both constraints \eqref{asympcond} and moduli \eqref{t_I} now determine the full spectral geometry.

It is also useful to define the holomorphic effective potential 
\be\label{hol_eff_pot}
V_{\mathrm{h;eff}}'(z) = y(z).
\ee
\noindent
In this case, the effective potential is given by the real part of the holomorphic effective potential, in such a way that
\be
V_{\mathrm{eff}}(\lambda) = \re \int^\lambda \rmd z\, y(z).
\ee

\subsection{The Approach via Orthogonal Polynomials}\label{sec:op}

While saddle--point analysis is the appropriate framework to describe the spectral geometry of multi--cut configurations, it gets a bit more cumbersome when one wishes to address the computation of the full free energy. In the 't~Hooft limit, where $N \to + \infty$ with $t=g_s N$ held fixed, the perturbative, large $N$, topological expansion of the free energy is given by\footnote{Throughout this paper we shall use the symbol $\simeq$ to signal when in the presence of an asymptotic series \cite{asv11}.}
\be\label{topF}
F (g_s, \{ t_I \}) = \log Z \simeq \sum_{g=0}^{+\infty} g_s^{2g-2} F_g (t_I).
\ee
\noindent
Computing this genus expansion out of a hyperelliptic spectral curve has a long history---starting in \cite{ackm93, a96}, passing through \cite{bde00}, and recently culminating in the recursive procedure of \cite{eo07}---and it is in fact an intricate problem in algebraic geometry \cite{eo08}.

An easier approach to computing the free energy of a matrix model is to use the method of orthogonal polynomials; see, \textit{e.g.}, \cite{biz80, fgz93, m04, asv11}. On the other hand, this method is less general as it is not applicable to arbitrary multi--cut  configurations. However, as we shall also see in the course of this paper, orthogonal polynomials do work when addressing multi--cut Stokes phases. As such, let us swiftly review this method in the context of the one--cut solution (the multi--cut extension will be addressed later). Considering the partition function \eqref{Z multi cut} with a single cut, one may consider the positive--definite measure on $\BR$ given by
\be\label{orthomeasure}
\rmd\mu (z) = \rme^{- \frac{1}{g_s} V(z)}\, \frac{\rmd z}{2\pi}.
\ee
\noindent
Normalized orthogonal polynomials with respect to this measure are introduced as $p_n (z) = z^n + \cdots$, with inner product
\be\label{orthpol}
\int_{\BR} \rmd \mu(z)\, p_n (z) p_m (z) = h_n \delta_{nm}, \qquad n \ge 0.
\ee
\noindent
As the Vandermonde determinant may be written $\Delta(\lambda_i) = \det p_{j-1} (\lambda_i)$, the partition function of our matrix model may be computed as
\be
Z = \prod_{n=0}^{N-1} h_n = h_0^N \prod_{n=1}^N r_n^{N-n}, \label{Zorth}
\ee
\noindent
where we have defined $r_n = \frac{h_n}{h_{n-1}}$ for $n \ge 1$. These $r_n$ coefficients further appear in the recursion relations
\be\label{oprecursion}
p_{n+1} (z) = \left( z+s_n \right) p_n (z) - r_n\, p_{n-1} (z),
\ee
\noindent
together with coefficients $\{ s_n \}$ which will vanish for an even potential. Plugging the above \eqref{oprecursion} in the inner product \eqref{orthpol} one obtains a recursion relation directly for the $r_n$ coefficients \cite{biz80}. 

One example of great interest to us in the present work is that of the quartic potential $V(z) = \frac{\mu}{2} z^2 + \frac{\lambda}{4!} z^4$. In this case it follows that $s_n=0$ and \cite{biz80}
\be\label{4stringeq}
r_n \left( \mu + \frac{\lambda}{6}\, \big( r_{n-1} + r_n + r_{n+1} \big) \right) = n g_s.
\ee
\noindent
The free energy of the quartic matrix model (normalized against the Gaussian weight $V_{\mathrm{G}} (z) = \frac{1}{2} z^2$, as usual) then follows straight from the definition of the partition function \eqref{Zorth}
\be\label{prefree}
\CF \equiv F-F_{\mathrm{G}} = \log \frac{Z}{Z_{\mathrm{G}}} \simeq \sum_{g=0}^{+\infty} g_s^{2g-2} \CF_g (t) = \frac{t}{g_s} \log \frac{h_0}{h^{\mathrm{G}}_0} + \frac{t^2}{g_s^2}\, \frac{1}{N} \sum_{n=1}^N \left( 1-\frac{n}{N} \right) \log \frac{r_n}{r^{\mathrm{G}}_n}.
\ee
\noindent
This genus expansion is made explicit by first understanding the large $N$ expansion of the $r_n$ recursion coefficients. Changing variables as $x \equiv n g_s$, with $x \in [0,t]$ in the 't~Hooft limit, and defining $\CR(x) = r_n$ with $\CR^{\mathrm{G}}(x) = x$, the above example of the quartic potential (\ref{4stringeq}) becomes \cite{biz80, asv11}
\be\label{largeN4stringeq}
\CR (x) \left\{ \mu + \frac{\lambda}{6}\, \big( \CR (x-g_s) + \CR (x) + \CR (x+g_s) \big) \right\} = x.
\ee
\noindent
As $\CR(x)$ is even in the string coupling, it admits the usual asymptotic large $N$ expansion
\be\label{R pert}
\CR(x) \simeq \sum_{g=0}^{+\infty} g_s^{2g} R_{2g} (x),
\ee
\noindent
allowing for a recursive solution for the $R_{2g} (x)$. In particular, in the continuum limit the sum in (\ref{prefree}) may be computed via the Euler--Maclaurin formula (with $B_{2k}$ the Bernoulli numbers and $x = t\, \xi$)
\be\label{Euler Mac}
\lim_{N \to +\infty} \frac{1}{N} \sum_{n=1}^N \Psi \left( \frac{n}{N} \right) \simeq \int_0^1 \rmd \xi\, \Psi(\xi) + \left. \frac{1}{2N} \Psi(\xi) \right|_{\xi=0}^{\xi=1} + \sum_{k=1}^{+\infty} \left. \frac{1}{N^{2k}}\, \frac{B_{2k}}{(2k)!}\, \Psi^{(2k-1)} (\xi) \right|_{\xi=0}^{\xi=1},
\ee
\noindent
yielding
\bea
\CF (t, g_s) &\simeq& \frac{t}{2g_s} \left( 2 \log \frac{h_0}{h^{\mathrm{G}}_0} - \left. \log \frac{\CR(x)}{x} \right|_{x=0} \right) + \frac{1}{g_s^2} \int_0^t \rmd x \left( t-x \right) \log \frac{\CR(x)}{x} + \nonumber \\
&&
+ \sum_{g=1}^{+\infty} \left. g_s^{2g-2}\, \frac{B_{2g}}{(2g)!}\, \frac{\rmd^{2g-1}}{\rmd x^{2g-1}} \left[ \left( t-x \right) \log \frac{\CR(x)}{x} \right] \right|_{x=0}^{x=t}.
\label{FEMtoda}
\eea
\noindent
This analysis was first presented in \cite{biz80} and was recently extended to a full resurgent transseries analysis in \cite{asv11}, and we refer the reader to these references for further details. We shall later see how it generalizes to accommodate the two--cut Stokes phase of the quartic matrix model.

\subsection{Transseries and Resurgence: Basic Formulae}\label{sec:ts}

The discussion so far has focused upon the large $N$, perturbative construction of the matrix model free energy \eqref{topF}. If, on the other hand, one wishes to go beyond the perturbative analysis in order to build a fully nonperturbative solution to a given matrix model, one needs to make use of resurgent transseries. This subject was recently thoroughly addressed in \cite{asv11}, and we refer the reader to this reference for full details on these techniques and their origins. In here, we shall nonetheless cover just enough background to make the present paper a bit more self--contained.

Resurgent transseries essentially encode the full (generalized) multi--instanton content of a given non--linear system and, as such, yield nonperturbative solutions to these problems as expansions in both powers of the coupling constant and the (generalized) multi--instanton number(s). In general, many distinct instanton actions may appear and, as such, transseries will depend upon as many free parameters\footnote{Free parameters which are essentially parameterizing the corresponding nonperturbative ambiguities.} as there are distinct instanton actions. For most of this paper, and similarly to what was found in \cite{asv11} for the one--cut quartic matrix model and the Painlev\'e I equation \cite{gikm10}, a two--parameter transseries will be sufficient to describe the two--cut quartic matrix model and the Painlev\'e II equation. These two--parameter transseries generalize the one--parameter cases which were first introduced in the matrix model context in \cite{m08}.

Similarly to what happened in \cite{asv11}, we shall only need to consider the special case of a two--parameter transseries \textit{ansatz} with instanton action $A$ and ``generalized instanton'' action $-A$. This may be written as
\be\label{transseries}
F (z, \sigma_1, \sigma_2) = \sum_{n=0}^{+\infty} \sum_{m=0}^{+\infty} \sigma_1^n \sigma_2^m F^{(n|m)} (z),
\ee
\noindent
where $z$ is the coupling parameter (here chosen $\sim 1/g_s$) and $\sigma_1, \sigma_2$ are the transseries parameters. Further, the above $(n|m)$ sectors label generalized multi--instanton contributions of the form
\be\label{expnm}
F^{(n|m)}(z) \equiv \rme^{-(n-m) A z}\, \Phi_{(n|m)}(z) \simeq \rme^{-(n-m) A z}\, \sum_{g=1}^{+\infty} \frac{F^{(n|m)}_g}{z^{g + \beta_{nm}}}. 
\ee
\noindent
In this expression $\beta_{nm}$ is a characteristic exponent, to which we shall later return when needed. Resurgent transseries are defined along wedges in the complex $z$--plane (upon Borel resummation, see, \textit{e.g.}, \cite{asv11} for details) and they are ``glued'' along Stokes lines in order to construct the full analytic solution. This ``gluing'' is achieved via the Stokes automorphism $\underline{\frak{S}}_\theta$ which essentially acts upon the transseries \eqref{transseries} by shifting its parameters. For instance, given a one--parameter transseries with Stokes line on the positive real axis, the gluing is achieved by shifting $\sigma \to \sigma + S_1$ as one crosses from the upper to the lower positive--half--plane, where $S_1$ is the Stokes constant associated to that particular Stokes line---although, generically, there may be an infinite number of distinct Stokes constants. In our two--parameter case, there are two sets of Stokes coefficients, $S^{(k)}_\ell$ and $\widetilde{S}^{(k)}_\ell$, labeled by integers $k$ and $\ell$ with $k \ge 0$. Do notice that not all of these coefficients are independent and in \cite{asv11} some empirical relations between them have been found, in the Painlev\'e I context. We refer the reader to that reference for further details.

The main point of interest to us in this subsection concerns large--order analysis \cite{z81}, and how resurgent analysis improves it \cite{gikm10, asv11}. Recall that if a given function $F(z)$ has a branch--cut in the complex plane along some direction $\theta$, being analytic elsewhere, then
\be\label{Cauchy}
F(z) = \frac{1}{2\pi\rmi} \int_0^{\rme^{\rmi\theta} \cdot \infty} \rmd w\, \frac{\disc_\theta\, F(w)}{w-z},
\ee
\noindent
where we have assumed that there is no contribution arising from infinity \cite{z81, bw73}. The key point now is that the aforementioned Stokes automorphism $\underline{\frak{S}}_\theta$, which may be expressed as a multi--instanton expansion \cite{asv11}, relates to this branch--cut discontinuity as
\be
\underline{\frak{S}}_\theta = \1 - \disc_\theta,
\ee
\noindent
in such a way that the discontinuity itself may be written in terms of multi--instanton data. For instance, starting with the perturbative sector, it was shown in \cite{asv11} that in the two--parameter transseries set--up \eqref{transseries} there will be two branch--cuts, along $\theta=0$ and $\theta=\pi$, such that with $\beta_{00} = 0$ one finds
\bea
\disc_{0}\, \Phi_{(0|0)} &=& - \sum_{k=1}^{+\infty} \left( S_{1}^{(0)} \right)^{k} \rme^{-kAz}\, \Phi_{(k|0)}, \\
\disc_{\pi}\, \Phi_{(0|0)} &=& - \sum_{k=1}^{+\infty} \left( \widetilde{S}_{-1}^{(0)} \right)^{k} \rme^{kAz}\, \Phi_{(0|k)}.
\eea
\noindent
Using \eqref{expnm} and \eqref{Cauchy}, we then find the perturbative asymptotic coefficients to be given by \cite{asv11}
\bea
F_g^{(0|0)} &\simeq& \sum_{k=1}^{+\infty} \frac{\left( S_1^{(0)} \right)^k}{2\pi\rmi}\, \frac{\Gamma \left( g-\beta_{k,0} \right)}{\left( k A \right)^{g-\beta_{k,0}}}\, \sum_{h=1}^{+\infty} \frac{\Gamma \left( g-\beta_{k,0}-h+1 \right)}{\Gamma \left( g-\beta_{k,0} \right)}\, F_h^{(k|0)} \left( k A \right)^{h-1} + \nonumber \\
&&
+ \sum_{k=1}^{+\infty} \frac{\left( \widetilde{S}_{-1}^{(0)} \right)^k}{2\pi\rmi}\, \frac{\Gamma \left( g-\beta_{0,k} \right)}{\left( - k A \right)^{g-\beta_{0,k}}}\, \sum_{h=1}^{+\infty} \frac{\Gamma \left( g-\beta_{0,k}-h+1 \right)}{\Gamma \left( g-\beta_{0,k} \right)}\, F_h^{(0|k)} \left( - k A \right)^{h-1}.
\label{exp00}
\eea
\noindent
What this expression shows is that the asymptotic coefficients of the perturbative sector, for large $g$, are precisely controled by the coefficients of the (generalized) multi--instanton sectors, $(n|0)$ and $(0|n)$. Of course that besides the coefficients $F_g^{(n|0)}$ and $F_g^{(0|n)}$, the perturbative coefficients also depend on the two Stokes constants, $S_1^{(0)}$ and $\widetilde{S}_{-1}^{(0)}$, and these still need to be determined. For the moment, let us just note that the leading large--order growth is dictated by the Stokes constants and the one--loop (generalized) one--instanton coefficients $F_1^{(1|0)}$ and $F_1^{(0|1)}$. Higher loop coefficients in the $(1|0)$ and $(0|1)$ sectors yield corrections in $1/g$, whereas the higher $(n|0)$ and $(0|n)$ sectors yield corrections which are suppressed as $1/n^{g}$.

As we turn to the models of interest to us in the present work---such as matrix models or topological strings---there are a few extra points to consider. First, the perturbative sector \eqref{topF} is given by a topological genus expansion, in powers of $2g-2$, where the string coupling is $z=1/g_s$. Secondly, as we address matrix models or topological strings, one needs to consider a version of the multi--instanton sectors \eqref{expnm} where both the action $A$ and the perturbative coefficients $F_g^{(n|m)}$ become functions of the partial 't~Hooft moduli (or geometric moduli) $t_I$. But, more importantly, due to resonance effects which will later appear in either the quartic matrix model or the Painlev\'e II equation, one also needs to consider the inclusion of logarithmic sectors as \cite{gikm10, asv11}:
\be\label{expnmstring}
F^{(n|m)} (g_s, \{ t_I \}) \simeq \rme^{-(n-m) \frac{A (t_I)}{g_{s}}}\, \sum_{k=0}^{k_{nm}}\log^{k} g_{s} \sum_{g=0}^{+\infty} g_{s}^{g+\beta_{nm}^{[k]}}\, F_{g}^{(n|m)[k]} (t_I) \equiv \rme^{-(n-m) \frac{A (t_I)}{g_s}}\, \Phi_{(n|m)} (g_s, \{ t_I \}).
\ee
\noindent
We shall later uncover that the maximum logarithmic power is $k_{nm} = k_{mn} = \min(n,m) - m\, \delta_{nm}$ and that $\beta_{nm}^{[k]} = \beta_{mn}^{[k]} = \beta ( m+n ) - \left[ ( k_{nm}+k ) / 2 \right]_{I}$, where $[\bullet]_{I}$ is the integer part of the argument and where $\beta=1/2$. In practice, this essentially means that all we have done up to now was for the $k=0$ ``sector'', and that the $\beta_{nm}^{[k]}$ coefficients take into account the fact that the perturbative expansions may in fact begin at some negative integer. Going back to the perturbative $(0|0)$ sector in \eqref{topF}, we know that $F^{(0|0)}$ is given by a genus expansion containing only powers of the closed\footnote{One has to be a bit careful with the precise meaning of the labels: in full rigor, the coefficients $F_g$ in \eqref{topF} precisely stand for $F_{2g}^{(0|0)}$ in the present transseries language, as can be seen by comparing against \eqref{expnm}.} string coupling $g_s^2$. Thus, one needs to impose $F_{2n+1}^{(0|0)} = 0$ in \eqref{exp00}, which will produce a series of relations between the $(0|k)$ and $(k|0)$ contributions since its right--hand--side must vanish order by order in both $\frac{1}{g}$ and $k^{-g}$. As further explained in \cite{asv11}, in the end we find that for all $k$ and $g$,
\be\label{condtop00}
\left( S_{1}^{(0)} \right)^{k} F_{g}^{(k|0)[0]} = (-1)^{g+\beta_{0,k}^{[0]}} \left( \widetilde{S}_{-1}^{(0)} \right)^{k} F_{g}^{(0|k)[0]}. 
\ee
\noindent
When working out the full details of either the two--cut quartic matrix model or the Painlev\'{e} II equation, we shall find further relations between different coefficients $F_{g}^{(n|m)[k]}$, either when $m$ and $n$ are exchanged, or relating the $k \neq 0$ coefficients to the $k = 0$ coefficients. In some cases, these will imply further relations between different Stokes constants.

Finally, using the above relations \eqref{condtop00} back in the large--order formula for the perturbative sector \eqref{exp00}, we obtain the asymptotic large--order behavior of the perturbative coefficients in the string genus expansion \eqref{topF} as
\be\label{F00largeorder}
F_{2g}^{(0|0)} \simeq \sum_{k=1}^{+\infty} \frac{\left( S_{1}^{(0)} \right)^{k}}{\rmi\pi}\, \frac{\Gamma \left(2g-\beta_{k,0}^{[0]} \right)}{\left( k A \right)^{2g-\beta_{k,0}^{[0]}}}\, \sum_{h=0}^{+\infty} \frac{\Gamma \left(2g-h-\beta_{k,0}^{[0]} \right)}{\Gamma \left(2g-\beta_{k,0}^{[0]} \right)}\, F_{h}^{(k|0)[0]} \left( k A \right)^{h}.
\ee

This procedure may be extended in order to find the large--order behavior of all (generalized) multi--instanton sectors. In particular, we are here interested in the large--order behavior of the physical multi--instanton series $F^{(n|0)}$. The precise calculation is a bit more cumbersome due to the logarithmic sectors appearing in \eqref{expnmstring}, and we refer the reader to \cite{asv11} for full details. The final result is
\bea
F_{g}^{(n|0)[0]} &\simeq& \sum_{k=1}^{+\infty} \binom{n+k}{n}\, \frac{(S_{1}^{(0)})^{k}}{2\pi\rmi}\, \frac{\Gamma (g+\beta_{n,0}^{[0]}-\beta_{n+k,0}^{[0]})}{\left( k A \right)^{g+\beta_{n,0}^{[0]}-\beta_{n+k,0}^{[0]}}}\, \sum_{h=1}^{+\infty} \frac{\Gamma (g+\beta_{n,0}^{[0]}-\beta_{n+k,0}^{[0]}-h)}{\Gamma (g+\beta_{n,0}^{[0]}-\beta_{n+k,0}^{[0]})}\, F_{h}^{(n+k|0)[0]} \left( k A \right)^{h} \nonumber \\
&&
\hspace{-10pt}
+ \sum_{k=1}^{+\infty} \left\{ \frac{1}{2\pi\rmi} \sum_{m=1}^{k} \frac{1}{m!}\, \sum_{\ell=0}^{m}\,\sum_{\gamma_{i}\in\Gamma(m,k)}\, \sum_{\delta_{j}\in\Gamma(m,m-\ell+1)} \left( \prod_{j=1}^{m} \Sigma(n,j) \right) \right\} \times \nonumber \\
&&
\hspace{-10pt}
\times\, \sum_{r=0}^{k_{n+\ell-k,\ell}} \frac{\Gamma (g+\beta_{n,0}^{[0]}-\beta_{n+\ell-k,\ell}^{[r]})}{\left( - k A \right)^{g+\beta_{n,0}^{[0]}-\beta_{n+\ell-k,\ell}^{[r]}}}\, \sum_{h=0}^{+\infty} \frac{\Gamma (g+\beta_{n,0}^{[0]}-\beta_{n+\ell-k,\ell}^{[r]}-h)}{\Gamma (g+\beta_{n,0}^{[0]}-\beta_{n+\ell-k,\ell}^{[r]})}\, F_{h}^{(n+\ell-k|\ell)[r]} \left( - k A \right)^{h} \times \nonumber \\
&&
\hspace{-10pt}
\times \left.\left\{ \delta_{r0} + \Theta(r-1)\, \Big( B_{kA}(a) + \partial_{a} \Big)^{r-1} B_{kA}(a) \right\} \right|_{a=g+\beta_{n,0}^{[0]}-\beta_{n+\ell-k,\ell}^{[r]}-h-1} \,.
\label{asymn0}
\eea
\noindent
Let us define the many ingredients in this expression (but, again, we refer the reader to \cite{asv11} for the full details). The sums over $\gamma_i$ and $\delta_j$ are sums over Young diagrams, where a diagram $\gamma_i \in \Gamma(k,\ell) :\, 0 < \gamma_{1} \le \cdots \le \gamma_{k} = \ell$ has length $\ell(\Gamma) = k$, and where the maximum number of boxes for each $\gamma_i$ is $\ell(\Gamma^T) = \ell$. The sum over $\delta_s$ is similar, now with $0 < \delta_1 \leq \delta_2 \leq \cdots \leq \delta_k = k - m + 1$ and $0 < \delta_s \leq s+1$. These $\delta_s$ form a diagram $\Gamma(k,k-m+1)$ that has length $\ell(\Gamma) = k$ and $\ell(\Gamma^T) = k-m+1$, with an extra condition that each component $\delta_s \in \Gamma(k,k-m+1)$ has at most $s+1$ boxes. For these definitions to be consistent we still have to set $\gamma_0 \equiv 0$, $\delta_0 \equiv 1$, as well as the Stokes constants $S_0^{(k)} = \widetilde{S}_0^{(k)} = S_{-k}^{(k)} \equiv 0$. Next, defining $\mathbf{d}\gamma_{j} \equiv \gamma_j - \gamma_{j-1}$, and similarly for $\mathbf{d}\delta_{s}$, one has
\be 
\Sigma(n,j) \equiv \left( \left( j+1-\delta_{j} \right) \widetilde{S}_{-\mathbf{d}\gamma_{j}}^{(\mathbf{d}\delta_{j})} + \left( n-\gamma_{j}+j+1-\delta_{j} \right) S_{-\mathbf{d}\gamma_{j}}^{(\mathbf{d}\gamma_{j}+\mathbf{d}\delta_{j})} \right) \Theta \left( j+1-\delta_{j} \right),
\ee
\noindent
where $\Theta (x)$ is the Heaviside step--function. Finally, we have introduced the function
\be
B_{s} (a) \equiv \psi ( a+1 ) - \log ( -s ) \equiv \widetilde{B}_{s}(a) - \rmi\pi,
\ee
\noindent
with $\psi (z) = \frac{\Gamma' (z)}{\Gamma (z)}$ the digamma function.

There are a few relevant features to be found in \eqref{asymn0}. Besides the multitude of (generalized) multi--instanton sectors and Stokes constants that now play a role, there is also a new type of large--order effect. In fact, and unlike the usual perturbative case which had a leading large--order growth of $g!$, essentially  arising from the gamma function dependence, we now find a large--order growth of the type $g! \log g$, arising from the digamma function, and this is actually a \textit{leading} effect as compared to the $g!$ growth. The first signs of this effect were found in \cite{gikm10}, in the context of the Painlev\'{e} I equation, and further studied in \cite{asv11}.

\section{Multi--Instanton Analysis for $\mathbb{Z}_2$--Symmetric Systems}\label{sec:inst}

Having reviewed the main background ingredients required for our analysis, we may now proceed with our main goal and address the nonperturbative study of Stokes phases associated to multi--cut configurations. These phases arise when all cuts are equally filled and, to be very concrete and present fully explicit formulae, we shall next focus on two--cut set--ups (see \cite{msw08} as well). In this case, equal filling also implies that the configuration is $\mathbb{Z}_2$--symmetric. As we shall see in detail throughout this section, this symmetry implies that hyperelliptic integrals which appear in the calculation will reduce to elliptic integrals, and that, physically, the system will be found in a Stokes phase. Notice that, strictly within the orthogonal polynomial framework, it was already noticed in \cite{l92} that equal filling of the cuts would lead to a Stokes phase.

\subsection{Computing the Multi--Instanton Sectors}\label{sec:oneinst}

Let us begin by considering the multi--instanton sectors of a two--cut matrix model. We shall do this by following the strategy in \cite{msw08}, \textit{i.e.}, we shall consider the two--cut spectral geometry as a degeneration from a three--cut configuration. In principle one could also consider degenerations from more complicated configurations if one were to introduce several distinct instanton actions, but for our purposes degenerations from three cuts will suffice. In this case, a reference filling of eigenvalues across the cuts is of the form $(N_1, N_2, N_3)$, with $N_1+N_2+N_3=N$, the degeneration will simply be $N_2 \to 0$, and the $\mathbb{Z}_2$ symmetry will eventually demand $N_1=N_3$.

\FIGURE[ht]{
\label{3cutsmultiinst}
\centering
\includegraphics[width=9.5cm]{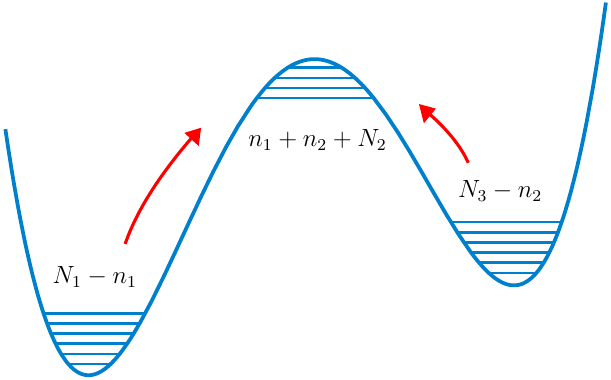}
\caption{Eigenvalue tunneling as the multi--instanton sectors of a three--cut matrix model.}
}

In matrix models, (multi) instantons are associated to (multiple) eigenvalue tunneling \cite{d91, d92, msw07, msw08} and, as such, the multi--instanton sectors are described by tunneling eigenvalues in--between the three cuts, as shown in figure \ref{3cutsmultiinst} (it is simple to see that two integers, $n_1$ and $n_2$, are enough to parameterize all possible exchanges of eigenvalues between three cuts, \textit{i.e.}, all possible background choices). In the particular case of the $\mathbb{Z}_2$--symmetric two--cut configuration, the reference background is of the form
\be
\left( \frac{N}{2}, 0, \frac{N}{2} \right).
\ee
\noindent
As we shall see later on, the one--instanton sector will correspond to summing over all configurations which leave a single eigenvalue on the middle--cut. From the spectral geometry viewpoint, the $\mathbb{Z}_2$ symmetry essentially places the cuts at $[-b, -a] \cup [a, b]$ and the spectral curve \eqref{sp_curve} becomes
\be
y(z) = M(z) \sqrt{\left( z^2-a^2 \right) \left( z^2-b^2 \right)},
\ee
\noindent
where $M(z)$ is given by \eqref{M multicut}. In this configuration, the pinching cycle will be found at $z=0$. The action associated to eigenvalue tunneling essentially measures their energy difference in--between cuts \cite{d91, d92, msw07, msw08}, as given by the holomorphic effective potential \eqref{hol_eff_pot}, and in the particular case of this $\mathbb{Z}_2$--symmetric configuration with equal filling it is simple to check that the equal filling essentially translates to
\be
\int_{-a}^{a} \rmd z\, y(z) = 0.
\ee
\noindent
This condition will further imply that one may completely evaluate all data in the spectral geometry just by using the asymptotics of the resolvent \eqref{asympcond}. One is left with one instanton action to evaluate, describing tunneling from each of the (equal) cuts up to the pinched cycle\footnote{This is the non--trivial saddle located outside the cut, where eigenvalues may tunnel to \cite{msw07}.} located at $x_0$ such that $M(x_0) = 0$ \cite{msw07}. In here $x_0=0$ and
\be\label{instantonaction0a}
A = \int_a^0 \rmd z\, y(z).
\ee

Having briefly explained the set--up, one may proceed and compute the partition functions associated to the relevant configurations along the lines in \cite{msw08}. Let now $y(z)$ be the spectral curve \eqref{sp_curve} of the \textit{three}--cut configuration, with cuts $[x_1, x_2] \cup [x_3, x_4] \cup [x_5, x_6]$. Let us consider the aforementioned set--up with $N_1 - n_1$, $N_2 + (n_1 + n_2)$ and $N_3 - n_2$ eigenvalues in the first, second and third cuts, respectively, and let us consider the associated multi--instanton amplitude written in terms of the 't~Hooft moduli \eqref{t_I}
\be\label{Zn1n2}
Z^{(n_1, n_2)} \equiv \frac{Z \left( t_1 - n_1 g_s, t_2 + n_1 g_s + n_2 g_s, t_3 - n_2 g_s \right)}{Z \left( t_1, t_2, t_3 \right)},
\ee
\noindent
with $t_1 + t_2 + t_3 = t$. For convenience we introduce the variables
\bea 
\label{s_ia}
s_1 &=& \frac{1}{2} \left( t_1 - t_2 - t_3 \right), \\
s_2 &=& \frac{1}{2} \left( t_3 - t_2 - t_1 \right),
\label{s_ib}
\eea
\noindent
and use them to expand the exponent of \eqref{Zn1n2} above (\textit{i.e.}, the difference of free energies between the ``eigenvalue--shifted'' configuration and the reference background), around $g_s = 0$ and for $n_1, n_2 \ll N$. One simply finds\footnote{For shortness we shall many times omit the arguments; it should be clear that whenever we write $F_0$ we always mean the reference configuration $F_0(t_1, t_2, t_3)$, and similarly in other cases.}
\be\label{Zn1n2F}
Z^{(n_1, n_2)} = \exp \left( - \frac{1}{g_s} \sum_{i=1}^2 n_i\, \partial_{s_i} F_0 \right) \exp \left( \frac{1}{2} \sum_{i,j=1}^2 n_i n_j\, \partial_{s_i} \partial_{s_j} F_0 \right) \bigg\{ 1 + \mathcal{O} (g_s) \bigg\}.
\ee
\noindent
In this expression we find two, in general different, actions
\be 
A_i = \partial_{s_i} F_0, \qquad i=1,2,
\ee
\noindent
which may be computed in terms of geometric data if we use the special geometry relations
\be\label{specialgeo}
\frac{\partial F_0}{\partial t_I} = \oint_{B^I} \rmd z \, y(z).
\ee
\noindent
In the present three--cut configuration, the two actions are then given by
\bea 
A_1 = \frac{\partial F_0}{\partial s_1} &=& \int_{x_2}^{x_3} \rmd z \, y(z), \\
A_2 = \frac{\partial F_0}{\partial s_2} &=& \int_{x_5}^{x_4} \rmd z \, y(z),
\eea
\noindent
and they have the usual geometric interpretation appearing in figure \ref{spectralcurve}, generalizing the one--cut case appearing in \cite{msw07, msw08}. 
The extension to an arbitrary number of cuts is straightforward. The other feature we find in \eqref{Zn1n2F} are the second derivatives of $F_0$, and for those it is convenient to introduce the (symmetric) period matrix
\be\label{matrixtau}
\tau_{ij} \equiv \frac{1}{2\pi\rmi}\, \frac{\partial^2 F_0}{\partial s_i \partial s_j}.
\ee

\FIGURE[ht]{
\label{spectralcurve}
\centering
\includegraphics[width=12cm]{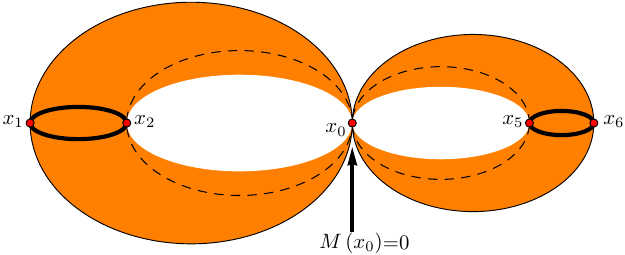}
\caption{The two--cut spectral curve $y(z)$ is a genus one curve, with a pinched cycle at the non--trivial saddle $x_0$ which is obtained by taking $x_3 \to x_4$ in the three--cut geometry (in the $\BZ_2$ symmetric scenario $x_0 = 0$). The instanton actions $A_1$ and $A_2$ naturally appear as $B$--cycles in this spectral geometry.}
}

Having understood the general form of the multi--instanton amplitudes, we still need to understand the precise nature of the multi--instanton expansion. The grand--canonical partition function is obtained as a sum over all possible eigenvalue distributions into the multiple cuts, with their total number fixed. In our case, and making use of the multi--instanton amplitudes \eqref{Zn1n2}, this translates to
\be\label{Zgeneral}
Z(N) = \sum_{n_1 = - N_2 + N_3}^{N_1} \sum_{n_2 = -n_1 - N_2}^{N_3} Z^{(n_1, n_2)}. 
\ee
\noindent
Let us now consider the reference background of interest to us, \textit{i.e.}, the $\mathbb{Z}_2$--symmetric two--cut configuration describing a multi--cut Stokes phase. This background has moduli $t_1 = t_3 = \frac{t}{2}$ and $t_2 = 0$, in which case both instanton actions will be equal $A_1 = A_2 \equiv A$, as well as $\tau_{11}=\tau_{22}$. Changing variables from $n_1$ and $n_2$ to $\ell = n_1 + n_2$ and $m = n_1 - n_2$, we may write the multi--instanton amplitudes \eqref{Zn1n2F} as
\be
Z^{(\ell, m)} = \exp \left(- \frac{\ell A}{g_s} \right) \exp \left( \frac{\rmi\pi}{2} \left( \tau_{11} + \tau_{12} \right) \ell^2 \right) \exp \left( \frac{\rmi\pi}{2} \left( \tau_{11} - \tau_{12} \right) m^2 \right) \bigg\{ 1 + \mathcal{O} (g_s) \bigg\},
\ee
\noindent
where it now becomes clear that it is $\ell = n_1 + n_2 \geq 0$ which will label the multi--instanton sectors. Of course this further implies that we still need to sum over the ``relative'' index $m$ in order to obtain the ``purely'' $\ell$--instanton amplitude: it is this sum over $m$ which essentially moves our calculation to the grand--canonical ensemble. In other words, the grand--canonical partition function \eqref{Zgeneral} is of the schematic form
\be\label{Zgrandextra}
Z(N) = {\cal Z}^{(\ell=0)} + {\cal Z}^{(\ell=1)} + {\cal Z}^{(\ell=2)} + \cdots = {\cal Z}^{(\ell=0)} \left( 1 + \frac{{\cal Z}^{(\ell=1)}}{{\cal Z}^{(\ell=0)}} + \frac{{\cal Z}^{(\ell=2)}}{{\cal Z}^{(\ell=0)}} + \dots \right),
\ee
\noindent
where now each term ${\cal Z}^{(\ell)}$ contains a sum over all possible values of $m = n_1 - n_2$ that satisfy $n_1 + n_2 = \ell$. Fixing $\ell$ eigenvalues on the middle--cut implies that we only have $N-\ell$ available eigenvalues to place in each of the two side--cuts, which yields the limits on the $m$--sum. But because $m$ jumps by values of two, it turns out that it is actually more convenient to change variables and use as the ``relative'' index $m=2r-\ell$. Overall, we find
\be\label{Zlsum}
{\cal Z}^{(\ell)} = \exp \left(- \frac{\ell A}{g_s} \right) \exp \left( \frac{\rmi\pi}{2} \left( \tau_{11} + \tau_{12} \right) \ell^2 \right) \sum_{r= - N/2 + \ell}^{N/2} \exp \left( \frac{\rmi\pi}{2} \left( \tau_{11} - \tau_{12} \right) \left( 2r - \ell \right)^2 \right) \bigg\{ 1 + \mathcal{O}(g_s) \bigg\}. 
\ee
\noindent
With a certain abuse of notation, we shall immediately identify the $\ell$--th instanton amplitude as
\be\label{linst}
Z^{(\ell)} = \frac{{\cal Z}^{(\ell)}}{{\cal Z}^{(0)}},
\ee
\noindent
where all that is now missing is the explicit evaluation of the many different ingredients which appear above, in particular explicitly evaluating the sum.

Let us begin by addressing the period matrix \eqref{matrixtau}, \textit{i.e.}, the second derivatives of the planar free energy. Using the special geometry relation \eqref{specialgeo} and the explicit form of the spectral curve \eqref{sp_curve}, it follows that
\be\label{d2F0dsidsj}
\frac{\partial^2 F_0}{\partial s_i \partial s_j} = (-1)^{j+1} \int_{x_{2j}}^{x_{2j+1}} \rmd z\, (-2) \frac{\partial (t \omega_0(z))}{\partial s_i},
\ee
\noindent
where the derivative of the resolvent has the form\footnote{In order to check this relation one explicitly uses \eqref{sp_curve} and \eqref{M multicut} when taking derivatives, and this will yield the polynomial structure in $z$. In order to fix the degree of this polynomial, one compares the asymptotics as $z \to + \infty$ on both sides of the equation. Generically, the degree will depend on the number of cuts as $s-2$.}
\be 
\frac{\partial (t \omega_0(z))}{\partial s_i} = \frac{C_0^{(i)}(t, s_k) + C_1^{(i)}(t, s_k)\, z}{\sqrt{\sigma_3 (z)}}.
\ee
\noindent
The coefficients which appear in this expression, $C_0^{(i)}(t, s_k)$ and $C_1^{(i)}(t, s_k)$, may be fixed by taking derivatives of the partial 't~Hooft moduli \eqref{t_I}, and by using the definition of the variables $\{ s_i \}$, \eqref{s_ia} and \eqref{s_ib}, as
\be\label{dti}
\frac{\partial t_I}{\partial s_i} = \begin{pmatrix} +1 \\ 0 \\ -1 \end{pmatrix} = - \frac{1}{2\pi\rmi} \oint_{A^I} \rmd z\, \frac{C_0^{(i)}(t, s_k) + C_1^{(i)}(t, s_k)\, z}{\sqrt{\sigma_3 (z)}}, \qquad i=1,2, \quad I=1,2,3.
\ee
\noindent
Note that although this relation corresponds to a system of $6$ equations for $4$ unknowns, two of the equations are redundant as we can deform contours in order to find $\sum_I \oint_{A^I} = 0$ (there is no residue at infinity). If we now define the integrals
\be
\mathcal{K}_I = \oint_{A^I} \frac{\rmd z}{2\pi\rmi}\, \frac{1}{\sqrt{\sigma_3 (z)}} \quad \text{and} \quad \mathcal{L}_I =  \oint_{A^I} \frac{\rmd z}{2\pi\rmi}\, \frac{z}{\sqrt{\sigma_3 (z)}},
\ee
\noindent
then we can express all the coefficients $C_j^{(i)}$ in terms of these integrals as
\bea 
&& C_0^{(1)} = \frac{\mathcal{L}_1 + \mathcal{L}_2}{\mathcal{L}_1 \mathcal{K}_2 - \mathcal{L}_2 \mathcal{K}_1}, \qquad C_0^{(2)} = \frac{\mathcal{L}_2 + \mathcal{L}_3}{\mathcal{L}_3 \mathcal{K}_2 - \mathcal{L}_2 \mathcal{K}_3}, \\
&& C_1^{(1)} = \frac{\mathcal{K}_1 + \mathcal{K}_2}{\mathcal{L}_2 \mathcal{K}_1 - \mathcal{L}_1 \mathcal{K}_2}, \qquad C_1^{(2)} = \frac{\mathcal{K}_2 + \mathcal{K}_3}{\mathcal{L}_2 \mathcal{K}_3 - \mathcal{L}_3 \mathcal{K}_2}.
\eea
\noindent
So far these results are only formal: hyperelliptic integrals are hard to evaluate. However, they may in fact be explicitly evaluated when one imposes $\mathbb{Z}_2$ symmetry into the problem. In this case, one places the cuts as $[-b, -a] \cup [-c, c] \cup [a, b]$ (where we shall later be interested in the $c \to 0$ degeneration) and it immediately follows that
\bea
&& \mathcal{K}_1 = \mathcal{K}_3 = -\frac{1}{2} \mathcal{K}_2 \equiv -\mathcal{K}, \\
&& \mathcal{L}_1 = -\mathcal{L}_3 \equiv -\mathcal{L}, \quad \mathcal{L}_2 = 0,
\eea
\noindent
leading to the (simplified) coefficients
\bea 
C_0^{(1)} &=& C_0^{(2)} = \frac{1}{2 \mathcal{K}}, \\
-C_1^{(1)} &=& C_1^{(2)} = \frac{1}{2 \mathcal{L}}.
\eea
\noindent
As they will be needed in the following, let us also introduce the $B$--cycle integrals:
\bea
\label{kappatildeextra}
\mathcal{\widetilde{K}} &\equiv& \int_{-a}^{-c} \frac{\rmd z}{\sqrt{\sigma_3 (z)}} = \int_c^a \frac{\rmd z}{\sqrt{\sigma_3 (z)}}, \\
\mathcal{\widetilde{L}} &\equiv& -\int_{-a}^{-c} \rmd z\, \frac{z}{\sqrt{\sigma_3 (z)}} =  \int_c^a \rmd z\, \frac{z}{\sqrt{\sigma_3 (z)}}.
\label{elletildeextra}
\eea
\noindent
All these $A$ and $B$--cycle integrals may be explicitly evaluated, and expressed in terms of complete \textit{elliptic} integrals of the first kind, $K(k^2)$, with $k$ being the elliptic modulus. This is also the technical reason why one may find Stokes phases within multi--cut configurations: symmetries (in this case a $\mathbb{Z}_2$ symmetry) may effectively reduce hyperelliptic geometries to elliptic ones! The results are
\bea 
\mathcal{K} &=& - \int_a^b \frac{\rmd x}{\pi}\, \frac{1}{\sqrt{\left| \sigma_3 (x) \right|}} = - \frac{1}{\pi b \sqrt{a^2-c^2}}\, K \left( \frac{c^2 \left( b^2 - a^2 \right)}{b^2 \left( c^2 - a^2 \right)} \right), \\
\mathcal{L} &=& \int_{-b}^{-a} \frac{\rmd x}{\pi}\, \frac{x}{\sqrt{\left| \sigma_3 (x) \right|}} =- \frac{1}{\pi \sqrt{a^2 - c^2}}\, K \left( \frac{b^2 - a^2}{c^2 - a^2} \right),
\eea
\noindent
and, for \eqref{kappatildeextra} and \eqref{elletildeextra},
\bea
\mathcal{\widetilde{K}} &=& \frac{1}{a \sqrt{b^2 - c^2}}\, K \left( \frac{b^2 \left( c^2 - a^2 \right)}{a^2 \left( c^2 - b^2 \right)} \right), \\
\mathcal{\widetilde{L}} &=&  \frac{1}{\sqrt{b^2 - a^2}}\, K \left( \frac{a^2 - c^2}{a^2 - b^2} \right).
\eea

Having explicitly evaluated all integrals, we may now start assembling these results back into our original formulae and address the degeneration limit $c \to 0$. In order to do that, it is first important to notice that this limit must be taken carefully as the free energy is not analytic in the 't~Hooft modulus associated to the shrinking cycle \cite{msw08}. This may be explicitly seen by splitting the free energies as
\be
F_g ( t_1, t_2, t_3 ) = F_g^{\text{G}} (t_2) + \widehat{F}_g ( t_1, t_2, t_3 ),
\ee
\noindent
where $F_g^{\text{G}} (t_2)$ are the genus $g$ free energies of the Gaussian model depending on the vanishing 't~Hooft modulus, which, at genus $g=0$ and $g=1$, have a dependence as $\log t_2$. As explained in \cite{msw08}, for the $\ell$--instanton sector it is not appropriate to look at the integration over the $\ell$ eigenvalues in the collapsing cycle as a large $N$ approximation; rather one should \textit{exactly} evaluate the Gaussian partition function associated to this cycle, which is
\be 
Z_{\ell}^{\mathrm{G}} = \frac{g_s^{\ell^2/2}}{(2\pi)^{\ell/2}}\, G_2 \left( \ell + 1 \right)
\ee
\noindent
with $G_2 \left( \ell + 1 \right)$ the Barnes function. Then, the partition function around the $\ell$--instanton configuration should be properly written as
\be
Z^{(\ell)} = Z_{\ell}^{\mathrm{G}}\, \widehat{Z}^{(\ell)},
\ee
\noindent
where all ``hatted'' quantities in $\widehat{Z}^{(\ell)}$ are now regularized and analytic in the $t_2 \to 0$ limit.

The instanton action is the simplest quantity to evaluate as it is in fact regular in the $t_2 \to 0$ limit. One simply finds
\be\label{action_sp}
\widehat{A} = \int_a^c \rmd z\, \widetilde{M}(z) \sqrt{(z^2 - a^2)(z^2 - b^2)(z^2 - c^2)} \xrightarrow[c \to 0]{} \int_a^0 \rmd z\, M(z) \sqrt{(z^2 - a^2)(z^2 - b^2)},
\ee
\noindent
where $M(z) = z \widetilde{M}(z)$. To compute the period matrix we must first address the second derivatives of the planar free energy, \eqref{d2F0dsidsj}, which are given by
\be
\partial^2_{s_1} F_0 \equiv \partial^2_{s_2} F_0 = \frac{1}{\mathcal{K}} \int_c^a \frac{\rmd z}{\sqrt{\sigma_3 (z)}} + \frac{1}{\mathcal{L}} \int_c^a \rmd z\, \frac{z}{\sqrt{\sigma_3 (z)}} = \frac{\widetilde{\mathcal{K}}}{\mathcal{K}} + \frac{\widetilde{\mathcal{L}}}{\mathcal{L}},
\ee
\noindent
and by
\be 
\partial_{s_1} \partial_{s_2} F_0 = \frac{\widetilde{\mathcal{K}}}{\mathcal{K}} - \frac{\widetilde{\mathcal{L}}}{\mathcal{L}}.
\ee
\noindent
With these results, the period matrix follows immediately. In particular we obtain
\bea 
\tau_{11} + \tau_{12} &=& - \frac{\rmi}{\pi}\, \frac{\widetilde{\mathcal{K}}}{\mathcal{K}}, \label{11+12} \\
\tau_{11} - \tau_{12} &=& - \frac{\rmi}{\pi}\, \frac{\widetilde{\mathcal{L}}}{\mathcal{L}}. \label{11-12}
\eea
\noindent
The need for regulation of the shrinking cycle is now very clean. In fact, if one takes the $c \to 0$ limit in \eqref{11+12} above one obtains
\be 
\lim_{c \to 0} \rmi \pi \left(\tau_{11} + \tau_{12} \right) \sim 2 \lim_{c \to 0} \log c + \log \left( \frac{b^2 - a^2}{16 a^2 b^2} \right) + \cdots.
\ee
\noindent
However, as explained, this logarithmic divergence---which emerges in one of the elliptic integrals---will be precisely canceled by the ``Gaussian divergence'' arising from the shrinking cycle. The regulation is simply \cite{msw08}
\be 
\partial^2_s \widehat{F}_0 = \lim_{c \to 0} \left( \partial^2_s F_0 - \log t_2 \right),
\ee
\noindent
where the vanishing 't~Hooft modulus is, via \eqref{t_I},
\be 
t_2 = \frac{1}{2 \pi} \int_{-c}^{c} \rmd z\, \widetilde{M}(z) \sqrt{(z^2 - a^2)(z^2 - b^2)(z^2 - c^2)}.
\ee
\noindent
Changing variables $z \rightarrow z/c$, expanding the result in powers of $c$ and performing the integration, it follows\footnote{In the purely three--cut scenario it is simple to check that $\widetilde{M}$ is just a constant; more on this in the following.}
\be 
t_2 = \frac{1}{4} \widetilde{M}\, a b\, c^2 \left( 1 + \mathcal{O}(c^2) \right),
\ee
\noindent
which will indeed cancel the divergence above. As for the combination \eqref{11-12}, it has a regular $c \to 0$ limit. Using known properties of elliptic integrals \cite{wolfram} one may compute
\be 
\tau_{11} - \tau_{12} = \frac{\rmi}{2}\, \frac{K(1-k^2)}{K(k^2)} \equiv \frac{\rmi}{2} \frac{K'}{K},
\ee
\noindent
where the elliptic modulus in this $\BZ_2$--symmetric limit is simply given by $k = \frac{b-a}{b+a}$.

Finally, in order to obtain the multi--instanton amplitudes \eqref{linst}, all one has to do is evaluate the sums in \eqref{Zlsum}. When $\ell = 0$, the sum in \eqref{Zlsum} yields the Jacobi (elliptic) theta--function given by
\be\label{theta}
\vartheta_3 \left( z \, | \, q \right) = 1 + 2 \sum_{r=1}^{+\infty} q^{r^2} \cos \left( 2 r z \right). 
\ee
\noindent
In fact, using this definition it is straightforward to evaluate
\be
\lim_{N \to + \infty}\, \sum_{r=-N/2}^{N/2} \exp \left( \frac{\rmi\pi}{2} \left( \tau_{11} - \tau_{12} \right) (2 r)^2 \right) = \vartheta_3 \left( 0 \left| \, \rme^{- \pi \frac{K'}{K}} \right. \right).
\ee
\noindent
When $\ell \neq 0$, and using simple properties of theta--functions \cite{wolfram}, one may obtain instead\footnote{The periodicity of the theta--function $\vartheta_3 \left( z + n\, \pi \, | \, q \right) = \vartheta_3 \left( z \, | \, q \right)$ implies that only the parity of $\ell$ is relevant.}
\be
\lim_{N \to + \infty}\, \sum_{r=-N/2+\ell}^{N/2} \exp \left( \frac{\rmi\pi}{2} \left( \tau_{11} - \tau_{12} \right) \left( 2 r - \ell \right)^2 \right) = k^{\frac{1-(-1)^{\ell}}{4}}\, \vartheta_3 \left( 0, \rme^{- \pi \frac{K'}{K}} \right).
\ee
\noindent
As we use both results above in the ratio \eqref{linst} for the $\ell$--instanton partition function, we observe the remarkable cancelation of the elliptic/theta function contribution: the only trace of their existence which remains is that the result will have a different $k$--dependence, depending on whether the instanton number is even or odd. That neither elliptic nor theta functions should be present in the final result is of course what one would have expected, when addressing a Stokes phase of a given matrix model. As such, our final result is 
\be\label{Zlresult!}
Z^{(\ell)} = \frac{g_s^{\ell^2/2}}{\left( 2\pi \right)^{\ell/2}}\, G_2 \left( \ell + 1 \right) k^{\frac{1-(-1)^{\ell}}{4}}\, \hat{q}^{\frac{\ell^2}{2}}\, \exp \left( - \frac{\ell \widehat{A}}{g_s} \right) \Big\{ 1 + \mathcal{O}(g_s) \Big\},
\ee
\noindent
where 
\be
\hat{q}^{\frac{1}{2}} \equiv \frac{\sqrt{b^2 - a^2}}{2 \sqrt{\widetilde{M}} \left( a b \right)^{3/2}}.
\ee
\noindent
In the following sections we shall test this result with great accuracy, by matching against large--order data. Besides the instanton action we shall give particular attention to testing the one--loop coefficient in the one--instanton sector (which also relates to one of the Stokes constants \cite{msw07, asv11}) which, written in terms of spectral geometry data, is very simply given by
\be\label{1inst}
S_1^{(0)} F_0^{(1|0)} = \frac{1}{2 \sqrt{2 \pi \widetilde{M}}}\, \frac{b-a}{\left( a b \right)^{3/2}}. 
\ee

\subsection{Stokes Phases and Background Independence}\label{sec:backgr}

In the previous subsection we used saddle--point analysis in order to explicitly find all multi--instanton amplitudes in a two--cut matrix model (at least to leading order in the string coupling). As we have seen, the situation with a multiple number of cuts is---as long as one can evaluate all hyperelliptic integrals---a straightforward extension from the single--cut case \cite{bde00, msw07, msw08}. Another interesting aspect of our line of work is that all these analytical results may be numerically tested to very high precision by making the match against large--order analysis; see, \textit{e.g.}, \cite{bw73, z81, m06, msw07, m08, msw08, ps09, mpp09, gikm10, kmr10, asv11}. As such, the obvious question to address now is whether obtaining large--order data for  all the (generalized) multi--instanton coefficients $F^{(n|m)}_g$ is feasible, and perhaps also a simple extension from the one--cut case. In general, this is \textit{not} the case and producing large--order data in multi--cut situations is a much harder problem; see, \textit{e.g.}, \cite{msw08, kmr10}.

While there are several approaches to constructing large--order data, in this paper we shall focus solely in the orthogonal polynomial method \cite{biz80} (more generally, the transseries approach as developed in \cite{m08, asv11}). As mentioned, in general this method is in fact \textit{not} applicable to multi--cut configurations and what we shall discuss now is how this situation changes if we focus on a given Stokes phase of our system. As we also discussed in the introduction, some of the earlier work done in the exploration of the phase spaces of matrix models with multi--welled potentials was carried out in the orthogonal polynomial framework; see, \textit{e.g.}, \cite{m88, ddjt90, j91, l92, s92, bdjt93}. Such works were mainly based on numerical computations of the recursion coefficients, $r_n$, appearing in the string equation (equation \eqref{4stringeq} in the case of the quartic model) and the main discovery concerned the appearance of multi--branch solutions at large $N$, as we illustrate in figure \ref{quartic_numerics}.

\FIGURE[ht]{
\label{quartic_numerics}
\centering
\includegraphics[width=7cm]{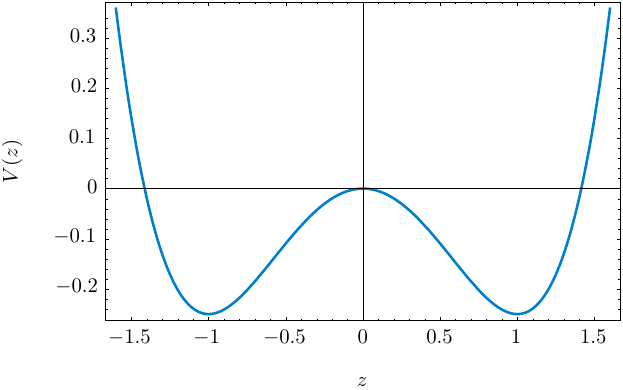}
$\qquad$
\includegraphics[width=6.75cm]{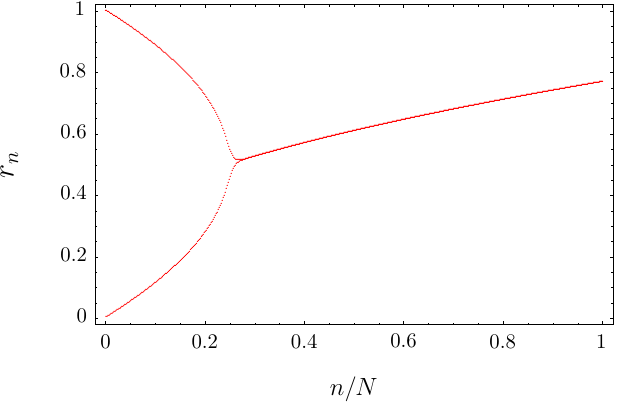}
\caption{Numerical data for the quartic potential. The first image shows the quartic potential $V(z) = \frac{\mu}{2} z^2 + \frac{\lambda}{4!} z^4$ with $\mu=-1$ and $\lambda=6$, while the second image displays the corresponding recursion coefficients $r_n$ recursively obtained from the string equation \eqref{4stringeq} (in the plots with choosing $N=1000$ and after $4000$ iterations in the numerical method described in \cite{j91, s92}).}
}

Let us consider the case of the quartic potential $V(z) = \frac{\mu}{2} z^2 + \frac{\lambda}{4!} z^4$ which, when $\mu=-1$ and $\lambda=6$, is depicted in the first image of figure \ref{quartic_numerics}. With a large $N$ choice of $N=1000$ eigenvalues, and given the string equation for this model presented in \eqref{4stringeq}, one may numerically iterate the recursion in order to compute the coefficients $r_n$ and the result is shown in the second image of figure \ref{quartic_numerics} (in here we have used the same numerical method as in \cite{j91, s92}). What this plot tells us is that, in some region of parameter space, the large $N$ behavior of the $r_n$ coefficients falls into a single branch, whereas in another region the even and odd coefficients actually split into alternating branches, with period two. As we shall show in the next section, this splitting of branches is telling us how the continuum limit should be taken in a multi--cut Stokes phase and, as such, how orthogonal polynomials may be used to generate large--order results. In other words, if the recursion coefficients have a periodic large $N$ behavior, the free energy will have a well--defined topological expansion with exponentially suppressed instanton corrections---characteristic of a Stokes phase---and orthogonal polynomials may be simply used. Furthermore, notice that the variable $n/N$ in the horizontal axis becomes the 't~Hooft parameter in the continuum limit. In this case, note that the two branches merge near $n/N = 1/4$ which in the continuum language corresponds to $\lambda t = 3/2$. This critical point actually occurs when the two cuts of the quartic matrix model collide, and at this point the system is described in the double--scaling limit by the Painlev\'e II equation. We shall have more to say about this in a later section.

\FIGURE[ht]{
\label{chaos}
\centering
\includegraphics[width=7cm]{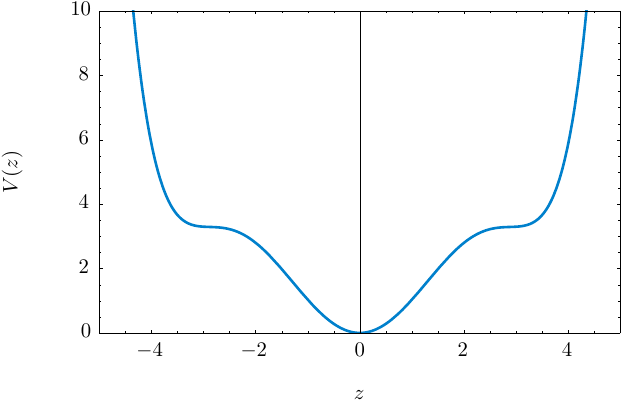}
$\qquad$
\includegraphics[width=6.75cm]{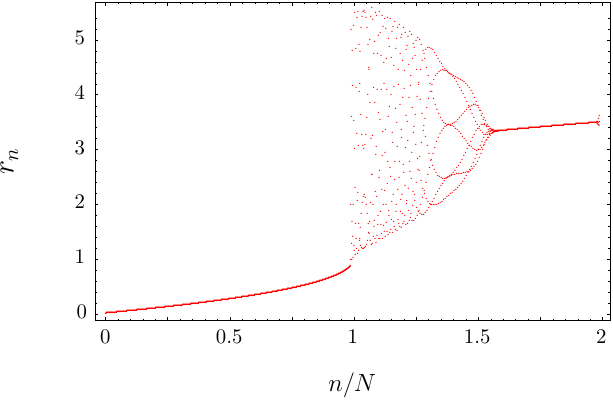}
\caption{Numerical data for a sixth--order potential. The first image shows a sixth--order potential, while the second image displays the corresponding recursion coefficients $r_n$ recursively obtained from its string equation (with $N=1000$ and after $4000$ iterations in the numerical method described in \cite{j91, s92}).}
}

It is important to distinguish the Stokes phase, where the free energy has a ``good'' large $N$ 't~Hooft expansion, from more complicated cases which may also appear as transitions occur to other phases. For instance, a different behavior is shown in figure \ref{chaos}, obtained from the string equation of a sixth--order potential. We no longer find just periodic behavior, but also regions of \textit{quasi--periodic} behavior (as shown in \cite{bde00}): this quasi--periodicity is a sign of the theta--functions which control the recursion coefficients in this phase and which appear as one constructs the grand--canonical partition function of the matrix model as a sum over all choices of filling fractions \cite{bde00}. This was recently made explicit in \cite{e08, em08}, with the construction of general, nonperturbative, background independent partition functions for matrix models and topological strings in terms of theta functions. In this case, the free energy has an asymptotic large $N$ behavior which is also controled by theta functions and a \textit{na\"\i ve} use of orthogonal polynomials will not work; rather one has to use the full power of resurgent transseries.

In summary, one may be faced with at least two different phases or backgrounds when addressing multi--cut configurations: either periodic or quasi--periodic behavior of the recursion coefficients, corresponding to either Stokes or anti--Stokes phases. In the Stokes phase the large $N$ asymptotics is essentially given by an 't~Hooft topological genus expansion, while in the anti--Stokes phase the asymptotics is of theta--function type. These issues were addressed in \cite{mpp09} and we refer the reader to their excellent discussion (where the authors of \cite{mpp09} used the terminology of ``boundary'' and ``interior'' points to denote what we here call Stokes and anti--Stokes regions). In particular, an expansion around a given background is well--defined when either \cite{mpp09}:
\begin{enumerate}
\item In a Stokes region, one will find an admissible large $N$ 't~Hooft genus expansion in powers of $1/N^2$, with exponentially suppressed multi--instanton corrections, if
\be 
\re \left( \frac{A(t)}{g_s} \right) > 0.
\ee
\item In an anti--Stokes region, the free energy will display theta--function asymptotics \cite{e08, em08}. This expansion will be admissible if the filling fractions are real, $\frac{N_i}{N} \in \mathbb{R}$, and if
\be\label{Boutroux2}
\re \left( \frac{\partial_{s_i} F_0}{g_s} \right) = 0.
\ee
\end{enumerate}
\noindent
The conditions of admissibility were first discussed in \cite{d91, d92}, and later further addressed in \cite{bm06, b07, bt11} where they were shown to be equivalent to having the spectral curve as a Boutroux curve. Let us now stress that our construction in the previous subsection precisely fulfils the first condition above. In fact we were able to find a well--defined (exponentially suppressed) multi--instanton expansion, which is clear both from the general structure of \eqref{Zgrandextra} as well as from our final result \eqref{Zlresult!}. In this process, the $\mathbb{Z}_2$ symmetry plays an important role since it is the equality of the two instanton actions what allows us to write down a multi--instanton expansion for the (grand--canonical) partition function. Of course we still must make sure that the examples we shall address next also satisfy this condition.

\section{Large--Order Behavior of $\mathbb{Z}_2$--Symmetric Systems}\label{sec:lo}

Our next goal is to illustrate how the multi--instanton effects we have uncovered in the previous section make their appearance in different examples, and how we may test them by comparing against large--order analysis. We shall first address the quartic matrix model in its two--cut Stokes phase, as this is a particularly clean application of all our nonperturbative machinery. However, it is also important to have in mind that not all nonperturbative effects arise from what we may call $B$--cycle instantons \cite{msw07}, \textit{i.e.}, instantons whose action is given by a $B$--cycle integration of the spectral curve one--form as in figure \ref{spectralcurve}. In fact, in some cases one needs to consider $A$--cycle instantons instead \cite{ps09}, \textit{i.e.}, instantons whose action arises from integrating the spectral curve one--form along an $A$--cycle and thus, because of \eqref{t_I}, instantons which have an almost ``universal'' structure. As such, we shall illustrate this possibility with another example: the ``triple'' Penner matrix model which appears in the context of studying four--point correlation functions in the AGT set--up. Finally, notice that one of the key points that allowed us to solve for the nonperturbative structure of a multi--cut configuration in the previous section was its $\BZ_2$ symmetry and, as such, this will be a required ingredient also for our following examples.

\subsection{The Two--Cut Quartic Model in the Stokes Phase}\label{sec:quartic}

Let us begin by addressing the quartic matrix model in its two--cut Stokes phase. This is accomplished by considering the matrix model partition function \eqref{Z multi cut}  with quartic potential 
\be\label{quarticpotentialmu}
V(z) = \frac{\mu}{2} z^2 + \frac{\lambda}{4!} z^4, \qquad \mu<0, \quad \lambda>0,
\ee
\noindent
where we shall choose $\mu = -1$ without any loss of generality (this potential was depicted earlier, in figure \ref{quartic_numerics}). We shall first fully work out its two--cut spectral geometry and use this data to obtain explicit formulae for all the nonperturbative quantities we addressed earlier in subsection \ref{sec:oneinst}. Then, we will use orthogonal polynomials and resurgent transseries in order to, on one hand, readdress the results of subsection \ref{sec:oneinst}, and, on the other hand, produce large--order data that will be used to test and confirm our overall nonperturbative picture.

Beginning with the spectral curve \eqref{sp_curve}, it is simple to compute
\be 
M(z) = \frac{\lambda}{6} z
\ee
\noindent
from \eqref{M multicut}, and the endpoints of the cuts follow from the asymptotic constraints \eqref{asympcond} as
\be
a^2 = \frac{6}{\lambda} \left( 1 - \sqrt{\frac{2 \lambda t}{3}} \right) \quad \text{and} \quad b^2 = \frac{6}{\lambda} \left( 1 + \sqrt{\frac{2 \lambda t}{3}} \right).
\ee
\noindent
Integrating the spectral curve, the holomorphic effective potential \eqref{hol_eff_pot} follows:
\be\label{eff pot quartic}
V_{\mathrm{h;eff}} (z) = \frac{\lambda}{48} \left\{ \left( 2z^2 - a^2 - b^2 \right) \sqrt{\left( z^2 - a^2 \right) \left( z^2 - b^2 \right)} - \left( b^2 - a^2 \right)^2 \log \left( \frac{\sqrt{z^2-a^2} + \sqrt{z^2-b^2}}{\sqrt{b^2-a^2}} \right) \right\}. 
\ee
\noindent
The real part of this potential is shown in figure \ref{Eff Pot Quartic} where the symmetric cuts and the pinched cycle are very clearly identifiable. Given this result, one may immediately compute the instanton action, with either \eqref{instantonaction0a} or \eqref{action_sp}, as
\be\label{A SP quartic}
A (\lambda,t) = V_{\mathrm{h;eff}} (0) - V_{\mathrm{h;eff}} (a) = 
\frac{3}{2\lambda} \sqrt{1-\frac{2\lambda t}{3}} - t \log \left( \frac{\sqrt{3}+\sqrt{3-2\lambda t}}{\sqrt{2\lambda t}} \right).
\ee
\noindent
In its domain of validity, $0 < \lambda t < \frac{3}{2}$, this action is indeed real positive as expected.

\FIGURE[ht]{
\label{Eff Pot Quartic}
\centering
\includegraphics[width=8cm]{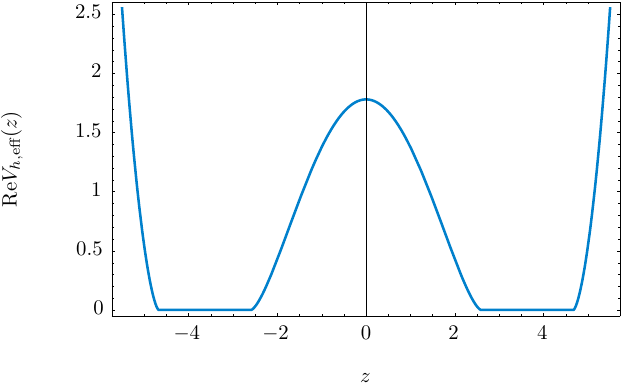}
$\qquad$
\includegraphics[width=5.75cm]{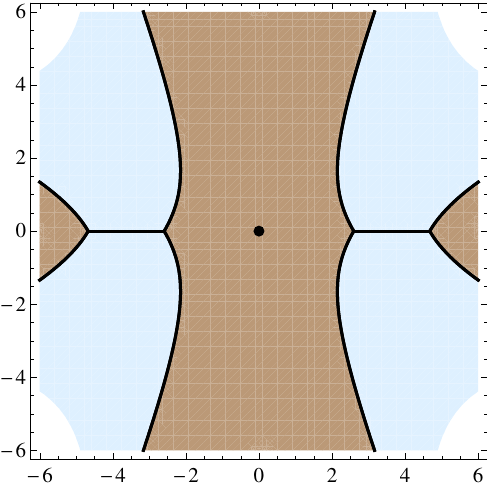}
\caption{The real part of the holomorphic effective potential for the two--cut quartic matrix model, \eqref{eff pot quartic}, both on the real axis (left) and on the complex plane (right), when $t=1$ and $\lambda=0.42$. The brown areas indicate when $\re\, V_{\mathrm{h;eff}}(z) > 0$ and the blue ones when $\re\, V_{\mathrm{h;eff}}(z) < 0$. The horizontal black lines are precisely the cuts of the spectral curve and the black dot the pinched cycle.}
}

Similarly to what was done in the one--cut case with the quartic matrix model \cite{msw07, asv11}, one may now test all our nonperturbative formulae against large--order data in a simple and explicit example. Of course one first needs to generate the large--order data itself and, for the present two--cut scenario, the procedure will be slightly more involved than the one in \cite{msw07, asv11} (which, on what concerned the perturbative sector, was a simple extension of the pioneering work in \cite{biz80}). Let us also stress that because this data precisely constructs the large $N$ expansion in this phase, it will further confirm that it is in fact of 't~Hooft type, \textit{i.e.}, a Stokes phase. The analysis starts by addressing orthogonal polynomials in this model, whose string equation \eqref{4stringeq} is currently written as
\be\label{rec eq2}
r_n \left\lbrace -1 + \frac{\lambda}{6} \left( r_{n-1} + r_{n} + r_{n+1} \right) \right\rbrace = n g_s.
\ee
\noindent
Recall from our review in subsection \ref{sec:op} that, in the one--cut case, the recursion coefficients $r_n$ approach a single function $\CR(x)$ with genus expansion \eqref{R pert} in its perturbative sector. This function satisfies a finite difference equation, \eqref{largeN4stringeq}, which was solved using resurgent transseries in \cite{m08, asv11}. The key point here is that transseries solutions allow for an inclusion of \textit{all} multi--instanton sectors, as we briefly mentioned in \eqref{transseries}, going beyond the usual large $N$ expansion. Furthermore, the free energy follows as \eqref{FEMtoda}. This time around, with two cuts, as we discussed previously and plotted in figure \ref{quartic_numerics}, a numerical solution of the above recursive equation \eqref{rec eq2}, approaches, in the large $N$ limit, \textit{two} distinct functions. Thus, what one now has to do is to generalize the aforementioned framework into a period two \textit{ansatz}, as first suggested in \cite{m88, ddjt90, cm91, bdjt93}. As such, we shall consider
\bea
\label{2 limitsa}
r_n &\to& \CP(x), \qquad n\,\, \text{even}, \\
r_n &\to& \CQ(x), \qquad n\,\, \text{odd}.
\label{2 limitsb}
\eea
\noindent
In this case, the large $N$ limit of our recursion \eqref{rec eq2} will split into two coupled equations
\bea
\label{rec eq quartic1}
\CP(x) \left\lbrace -1 + \frac{\lambda}{6} \left( \CQ(x-g_s) + \CP(x) + \CQ(x+g_s) \right) \right\rbrace &=& x, \\
\CQ(x) \left\lbrace -1 + \frac{\lambda}{6} \left( \CP(x-g_s) + \CQ(x) + \CP(x+g_s) \right) \right\rbrace &=& x,
\label{rec eq quartic2}
\eea
\noindent
and these are the equations we wish to solve via transseries methods, following the work in \cite{asv11}.

\subsubsection*{Two--Parameter Transseries Solution to the String Equations}

The simplest approach to solving the above string equations, \eqref{rec eq quartic1} and \eqref{rec eq quartic2}, is to start with a perturbative \textit{ansatz} for both $\CP(x)$ and $\CQ(x)$ of the type \eqref{R pert}, generalizing the work in \cite{biz80}, as
\be
\CP (x) \simeq \sum_{g=0}^{+\infty} g_s^{2g} P_{2g} (x), \qquad
\CQ (x) \simeq \sum_{g=0}^{+\infty} g_s^{2g} Q_{2g} (x).
\ee
\noindent
At genus zero, for instance, it is then simple to obtain
\bea
P_0 (x) &=& \frac{3}{\lambda} \left( 1 - \sqrt{1-\frac{2 \lambda x}{3}} \right), \\
Q_0 (x) &=& \frac{3}{\lambda} \left( 1 + \sqrt{1-\frac{2 \lambda x}{3}} \right),
\eea
\noindent
where we have assumed that $\CP \not= \CQ$, \textit{i.e.}, explicitly imposed the period--two \textit{ansatz} \cite{ddjt90, cm91, bdjt93}. In the domain of validity of the two--cut phase, $0 < \lambda x < \frac{3}{2}$, this in fact corresponds to two distinct (real) functions which meet at the (critical) point $\lambda x = \frac{3}{2}$, where $P_0 = \frac{3}{\lambda} = Q_0$.

Going beyond the perturbative large $N$ expansion, one is first required to include all multi--instanton sectors via an one--parameter transseries \textit{ansatz} \cite{m08, asv11},
\bea
\CP (x) &=& \sum_{n=0}^{+\infty} \sigma^n P^{(n)} (x), \qquad P^{(n)} (x) \simeq \rme^{- n A(x)/ g_s}\, \sum_{g=0}^{+\infty} g_s^g\, P^{(n)}_g (x), \\
\CQ (x) &=& \sum_{n=0}^{+\infty} \sigma^n Q^{(n)} (x), \qquad Q^{(n)} (x) \simeq \rme^{- n A(x)/ g_s}\, \sum_{g=0}^{+\infty} g_s^g\, Q^{(n)}_g (x),
\eea
\noindent
where we have imposed that both transseries expansions have the same structure, in particular that they have the same instanton action. This may \textit{a priori} seem as an unnecessary assumption, but it is justified on two levels. On the one hand, this is required so that we may actually find non--trivial solutions to the string equations \eqref{rec eq quartic1} and \eqref{rec eq quartic2} (which are being solved ``perturbatively'', \textit{i.e.}, as an expansion both in powers of the string coupling \textit{and} in powers of the transseries parameter which corresponds to the instanton number). On the other hand, as we shall see later on, our large--order analysis will show that the perturbative sectors $P^{(0)}(x)$, $Q^{(0)}(x)$ are indeed governed by the same instanton action, thus ``experimentally'' confirming this assumption. Plugging these expressions back into the string equations, \eqref{rec eq quartic1} and \eqref{rec eq quartic2}, one finds, at first order in instanton number and zeroth order in the string coupling, an equation for the instanton action as
\be
\cosh^2 \left( A'(x) \right) = \frac{3}{2\lambda x}.
\ee
\noindent
Notice that there are four sign ambiguities in this equation: two from the quadratic power and two from the (even) hyperbolic cosine function. For the moment we shall assume the quadratic sign ambiguity arises as an artifact of the period--two \textit{ansatz}, and thus only address the $\cosh z$ sign ambiguity (which is now equivalent to the one in the one--cut case \cite{m08, asv11}), leaving the complete exploration of the four sign ambiguities for future work. In this case one obtains for the instanton action:
\be\label{transseriesinstact}
A(x) = \pm \frac{\sqrt{9 - 6 \lambda x}}{2 \lambda} \mp x\, \text{arccosh} \left( \sqrt{\frac{3}{2 \lambda x}} \right) + 2 \pi \rmi\, x\, p + c_{\text{int}},
\ee
\noindent
where $p \in \BZ$. We shall set both the integer ambiguity $p$ and the integration constant $c_{\text{int}}$ to zero so that later this result will yield the Painlev\'e II instanton action, in the corresponding double--scaling limit. As to the sign ambiguity, notice that choosing the upper sign makes this expression precisely match the instanton action as computed via spectral methods, \eqref{A SP quartic}.

However, as shown in \cite{asv11} in the one--cut case, both signs of the instanton action \eqref{transseriesinstact} are important when performing the fully nonperturbative resurgent transseries analysis. A similar situation will happen in the present two--cut scenario, as we shall adopt the following two--parameter transseries \textit{ans\"atze} for the full nonperturbative content of the two--cut quartic matrix model:
\be\label{expansion P and Q}
\CP (x) = \sum_{n=0}^{+\infty} \sum_{m=0}^{+\infty} \sigma_1^{n} \sigma_2^{m}\, P^{(n|m)} (x), \qquad \CQ (x) = \sum_{n=0}^{+\infty} \sum_{m=0}^{+\infty} \sigma_1^{n} \sigma_2^{m}\, Q^{(n|m)}(x), 
\ee
\noindent
where each $P^{(n|m)} (x)$ sector (and similarly for $Q^{(n|m)} (x)$) has an expansion of the form:
\be\label{(n|m)}
P^{(n|m)} (x) \simeq \rme^{-(n-m) A(x)/g_s} \sum_{g=\beta_{nm}}^{+\infty} g_s^g\, P^{(n|m)}_g (x). 
\ee
\noindent
As one plugs these expansions back into the string equations, \eqref{rec eq quartic1} and \eqref{rec eq quartic2}, one can equate the terms with given powers $\sigma_1^n$ and $\sigma_2^m$ and find the following two coupled equations
\bea
\label{eq(n|m)1}
x\, \delta_{n0}\, \delta_{m0} &=& - P^{(n|m)}(x) + \\
&&
\hspace{-20mm}
+ \frac{\lambda}{6} \sum_{n_1=0}^{n} \sum_{m_1=0}^{m} P^{(n_1|m_1)}(x) \left\{ Q^{(n-n_1|m-m_1)}(x-g_s) + P^{(n-n_1|m-m_1)}(x) + Q^{(n-n_1|m-m_1)}(x+g_s) \right\}, \nonumber \\
\label{eq(n|m)2}
x\, \delta_{n0}\, \delta_{m0} &=& - Q^{(n|m)}(x) + \\
&&
\hspace{-20mm}
+ \frac{\lambda}{6} \sum_{n_1=0}^{n} \sum_{m_1=0}^{m} Q^{(n_1|m_1)}(x) \left\{ P^{(n-n_1|m-m_1)}(x-g_s) + Q^{(n-n_1|m-m_1)}(x) + P^{(n-n_1|m-m_1)}(x+g_s) \right\}. \nonumber
\eea
\noindent
If one next expands these equations in powers of the string coupling, $g_s$, this will produce---at each order---systems of either algebraic or (linear) differential equations which allow us to find the coefficients $P^{(n|m)}_g (x)$ and $Q^{(n|m)}_g (x)$ in terms of the ``earlier'' ones $P^{(n'|m')}_{g'} (x)$ and $Q^{(n'|m')}_{g'}(z)$ with $n' \leq n$, $m' \leq m$ and $g' \leq g$ (and their derivatives). As a technical aside, let us also note that the many exponentials appearing in \eqref{eq(n|m)1} and \eqref{eq(n|m)2} via \eqref{(n|m)} will bring down extra powers of the string coupling. In fact, we shall always have in mind the following expansions:
\be 
\exp \left(-n\, \frac{A (x \pm g_s)}{g_s} \right) = \exp \left(-n\, \frac{A (x)}{g_s} \right) \times \rme^{\mp n A'(x)}\, \sum_{\ell' =0}^{+\infty} \frac{1}{\ell'!} \left( - n \sum_{\ell=2}^{+\infty} \left( \pm 1 \right)^{\ell} g_s^{\ell-1}\, \frac{A^{(\ell)} (x)}{\ell!} \right)^{\ell'}.
\ee
\noindent
From here on, the extraction of the $P^{(n|m)}_g$ and $Q^{(n|m)}_g$ coefficients is absolutely straightforward with the help of a computer, very much in line with the strategy used in \cite{asv11}. Most of our explicit results are collected in appendix \ref{ap2}, but for completeness we next discuss a couple of examples.

Consider the purely perturbative sector, corresponding to $n=0=m$, which we have also addressed a few paragraphs above. At order $g_s^0$ it is simple to see that, once again, one finds the solution
\bea
P^{(0|0)}_0 (x) &=& \frac{3 - \sqrt{9 - 6 \lambda x}}{\lambda} \equiv \frac{3 - p}{\lambda}, \label{p0} \\
Q^{(0|0)}_0 (x) &=& \frac{3 + \sqrt{9 - 6 \lambda x}}{\lambda} \equiv \frac{3 + p}{\lambda}. \label{q0}
\eea
\noindent
Here we have defined $p \equiv \sqrt{9 - 6 \lambda x}$, as rewriting and solving most equations in terms of this variable will make life much easier. The remaining perturbative coefficients are recursively obtained from algebraic equations and this is generically the case for most of the $(n|m)$ sectors (see the appendix \ref{ap2} for further details and explicit expressions).

One exception to the aforementioned straightforward algebraic procedure is when $n=m\pm1$. In this case one finds the phenomenon of resonance, also discussed in the present context in \cite{gikm10, asv11}, and one needs to solve a (linear) differential equation instead. Let us illustrate this situation in the one--instanton sector $(1|0)$. One finds, at order $g_s^0$,
\bea
P^{(1|0)}_0 + \frac{3-p}{3} \cosh \left( A'(x) \right) Q^{(1|0)}_0  &=& 0, \\	
Q^{(1|0)}_0 + \frac{3+p}{3} \cosh \left( A'(x) \right) P^{(1|0)}_0  &=& 0.
\eea
\noindent
These two equations do not allow us to solve for both $P^{(1|0)}_0$ and $Q^{(1|0)}_0$, but only for their ratio $P^{(1|0)}_0/Q^{(1|0)}_0$. On the other hand, eliminating $P^{(1|0)}_0$ and $Q^{(1|0)}_0$, one may instead find a differential equation for the instanton action---which we have solved earlier in \eqref{transseriesinstact}. Proceeding to next order, $g_s^1$, the equations read\footnote{Notice that these equations involve derivatives of $P^{(1|0)}_0 (x)$ and $Q^{(1|0)}_0 (x)$.} 
\bea
\hspace{-0.9cm}
P^{(1|0)}_1 + \frac{\left( 3-p \right) p}{9\lambda} \sinh \left( A'(x) \right) Q^{(1|0) \prime}_0 (x) + \frac{3-p}{6} \cosh \left( A'(x) \right) \left( 2 Q^{(1|0)}_1 - Q^{(1|0)}_0 A''(x) \right) &=& 0, \\
\hspace{-0.9cm}
Q^{(1|0)}_1 + \frac{\left( 3+p \right) p}{9\lambda} \sinh \left( A'(x) \right) P^{(1|0) \prime}_0 (x) + \frac{3+p}{6} \cosh \left( A'(x) \right) \left( 2 P^{(1|0)}_1 - P^{(1|0)}_0 A''(x) \right) &=& 0.
\eea
\noindent
The situation is the same as in the $(1|0)$ sector at order $g_s^0$. All we can now do is to eliminate the ratio $P^{(1|0)}_1/Q^{(1|0)}_1$ and use our knowledge of the lower sectors---namely the relation between $P^{(1|0)}_0$ and $Q^{(1|0)}_0$, and the result for the instanton action---in order to obtain a linear differential equation yielding
\be\label{10PQ}
Q^{(1|0)}_0 = \sqrt{\frac{3+p}{p}} \qquad \text{and} \qquad P^{(1|0)}_0 = -\sqrt{\frac{3-p}{p}}.
\ee
\noindent
These examples show a feature which is characteristic of resonance and of the $n=m\pm1$ sectors, namely, that the equations we obtain at order $g_s^k$ produce differential equations whose solutions yield the instanton coefficients at order $k-1$. At this stage the reader may object that the differential equations alone are not enough if one does not specify boundary conditions. In fact, all integration constants involved in this procedure must be fixed by using data available in the double--scaling limit and we shall postpone that discussion for the next section (although we have already used this fact in fixing the integration constants in \eqref{10PQ} above).

Other interesting features appear in the higher multi--instanton sectors, and many of these were first uncovered in the one--cut example studied in \cite{asv11}. For example starting in the $(2|1)$ sector, logarithms make their appearance into the game and they recursively propagate to the ensuing higher sectors. Akin to what happened in \cite{asv11}, these logarithms are indeed expected in the construction of the transseries solution and, again, we shall further discuss this issue in the next section, within the analysis of the Painlev\'e II equation. Another interesting feature happens when $n=m$ (and the exponential term cancels). In this case, we find that all the coefficients $P^{(n|n)}_g$ (respectively $Q$) with \textit{odd} $g$ vanish, and the perturbative expansion in \eqref{(n|m)} contains only powers of $g_s^2$, \textit{i.e.}, it is an expansion in the \textit{closed} string coupling. As aforementioned, further data is presented in appendix \ref{ap2}, where we also find general patterns for the multi--instanton coefficients and relate the logarithmic sectors with the non--logarithmic ones.

\subsubsection*{The Nonperturbative Free Energy and Large--Order Analysis}

In order to test the multi--instanton results obtained in section \ref{sec:inst}, one needs to match them against the large--order behavior of the free energy, and this is what we shall now address. As such, we will derive the nonperturbative free energy of the two--cut quartic matrix model out of the transseries solution to the string equations \eqref{rec eq quartic1} and \eqref{rec eq quartic2} we have just obtained, even though we will not be interested in extracting as much data. The starting point in this construction is expression \eqref{Zorth}, which yields the partition function in terms of the orthogonal--polynomial recursion coefficients $r_n$. Since in the present configuration these recursion coefficients split into two different branches at large $N$, it is useful to first rewrite \eqref{Zorth} for $2N$ eigenvalues (and thus with 't~Hooft coupling $t = 2 N g_s$) as
\be\label{Z_2limits}
Z = h_0^{2N} \prod_{i=1}^{2N} r_i^{2N-i} = h_0^{2N} \prod_{i=1}^{N} r_{2i}^{2N-2i} \prod_{j=1}^{N} r_{2j-1}^{2N-(2j-1)}.
\ee
\noindent
Similarly to what was done in \eqref{prefree}, the free energy follows by taking the logarithm of the above expression (and normalizing against the Gaussian weight, as usual). One finds:
\be\label{F_2limits}
\CF = \frac{t}{g_s} \log \frac{h_0}{h_0^\text{G}} + \frac{t^2}{g_s^2}\, \frac{1}{2N} \sum_{n=1}^N \left( 1 - \frac{n}{N} \right) \log \frac{r_{2n}}{r_{2n}^\text{G}} + \frac{t^2}{g_s^2}\, \frac{1}{2N} \sum_{n=1}^N \left( 1 - \frac{n-\frac{1}{2}}{N} \right) \log \frac{r_{2n-1}}{r_{2n-1}^\text{G}}.
\ee
\noindent
It is now clear the reason why we rewrote the partition function \eqref{Zorth} as \eqref{Z_2limits} above: because of the even/odd split in \eqref{2 limitsa} and \eqref{2 limitsb}, the large $N$ limit of \eqref{F_2limits} will precisely construct the free energy out of $\CP(x)$ and $\CQ(x)$. In the continuum limit the first sum in \eqref{F_2limits}, which we will denote by the ``even'' sum,  is essentially the same as the sum in \eqref{prefree} and thus may be computed via the Euler--Maclaurin formula \eqref{Euler Mac}. The second sum in \eqref{F_2limits}, the ``odd'' sum, is a bit more subtle and requires slight modifications. In fact, from \eqref{q0}, recall that $\lim_{x \to 0} Q^{(0|0)}_0 (x) \not = 0$ making $Q^{(0|0)}_0 (x)/x$ ill--defined at the origin (alongside with its derivatives), but this problem is solved by simply considering the Gaussian contribution separately in the ``odd'' sector. Furthermore, the ``odd'' Euler--Maclaurin formula is now written as (following a similar analysis in \cite{mp09})
\be
\lim_{N \to + \infty} \frac{1}{N} \sum_{n=1}^N \Phi \left( \frac{n-\frac{1}{2}}{N} \right) \simeq \int_0^1 \rmd\xi\, \Phi(\xi) - \left. \sum_{k=1}^{+ \infty} \frac{1}{N^{2k}}\, \frac{\left( 1 - 2^{1-2k} \right) B_{2k}}{(2k)!}\, \Phi^{(2k-1)} (\xi) \right|_{\xi=0}^{\xi=1}.
\ee
\noindent
Assembling all contributions together, our formula for the free energy finally takes a familiar form \cite{biz80, m04, m08}
\bea
\CF (t,g_s) &\simeq& \frac{t}{2 g_s} \left( 2 \log \frac{h_0}{h_0^\text{G}} - \left. \log \frac{\CP (x)}{x} \right|_{x=0} \right) + \frac{1}{g_s^2}\, \frak{G} (t,g_s) + \frac{1}{2 g_s^2} \int_0^t \rmd x \left( t-x \right) \log \frac{\CP (x)}{x} + \nonumber \\
&&
+ \frac{1}{2 g_s^2} \int_0^t \rmd x \left( t-x \right) \log \CQ (x) + \frac{1}{2} \sum_{g=1}^{+\infty} g_s^{2g-2}\, \frac{2^{2g}\, B_{2g}}{(2g)!}\, \frac{\rmd^{2g-1}}{\rmd x^{2g-1}} \left. \left[ \left( t-x \right) \log \frac{\CP (x)}{x} \right] \right|_{x=0}^{x=t} - \nonumber \\
&&
- \frac{1}{2} \sum_{g=1}^{+\infty} g_s^{2g-2}\, \frac{\left( 2^{2g}-2 \right) B_{2g}}{(2g)!}\, \frac{\rmd^{2g-1}}{\rmd x^{2g-1}} \bigg[ \left( t-x \right) \log \CQ (x) \bigg] \bigg|_{x=0}^{x=t}.
\label{F_EM}
\eea
\noindent
The function $\frak{G} (t,g_s)$ comes from the Gaussian normalization in the ``odd'' part and is given by
\be 
\frak{G} (t,g_s) \equiv - \sum_{k=1}^N \left( 2N-2i+1 \right) \log \left( \left( 2i-1 \right) g_s \right).
\ee
\noindent
When computing the free energy, this expression may be first evaluated exactly and then expanded in powers of the string coupling.

Let us note that while at the perturbative level, \textit{i.e.}, when $n=0=m$, the Euler--Maclaurin recipe \eqref{F_EM} is an efficient way to produce large--order data, the same is not valid when addressing the (generalized) multi--instanton sectors (more on this next). In any case, using the expansions \eqref{(n|m)} when $n=0=m$ (which we have described how to compute in the paragraphs above, and whose data we have presented in appendix \ref{ap2}) and inserting them into a \textit{Mathematica} script encoding the Euler--Maclaurin expansion, we have computed the coefficients $\CF^{(0|0)}_g$ in the perturbative free energy of the $\BZ_2$--symmetric two--cut quartic matrix model up to genus $g=20$ and some partial results are presented in greater detail in appendix \ref{ap1}.

In order to obtain data concerning the higher instanton sectors in an effective way, and while remaining within the orthogonal polynomial framework, one uses a small trick due to \cite{m08}. Starting off with the partition function, written as either \eqref{Zorth} or \eqref{Z_2limits}, it is simple to show that (subscripts in the partition function indicate the total number of eigenvalues considered)
\be
\frac{Z_{2(N+1)}\, Z_{2(N-1)}}{Z_{2N}^2} = r_{2N+1}\, r_{2N}^2\, r_{2N-1},
\ee
\noindent
which, at the free energy level, may be written as
\be\label{Ftoda}
\CF ( t + 2 g_s) - 2 \CF ( t ) + \CF ( t - 2 g_s ) = \log \left( \CQ \left( t + g_s \right) \CP^2 \left( t \right) \CQ \left( t - g_s \right) \right).
\ee
\noindent
This expression is, in fact, a rewriting of the Euler--Maclaurin formula \eqref{F_EM}, but from a computational point--of--view it also makes it much easier to extract large--order data.

We may now finally address tests of our multi--instanton formulae using large--order analysis, and further compute Stokes coefficients for the problem at hand. The main quantity we wish to focus upon is the one--instanton, one--loop coefficient $\CF^{(1|0)}_0$. At this stage, its calculation is simple if we are to use \eqref{Ftoda} above: all one has to do is to plug in two--parameter transseries \textit{ans\"atze} for all quantities and it quickly follows that, for $n=1$, $m=0$ and at order $g_s^0$, one has
\be 
4 \sinh^2 \left( A'(x) \right) \CF^{(1|0)}_0 = 2 \left( \frac{P^{(1|0)}_0 (x)}{P^{(0|0)}_0 (x)} + \cosh \left( A'(x) \right) \frac{Q^{(1|0)}_0 (x)}{Q^{(0|0)}_0 (x)} \right).
\ee
\noindent
If we plug in our results for the perturbative contributions, \eqref{p0} and \eqref{q0}, for the one--instanton contributions, \eqref{10PQ}, and for the instanton action, \eqref{transseriesinstact}, we finally obtain
\be\label{newF1instTS}
\CF^{(1|0)}_0 = - \frac{\lambda}{2}\, \sqrt{\frac{3-p}{p^3}}.
\ee
\noindent
As we have discussed in detail in \ref{sec:ts}, a key point about this quantity is that it controls the leading large--order growth of the asymptotic perturbative expansion, as explicitly shown in \eqref{F00largeorder}. For completeness, let us just recall that expression in here:
\be\label{Fg large order}
\CF^{(0|0)}_g \sim \frac{S_1^{(0)}}{\rmi\pi}\, \frac{\Gamma \left( 2g+b \right)}{A^{2g+b}} \left\{ \CF^{(1|0)}_0 + \frac{A}{2g + b -1}\, \CF^{(1|0)}_1 + \cdots \right\}.
\ee
\noindent
Many large--order tests may now be carried out; let us here mention a few of those following \cite{msw07} (but, let us note, many more higher--precision tests may be carried through, as in \cite{asv11}, and these we leave for future work). One obvious test concerns the instanton action, which may be numerically extracted from the sequence:
\be\label{seqA}
\alpha_g^{(\CF)} = \frac{\CF^{(0|0)}_{g+1}}{4 g^2 \CF^{(0|0)}_g} \sim \frac{1}{A^2} \left(1 + \frac{2b + 1}{2g} + \cdots \right).
\ee
\noindent
The parameter $b$ will be equal to $-5/2$, but that can be tested as well, \textit{e.g.}, using the sequence:
\be\label{b_seq}
b \sim \frac{1}{2} \left( 2 g \left( A^2 \alpha_g^{(\CF)} - 1 \right) - 1 \right) + \cdots.
\ee
\noindent
Finally, one approach to testing the one--loop coefficient is to use the sequence:
\be\label{oneseq}
\beta_g^{(\CF)} = \frac{\rmi\pi}{S_1^{(0)}}\, \frac{A^{2g+b}\, \CF^{(0|0)}_g}{\Gamma \left( 2g + b \right)} \sim \CF^{(1|0)}_0 + \cdots.
\ee
\noindent
We should note that all sequences above have been built with free energy quantities but, of course, one may also perform the exact analogue large--order tests directly using the solutions to the string equations, $\CP(x)$ and $\CQ(x)$. In fact, all these quantities have their large--order behavior dictated by the very same instanton action\footnote{This was previously shown via the string equations, but we also checked it numerically to very high precision.} and, as such, we shall use either $\CP(x)$ or $\CQ(x)$ whenever possible as we have obtained far more large--order data for these quantities than for the free energy. We shall denote those corresponding sequences with the respective superscript. We also note that all these quantities have ``closed string'' expansions (\textit{i.e.}, in powers of $g_s^2$) in their $(0|0)$ sectors, so the sequences above are tested for \textit{even} $g$.

\FIGURE[ht]{
\label{Action}
\centering
\includegraphics[width=7cm]{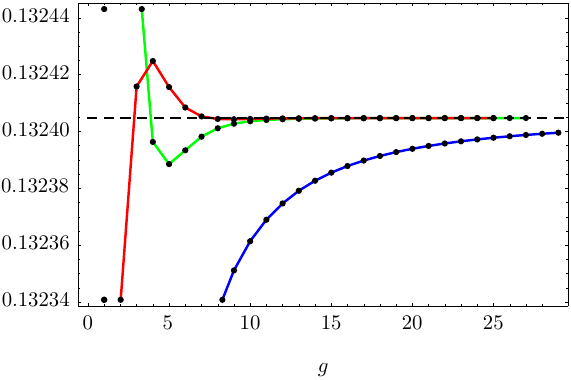}
$\qquad$
\includegraphics[width=7.2cm]{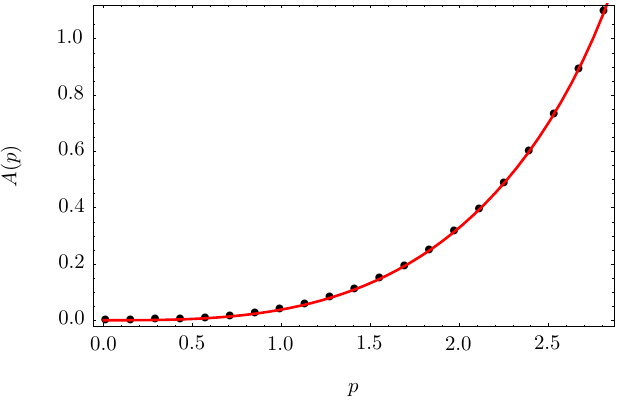}
\caption{The first image depicts a test of the instanton action using the sequence $\alpha_g^{(\CP)}$ and its first Richardson transforms, when $\lambda=0.5$ and $p=1.2$. The large--order convergence towards the correct result is clear (the dashed horizontal line is the analytical prediction), with an error of $10^{-6}\%$ after just four Richardson transforms. The second image shows a test of the instanton action at fixed $\lambda=1$ but with varying $p$, after implementing four Richardson transforms. Large--order data makes up the dots, while the analytical prediction is given as the solid red line. Again, the match is extremely clear.}
}

The first natural test to do concerns the instanton action, which is shown in figure \ref{Action}. Clearly, there is a very strong agreement between the ``theoretical'' prediction (be it from either saddle--point \eqref{A SP quartic} or transseries \eqref{transseriesinstact} approaches) and the ``numerical'' data. On the left of figure \ref{Action} we have plotted data at a particular point in moduli space\footnote{Recall the domain of validity of the two--cut Stokes phase, $0 < \lambda x < \frac{3}{2}$, or, equivalently, $0 < p < 3$.}, namely, $\lambda = 0.5$ and $p = 1.2$, concerning the sequence $\alpha_g^{(\CP)}$ and its first sequential Richardson extrapolations (see, \textit{e.g.}, \cite{msw07} for a short discussion of Richardson transforms and their role in accelerating the convergence of a given sequence, within the present matrix model context). That the large--order data approaches the analytic prediction is very clear: after just four Richardson transforms the error is already of the order $10^{-6}\%$ at genus $g=60$. On the right of figure \ref{Action} we have fixed $\lambda=1$ but vary $p$ over its full range. Once again we check that the numerical data (the black dots in the figure), after just four Richardson transforms, is never further than $10^{-6}\%$ away from the analytical prediction (the solid red line), thus fully validating our results.

As we move on to testing the one--instanton, one--loop coefficient, it is important to first recall that the transseries framework only predicts large--order behavior up to the Stokes factors---in this case up to the Stokes factor $S_1^{(0)}$, see \eqref{Fg large order}. However, we also have computed the same quantity via spectral curve analysis \eqref{1inst} (this was one of the main results in section \ref{sec:oneinst}) and, following \cite{msw07, m08, asv11}, the spectral curve result should provide for the full answer, Stokes factor included. In this case, the calculation of $S_1^{(0)} \, F_0^{(1|0)}$ in \eqref{1inst} and the calculation of $F_0^{(1|0)}$ in \eqref{newF1instTS} combine to predict the Stokes parameter as
\be\label{2cqmmsc}
S_1^{(0)} = - \rmi \sqrt{\frac{6}{\pi \lambda}}.
\ee
\noindent
It is quite interesting to compare the result for this ``simplest'' Stokes constant (at least that one constant which may be analytically computed from saddle--point analysis), in the present two--cut configuration, with the analogue Stokes constant for the one--cut configuration in \cite{m08, asv11}. For the quartic matrix model one thus finds:
\be
\left. S_1^{(0)} \right|_{\text{two-cut}}= - \sqrt{2} \left. S_1^{(0)} \right|_{\text{one-cut}}.
\ee
\noindent
With the knowledge of this Stokes constant (which we should more properly denote by $S_1^{(0) \CF}$ since it refers to the free energy), we can proceed to test the relation \eqref{oneseq} for the sequence $\beta_g$. Since besides the free energy the quantities $\CP(x)$ and $\CQ(x)$ also obey a relation similar to \eqref{oneseq}, a natural question to ask is whether the Stokes constant for these different quantities is the same. Indeed we find that it is the case, namely that 
\be 
S_1^{(0) \CF} = S_1^{(0) \CP} = S_1^{(0) \CQ} \equiv S_1^{(0)}.
\ee
\noindent
This is to say that, when testing the asymptotic relation \eqref{oneseq} for either $\beta_g^{(\CP)}$, $\beta_g^{(\CQ)}$ or $\beta_g^{(\CF)}$, we find that the relation holds to very high accuracy with the Stokes constants being the same in all three cases. On the other hand, the value of $b$ is different, with $b=-1/2$ for $\beta_g^{(\CP)}$ and $\beta_g^{(\CQ)}$ and $b=-5/2$ for $\beta_g^{(\CF)}$ (see \cite{asv11} for a discussion of this point). With this knowledge, we have tested our instanton predictions with the sequences $\beta_g^{(\CP, \CQ)}$, finding that the numerical data has an error smaller than $10^{-5} \%$ at genus $g=60$ as compared to the analytical prediction for $S_1^{(0)} P^{(1|0)}_0$ (or $Q$), within most of the allowed range for $\lambda$ and the variable $p$. Note, however, that $P^{(1|0)}_0$ (and also $Q$) diverges as one approaches $p \rightarrow 0$, making the convergence of numerical data to analytical prediction naturally a bit worse once we  get too close to $p=0$. These results are illustrated in figure \ref{oneloop}. On the left of this figure we have fixed $\lambda=0.5$ and $p=1.5$, and plotted the sequence $\beta_g^{(\CP)}$ alongside with its Richardson transforms. It is again very clear how the data approaches the analytical prediction (the horizontal dashed line). On the right of figure \ref{oneloop} we have fixed $\lambda=1$ and changed $p$ over its full range, plotting the fourth Richardson transform of the sequence $\beta_g^{(\CP)}$ (black dots) and the analytical prediction (solid red line). The agreement is, once again, evident. Let us mention that the very same tests may also be carried out for the free energy. In this case, we find an equally conclusive agreement, albeit with a smaller accuracy ($10^{-3} \%$) as we have less large--order data available.

\FIGURE[ht]{
\label{oneloop}
\centering
\includegraphics[width=7cm]{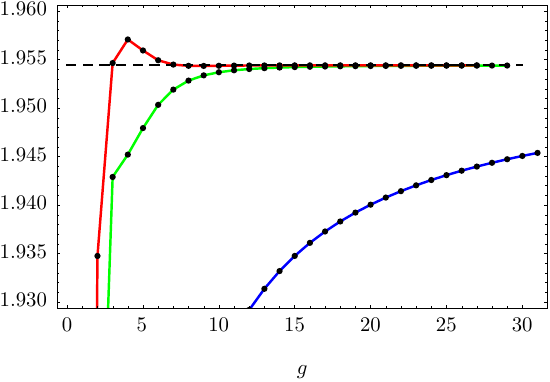}
$\qquad$
\includegraphics[width=7.6cm]{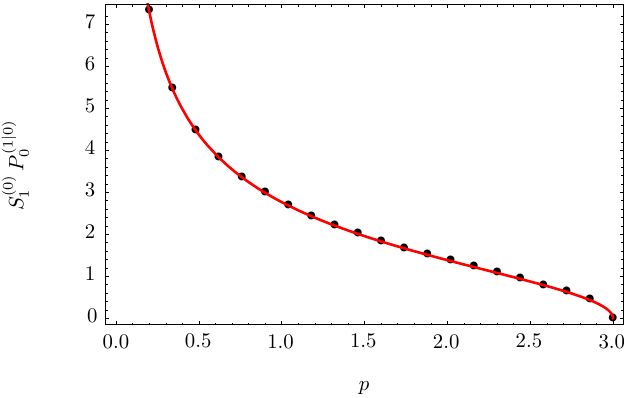}
\caption{The first image depicts a test of the one--instanton, one--loop coefficient using the sequence $\beta_g^{(\CP)}$ and its first Richardson transforms, when $\lambda=0.5$ and $p=1.5$. The large--order convergence towards the correct result is clear (the dashed horizontal line is the analytical prediction), with an error of the order of $10^{-6}\%$. The second image shows a test of the one--instanton, one--loop coefficient at fixed $\lambda=1$ but with varying $p$ over its full range, after implementing just four Richardson transforms on the sequence $\beta_g^{(\CP)}$. Large--order data makes up the dots, while the analytical prediction is given as the solid red line. As in previous cases, the agreement is extremely clear.}
}

At this stage one could proceed along the lines in \cite{asv11} and test both multi--instanton formulae as well as the validity of generalized multi--instanton sectors appearing via our resurgence formulae. This would involve techniques of Borel--Pad\'e resummation and, as such, within the context of the two--cut Stokes phase of the quartic matrix model, we shall leave these precision tests for future work. Do notice that we shall, nonetheless, test the validity of our multi--instanton formulae in the double--scaling limit towards the Painlev\'e II equation in a following section.

\subsection{The Triple Penner Potential and AGT Stokes Phenomena}\label{sec:penner}

As we address nonperturbative phenomena within Stokes phases of multi--cut solutions, it is important to note that it is not always the case that instanton effects arise from $B$--cycle eigenvalue tunneling (as discussed in \cite{msw07, msw08} and also as developed in section \ref{sec:inst} of the present paper). In some situations, one finds systems whose instanton effects are dictated by $A$--cycle eigenvalue tunneling instead \cite{ps09, sw09}. For completeness of our analysis, we shall now address an example along these lines. As before, we will remain within the simplified realm of two--cut configurations, with the equal filling of eigenvalues ensuring $\BZ_2$ symmetry of the spectral curve.

We shall address multi--Penner matrix models. The single Penner model was first introduced in \cite{p88, cdl91, akm94} and its nonperturbative effects were later addressed in \cite{ps09}. Extra motivation for studying this system arises within the framework of the AGT conjecture \cite{agt09, w09}, establishing a relation between  partition functions in $4d$ $\mathcal{N}=2$ superconformal quiver gauge theories and correlation functions in $2d$ conformal field theories (CFT). Within this set--up, we are particularly interested in the relation to matrix models following \cite{dv09}, where the quiver gauge theories are related to multi--Penner matrix models, and where the AGT relations follow from the interconnections betweens these matrix models and CFT \cite{k99, k04}. This was further studied in \cite{sw09}, in particular addressing the three--point correlation function as a Penner matrix model calculation. The results we shall obtain below follow in this very same spirit, as they similarly relate to the CFT four--point correlation function with a specific, symmetric choice of insertion points. However, all our computations are carried through exclusively from a matrix model point of view, and any possible applications within the AGT context will require further examination.

The multi--Penner potential is a sum over logarithms, as
\be\label{multipenner potential gen}
V(z) = \sum_{i=1}^k \mu_i \log \left( z - \theta_i \right).
\ee
\noindent
In order to obtain a $\BZ_2$--symmetric potential with two wells, we shall set $k=3$, $\theta_2 = \frac{1}{2} \left( \theta_1+\theta_3 \right)$, $\mu_1 = \alpha = \mu_3$ and $\mu_2 = \beta$. The potential now reads
\be\label{multipenner potential}
V(z) = \alpha \log \left( z - \theta_1 \right) + \beta \log \left( z - \frac{1}{2} \left( \theta_1+\theta_3 \right) \right) + \alpha \log \left( z - \theta_3 \right).
\ee
\noindent
An example of such a potential is shown in figure \ref{Pot multipenner} (where we plot the real part of the potential---the imaginary part just jumps by $\pi$ at each logarithmic singularity). The choice of parameters in figure \ref{Pot multipenner} is precisely the case we are going to address, with $\theta_1 = 0$, $\theta_2 = \frac{1}{2}$ and $\theta_3 = 1$. In this case the potential is symmetric with respect to $z=1/2$ and not $z=0$, but the motivation for this choice is clear: when studying four--point correlation functions on the sphere it is usual to make three of the insertions at $z=0,1$ and $\infty$, with the fourth varying between $0$ and $1$. Herein, the matrix model does not ``see'' the point at $\infty$, and placing the fourth insertion at $z=1/2$ gives us the $\BZ_2$ symmetry we are looking for. In the end, all that distinguishes the two cases is a change of variables: a rescaling and a horizontal shift.

\FIGURE[ht]{
\label{Pot multipenner}
\centering
\includegraphics[width=10cm]{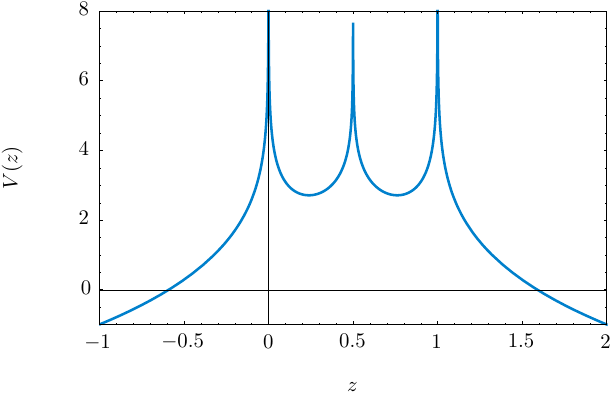}
\caption{Real part of the Penner potential \eqref{multipenner potential} for $\theta_1 = 0$, $\theta_2 = 1/2$, $\theta_3 = 1$, and $\mu_1 = \mu_2 = \mu_3 = -1$.}
}

The saddle--point analysis we introduced in subsection \ref{sec:sp} applies straightforwardly to this case, so we can proceed and compute the endpoints of the cuts, which we here denote by $C_1 \cup C_2 = \left[ 1/2 - b, 1/2 - a \right] \cup \left[ 1/2 + a, 1/2 + b \right]$. The asymptotic behavior of the resolvent gives us three conditions for the endpoints, one of them being redundant. The other two are
\bea 
\frac{2 \alpha}{\sqrt{\left( \frac{1}{4} - a^2 \right) \left( \frac{1}{4} - b^2 \right)}} + \frac{\beta}{\sqrt{a^2 b^2}} = 0, \\
2 \alpha + \beta + \frac{1}{\sqrt{\left( \frac{1}{4} - a^2 \right) \left( \frac{1}{4} - b^2 \right)}} - \frac{\beta}{4 \sqrt{a^2 b^2}} = 2t,
\eea
\noindent
from where we find the solutions 
\bea 
a^2 &=& \frac{2 t \left( t - 2 \alpha \right) + \beta \left( \beta + 2 \alpha - 2 t \right) - 2 \sqrt{t \left( t - 2 \alpha \right) \left( t - \beta \right) \left( t - 2 \alpha - \beta \right)}}{4 \left( \beta + 2 \alpha - 2 t \right)^2}, \\
b^2 &=& \frac{2 t \left( t - 2 \alpha \right) + \beta \left( \beta + 2 \alpha - 2 t \right) + 2 \sqrt{t \left( t - 2 \alpha \right) \left( t - \beta \right) \left( t - 2 \alpha - \beta \right)}}{4 \left( \beta + 2 \alpha - 2 t \right)^2}.
\eea
\noindent
From this point on, the picture is different from the one we discussed for the quartic potential. The main difference is that now there is no eigenvalue tunneling, in the sense that one eigenvalue gets removed from one of the cuts and displaced along a $B$--cycle to a non--trivial saddle outside of that cut \cite{msw07}. We can check this explicitly by looking for the zeroes of $M(z)$ (recall that the eigenvalues get displaced from their cut to a pinched cycle, $x_0$, such that $M(x_0) = 0$). For a general multi--Penner potential \eqref{multipenner potential gen}, the moment function $M(z)$ is given by \eqref{M multicut} as
\be 
M(z) = \sum_{i=1}^k \frac{\mu_i}{\left( z-\theta_i \right) \sqrt{\sigma(\theta_i)}}.
\ee 
\noindent
For our particular example \eqref{multipenner potential} we find that $M(z)$ only has two zeroes, lying inside the cuts. In this case there are no non--trivial saddle points, \textit{i.e.}, no ``hills'' to place eigenvalues on top of. Nonperturbative tunneling effects will thus have to be distinct from our previous discussion of the quartic potential\footnote{Further note that there is no tunneling from one cut to the other as $\int_{1/2-a}^{1/2+a} \rmd z\, y(z) = 0$.}. Let us see how they arise in the following.

One may next compute the holomorphic effective potential \eqref{hol_eff_pot} and find the intricate expression (we are using the shorthand $\bar{z} = z - 1/2$)
\bea 
V_{\mathrm{h;eff}} (z) &=& \sqrt{\sigma_2 (\bar{z})} \left( \frac{4 \alpha}{\sqrt{\left( 4 a^2 - 1 \right) \left( 4 b^2 - 1 \right)}} + \frac{\beta}{2 a b} \right) + \frac{1}{\sqrt{\left( 4 a^2 - 1 \right) \left( 4 b^2 - 1 \right)}}\, \frac{1}{\sqrt{b^2-a^2}} \times \nonumber \\
&&
\times \left\{ \left( 8 a \alpha \left( 2 a^2 + 2 b^2 - 1 \right) + \frac{2 \beta}{b} \left( a^2 + b^2 \right) \sqrt{\left( 4 a^2 - 1 \right) \left( 4 b^2 - 1 \right)}  \right) \Pi \left( \phi, -\sqrt{n_1 n_2}, m \right) - \right. \nonumber \\
&&
- 4 b \beta \sqrt{\left( 4 a^2 - 1 \right) \left( 4 b^2 - 1 \right)}\, \Pi \left( \phi, \sqrt{n_1 n_2}, m \right) + \nonumber \\
&&
+ \frac{1}{b} \left( b^2 - a^2 \right) F (\phi, m) \left( \beta \sqrt{\left( 4 a^2 - 1 \right)  \left( 4 b^2 - 1 \right)} + 8 a b \alpha \right) - \nonumber \\
&&
- 4 a \alpha \left( 4 b^2 - 1 \right) \left( \Pi (\phi, n_1, m) + \Pi (\phi, n_2, m) \right) \bigg\}.
\label{eff pot penner}
\eea
\noindent
In this expression, $F(\phi,m)$ is the incomplete elliptic integral of the first kind and $\Pi(\phi,n,m)$ the incomplete elliptic integral of the third kind, with $n$ the elliptic characteristic and $m=k^2$ with $k$ the elliptic modulus (see, \textit{e.g.}, \cite{wolfram}). Furthermore, we have introduced
\be 
\phi = \sqrt{\frac{(b-a)(\bar{z}+a)}{(b+a)(\bar{z}-a)}}, \quad m = \frac{(b+a)^2}{(b-a)^2}, \quad n_1 = \frac{(1-2a)(b+a)}{(1+2a)(b-a)}, \quad n_2 = \frac{(1+2a)(b+a)}{(1-2a)(b-a)}.
\ee
\noindent
In figure \ref{Eff pot penner} we show the above potential \eqref{eff pot penner} for the choice $\alpha=-1$, $\beta=-1/2$ and $t=1/2$.

\FIGURE[ht]{
\label{Eff pot penner}
\centering
\includegraphics[width=10cm]{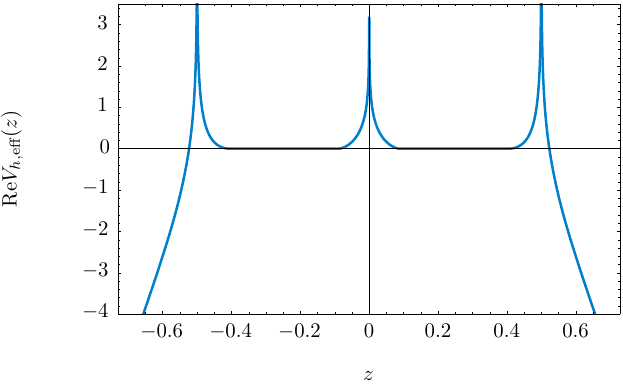}
\caption{The real part of the holomorphic effective potential for the two--cut ``triple'' Penner matrix model, along the real axis. In here we have chosen $t=1/2$, $\alpha = -1$ and $\beta=-1/2$.}
}

While expression \eqref{eff pot penner} may not be extremely insightful, it suffices to show that the holomorphic effective potential is a multi--valued function. This multi--sheeted structure arises from the branch cuts of the square roots but, more importantly and more non--trivially, from the branch cuts of the elliptic functions. As we shall see in detail next, this implies that in this case multi--instantons are associated to eigenvalue tunneling which removes one eigenvalue from the endpoint of a cut and then takes it back to this cut but on a \textit{different} sheet \cite{ps09}. In other words, multi--instantons are associated to $A$--cycles of the spectral curve \cite{ps09}. This may be shown in two ways. On the one hand one may analyze the branch--cut configurations of the elliptic integrals in \eqref{eff pot penner} and explicitly construct the multi--sheeted structure of this function in order to study its monodromy properties. On the other hand, one may use orthogonal polynomials to first exactly evaluate the partition function of the model and then perform a semiclassical expansion, where one will be able to identify the instanton actions with $A$--cycles via \eqref{t_I}. For simplicity, we shall choose the second approach where we will find many different instanton actions reflecting the many different spacings between the several sheets. All these actions will be multiples of $2\pi\rmi$.

As such, moving on to the orthogonal polynomial description, we first need to find the measure \eqref{orthomeasure} for our ``triple'' Penner potential \eqref{multipenner potential}. It is simple to find
\be
\rmd \mu (z) = \rme^{-\frac{1}{g_s} V(z)}\, \frac{\rmd z}{2\pi} = \left| z \right|^{-\frac{\alpha}{g_s}} \left| z - 1/2 \right|^{-\frac{\beta}{g_s}} \left| z - 1 \right|^{-\frac{\alpha}{g_s}}\, \frac{\rmd z}{2\pi}.
\ee
\noindent
If we now do the very simple change of variables
\be 
z \rightarrow \lambda = 2z-1, \qquad \rmd z \rightarrow \rmd \lambda = 2 \rmd z,
\ee
\noindent
the orthogonal polynomial measure becomes
\be\label{measure multipenner}
\rmd \mu (\lambda) = 2^{\frac{2 \alpha}{g_s} + \frac{\beta}{g_s} - 1} \left| 1 - \lambda^2 \right|^{-\frac{\alpha}{g_s}} \left| \lambda \right|^{-\frac{\beta}{g_s}} \frac{\rmd \lambda}{2 \pi}. 
\ee
\noindent
The reason for doing this change of variables is because orthogonal polynomials with respect to this last measure, \eqref{measure multipenner}, are known. In fact, in the same way that in the single Penner potential (which is $k=1$ in \eqref{multipenner potential gen}) we deal with the Laguerre polynomials (see, \textit{e.g.}, \cite{ps09}) in here the relevant orthogonal polynomials are the generalized Gegenbauer polynomials \cite{dx01, b01, b07op}. Their precise definition is
\be 
\int_{-1}^{1} \rmd \lambda\, w_{\rho, \sigma} (\lambda)\, C_n^{(\rho, \sigma)}(\lambda)\, C_m^{(\rho, \sigma)}(\lambda) = h_n\, \delta_{nm},
\ee
\noindent
with $w_{\rho, \sigma} (\lambda)$ being the weight function
\be 
w_{\rho, \sigma} (\lambda) = \frac{1}{B \left( \rho+1/2, \sigma+1/2 \right)} \left( 1 - \lambda^2 \right)^{\rho - 1/2} \left| \lambda \right|^{2 \sigma}, \qquad \rho, \sigma > -\frac{1}{2},
\ee
\noindent
and
\be 
B(x,y) = \frac{\Gamma(x) \Gamma(y)}{\Gamma(x+y)}.
\ee
\noindent
Going back to our multi--Penner measure \eqref{measure multipenner} one immediately identifies
\be\label{measure gegen}
\rmd \mu (\lambda) = 2^{\frac{2 \alpha}{g_s} + \frac{\beta}{g_s} - 1}\, B \left( 1 - \frac{\alpha}{g_s}, \frac{1}{2} - \frac{\beta}{2 g_s} \right) w_{\frac{1}{2} - \frac{\alpha}{g_s}, -\frac{\beta}{2 g_s}} (\lambda)\, \frac{\rmd \lambda}{2 \pi}.
\ee

In order to compute the partition function \eqref{Zorth} the next step is to address the coefficients $h_n$. This requires a few intermediate steps\footnote{In these intermediate computations that follow we shall work with $\rho = \frac{1}{2} - \frac{\alpha}{g_s}$ and $\sigma = - \frac{\beta}{2 g_s}$ for shortness. Then, when addressing the partition function, we will reintroduce the original expressions.}, for the polynomials need to be monic (\textit{i.e.}, the coefficient of the highest order term equals one). First, let us use the relation between the generalized Gegenbauer polynomials and the standard Jacobi polynomials $P_n^{(\mu, \nu)}$ \cite{dx01}
\bea
\label{Gegen Jacobi1}
C_{2n}^{(\rho, \sigma)} (\lambda) &=& \frac{(\rho + \sigma)_n}{(\sigma + 1/2)_n}\, P_n^{(\rho - 1/2, \sigma - 1/2)}(2 \lambda^2 - 1), \\
C_{2n+1}^{(\rho, \sigma)} (\lambda) &=& \frac{(\rho + \sigma)_{n+1}}{(\sigma + 1/2)_{n+1}}\, \lambda\, P_n^{(\rho - 1/2, \sigma + 1/2)}(2 \lambda^2 - 1),
\label{Gegen Jacobi2}
\eea
\noindent
where we used the Pochhammer symbol
\be 
(a)_m = a \left( a+1 \right) \cdots \left( a+m-1 \right) = \frac{\Gamma(a+m)}{\Gamma(a)}.
\ee
\noindent
Given this relation, one may immediately extract
\bea 
h_{2n} &=& \frac{\left( \rho + 1/2 \right)_n \left( \rho + \sigma \right)_n \left( \rho + \sigma \right)}{n! \left( \sigma + 1/2 \right)_n \left( \rho + \sigma + 2n \right)}, \\
h_{2n+1} &=& \frac{\left( \rho + 1/2 \right)_n \left( \rho + \sigma \right)_{n+1} \left( \rho + \sigma \right)}{n! \left( \sigma + 1/2 \right)_{n+1} \left( \rho + \sigma + 2n + 1 \right)}.
\eea
\noindent
These are not yet the coefficients we are looking for: as mentioned above, one must work with monic orthogonal polynomials and this is not the case for the $C_{n}^{(\rho, \sigma)}$ polynomials. But now one does know that Jacobi polynomials are normalized as
\be\label{Jacobi norm}
J_n^{\mu, \nu} (\lambda) \equiv \frac{2^n\, n!\, \Gamma ( n+\mu+\nu+1 )}{\Gamma ( 2n+\mu+\nu+1 )}\, P_n^{\mu, \nu} (\lambda) \sim \lambda^n + \cdots,
\ee
\noindent
which will allow us to normalize the generalized Gegenbauer polynomials. In fact, further taking into account the pre--factors in \eqref{Gegen Jacobi1} and \eqref{Gegen Jacobi2}, we finally define the adequately normalized version of these polynomials as\footnote{Note that one needs to divide by $2^n$ because $J_n (2x^2-1) \sim 2^n x^{2n} + \cdots$, so this cancels the $2^n$ factor in \eqref{Jacobi norm}.}
\bea 
G_{2n}^{(\rho, \sigma)} (\lambda) &=& \frac{(\sigma + 1/2)_n}{(\rho + \sigma)_n}\, \frac{n!\, \Gamma(n + \rho + \sigma)}{\Gamma(2n + \rho + \sigma)}\, C_{2n}^{(\rho, \sigma)} (\lambda) \sim \lambda^{2n} + \cdots, \\
G_{2n+1}^{(\rho, \sigma)} (\lambda) &=& \frac{(\sigma + 1/2)_{n+1}}{(\rho + \sigma)_{n+1}}\, \frac{n!\, \Gamma(n + \rho + \sigma + 1)}{\Gamma(2n + \rho + \sigma +1)}\, C_{2n+1}^{(\rho, \sigma)} (\lambda) \sim \lambda^{2n +1} + \cdots.
\eea
\noindent
We now have complete information to find the correctly normalized coefficients $h_n$. As should be clear from the above analysis, they naturally split into ``even'' and ``odd'', where one finds,
\bea 
h_{2n} &=& \frac{n!}{2\pi}\, 2^{\frac{2 \alpha}{g_s} + \frac{\beta}{g_s} - 1}\, \frac{\Gamma \left( n-\frac{\alpha}{g_s}+1 \right) \Gamma \left( n-\frac{\beta}{2g_s}+\frac{1}{2} \right) \Gamma \left( n-\frac{\alpha}{g_s}-\frac{\beta}{2g_s}+\frac{1}{2} \right)}{\Gamma \left( 2n-\frac{\alpha}{g_s}-\frac{\beta}{2g_s}+\frac{1}{2} \right) \Gamma \left( 2n-\frac{\alpha}{g_s}-\frac{\beta}{2g_s}+\frac{3}{2} \right)}, \\
h_{2n+1} &=& \frac{n!}{2\pi}\, 2^{\frac{2 \alpha}{g_s} + \frac{\beta}{g_s} - 1}\, \frac{\Gamma \left( n-\frac{\alpha}{g_s}+1 \right) \Gamma \left( n-\frac{\beta}{2g_s}+\frac{3}{2} \right) \Gamma \left( n-\frac{\alpha}{g_s}-\frac{\beta}{2g_s}+\frac{3}{2} \right)}{\Gamma \left( 2n-\frac{\alpha}{g_s}-\frac{\beta}{2g_s}+\frac{3}{2} \right) \Gamma \left( 2n-\frac{\alpha}{g_s}-\frac{\beta}{2g_s}+\frac{5}{2} \right)}.
\eea

Finally, one may compute the partition function for this model following \eqref{Zorth}. Due to the $\BZ_2$ symmetry of the spectral geometry, and similarly to what happened in the quartic model in \eqref{Z_2limits}, the partition function naturally splits into ``even'' and ``odd'' coefficients (the ones just above) as one writes
\be 
Z = \prod_{n=0}^{N/2 - 1} h_{2n} \prod_{n=0}^{N/2 - 1} h_{2n+1}.
\ee
\noindent
As we shall see when addressing the calculation of the free energy, it turns out that it is useful to write the many products of Gamma functions which appear in the partition function as Barnes functions. This may be done with a reorganization of the products which appear in the expression above and, besides its definition $\Gamma(z+1) = z\, \Gamma(z)$, the use of the property
\be
\Gamma \left( z + \frac{1}{2} \right) = 2^{1-2z}\, \sqrt{\pi}\, \frac{\Gamma \left( 2z \right)}{\Gamma \left( z \right)}.
\ee
\noindent
This property is particularly useful for the Gamma functions containing half--integer factors; indeed the two terms with a $\beta$ factor may now be rewritten as
\be 
\Gamma \left( n - \frac{\beta}{2g_s} + \frac{1}{2} \right) \Gamma \left( n - \frac{\beta}{2g_s} + \frac{3}{2} \right) = \pi\, 2^{- 4 n - 1 + \frac{2\beta}{g_s}}\,  \frac{\Gamma \left( 2n - \frac{\beta}{g_s} + 1 \right) \Gamma \left( 2n - \frac{\beta}{g_s} + 2 \right)}{\Gamma \left( n - \frac{\beta}{2g_s} + 1 \right)^2}.
\ee
\noindent
A similar reasoning may be applied to the terms whose numerators contain the combination $\frac{\alpha}{g_s}+\frac{\beta}{2g_s}$. Using $\eta \equiv n -\frac{\alpha}{2g_s}-\frac{\beta}{4g_s}$, the combination which appears in the respective denominators will become
\be 
\Gamma \left( 2\eta + \frac{1}{2} \right) \Gamma \left( 2\eta + \frac{3}{2} \right)^2 \Gamma \left( 2\eta + \frac{5}{2} \right) = \pi^2\, 2^{- 16\eta - 6}\, \frac{\Gamma \left( 4\eta + 1 \right) \Gamma \left( 4\eta + 2 \right) \Gamma \left( 4\eta + 3 \right) \Gamma \left( 4\eta + 4 \right)}{\Gamma \left( 2\eta + 1 \right)^2 \Gamma \left( 2\eta + 2 \right)^2}.
\ee
\noindent
The final required ingredient is the definition of the Barnes function, $G_2 (z)$, in terms of products of Gamma functions. Essentially, we shall use
\be
\prod_{n=0}^{N-1} \Gamma(n + x + 1) = \frac{G_2 (N + x + 1)}{G_2(x + 1)}
\ee
\noindent
alongside with its useful extension
\be
\prod_{n=0}^{N-1} \Gamma(k n + x + 1)\, \Gamma(k n + x + 2) \cdots \Gamma(k n + x + k) = \frac{G_2(k N + x + 1)}{G_2(x + 1)}.
\ee
\noindent
Assembling all different pieces together, and introducing the 't~Hooft coupling $t = N g_s$ as usual, the partition function of the ``triple'' Penner model finally follows as
\bea 
Z (t,g_s) &=& \frac{2^{\frac{t}{g_s} \left(\frac{t + \beta}{g_s} -2\right)}}{\pi^{\frac{t}{g_s}}}\, G_2 \left( 1 + \frac{t}{2g_s} \right)^2 G_2 \left( 1 - \frac{\alpha}{g_s} \right)^{-2} G_2 \left( 1 + \frac{t-2\alpha}{2g_s} \right)^2 \times \nonumber \\
&&
\times \frac{G_2 \left( 1 - \frac{\beta}{2g_s} \right)^2}{G_2 \left( 1 - \frac{\beta}{g_s} \right)}\, \frac{G_2 \left( 1 + \frac{t-\beta}{g_s} \right)}{G_2 \left( 1 + \frac{t-\beta}{2g_s} \right)^2}\, \frac{G_2 \left( 1 + \frac{t-2\alpha-\beta}{g_s} \right)}{G_2 \left( 1 + \frac{t-2\alpha-\beta}{2g_s} \right)^2}\, \frac{G_2 \left( 1 + \frac{2t-2\alpha-\beta}{2g_s} \right)^2}{G_2 \left( 1 + \frac{2t-2\alpha-\beta}{g_s} \right)}.
\label{pennerpartitionfunctionfinal}
\eea

This is, of course, an exact result; it encodes both perturbative and nonperturbative contributions. In order to analyze instantons in this model we will have to understand the usual large $N$ semi--classical expansion of the free energy. But this is actually a simple calculation given \eqref{pennerpartitionfunctionfinal}, as all one needs to know is the asymptotic expansion of the logarithm of the Barnes function, \textit{i.e.},
\be\label{Logbarnes exp}
\log G_2(z+1) \simeq \frac{1}{2} z^2 \log z - \frac{3}{4} z^2 + \frac{1}{2} z \log 2 \pi - \frac{1}{12} \log z + \zeta'(-1) + \sum_{g=2}^{+\infty} \frac{B_{2g}}{2g (2g-2)}\, \frac{1}{z^{2g-2}}.
\ee
\noindent
As always, we are interested in the normalized free energy $\CF = F - F_{\text{G}}$. The genus $g$ free energies, $\CF_g (t)$, then follow from the logarithm of the partition function \eqref{pennerpartitionfunctionfinal}, given the asymptotic expansion \eqref{Logbarnes exp}. As we are mainly interested in comparing the large--order behavior of the perturbative expansion against instanton data we shall only present results with genus $g \geq 2$, although $\CF_0$ and $\CF_1$ also follow straightforwardly from this procedure. One finds:
\bea\label{Fg penner}
\CF_g (t) &=& \frac{B_{2g}}{2g \left( 2g-2 \right)}\, \Bigg\{ 2^{2g-1} \left( t-2\alpha \right)^{2-2g} - 2\, \alpha^{2-2g} + \\
&&
+ \left( 2^{2g-1}-1 \right) \left( t^{2-2g} + \beta^{2-2g} - \left(t-\beta\right)^{2-2g} - \left( t-2\alpha-\beta \right)^{2-2g} + \left( 2t-2\alpha-\beta \right)^{2-2g} \right)  \Bigg\}. \nonumber
\eea
\noindent
This final result for the perturbative genus $g$ free energies confirms that indeed the tunneling effects are not associated to non--trivial saddle--points as in \cite{msw07} but rather to $A$--cycle effects as in \cite{ps09}: in fact, its large--order growth is essentially dictated by ``Bernoulli numbers growth'' which immediately indicates that all different actions will be multiples of $2\pi\rmi$ \cite{ps09}.

The final step we have to address is, thus, the explicit construction of the $A$--cycle instanton contributions. This again is done very much in line with the discussions in \cite{ps09, sw09} where each Barnes function in \eqref{pennerpartitionfunctionfinal}, or each Bernoulli component in \eqref{Fg penner}, leads to a discontinuity of the type 
\be 
\mathrm{Disc}\, \log G_2 (N+1) = \rmi \sum_{m=0}^{+\infty} \left( \frac{|N|}{m} + \frac{1}{2 \pi m^2} \right) \rme^{-2 \pi |N| m}
\ee
\noindent
at the Stokes line $N = \rmi |N|$. This is an \textit{exact} expression for the discontinuity of the Barnes function and, as we discussed earlier in subsection \ref{sec:ts}, it immediately yields the full multi--instanton content of the associated free energy. In particular, the full discontinuity of the free energy \eqref{Fg penner} is given by the sum of the discontinuities of the logarithms of Barnes functions, and this encodes its full multi--instanton structure. We finally find
\begin{align}
\mathrm{Disc}\, \CF =& \frac{1}{\pi g_s} \sum_{n=1}^{+\infty} \left\{ 
\left( \frac{\pi t}{n} - \frac{\rmi g_s}{n^2} \right) \rme^{-\frac{\pi\rmi t n}{g_s}} 
- \frac{1}{2} \left( \frac{2\pi t}{n} - \frac{\rmi g_s}{n^2} \right) \rme^{-\frac{2\pi\rmi t n}{g_s}} - \frac{1}{2} \left( \frac{2\pi\alpha}{n} - \frac{\rmi g_s}{n^2} \right) \rme^{-\frac{2\pi\rmi \alpha n}{g_s}} + \right. \nonumber \\
&
+ \left( \frac{\pi\beta}{n} - \frac{\rmi g_s}{n^2} \right) \rme^{-\frac{\pi\rmi \beta n}{g_s}} - \frac{1}{2} \left( \frac{2\pi\beta}{n} - \frac{\rmi g_s}{n^2} \right) \rme^{-\frac{2\pi\rmi \beta n}{g_s}} + \left( \frac{\pi (t-2\alpha)}{n} - \frac{\rmi g_s}{n^2} \right) \rme^{-\frac{\pi\rmi (t-2\alpha) n}{g_s}} - \nonumber \\
&
- \left( \frac{\pi (t-\beta)}{n} - \frac{\rmi g_s}{n^2} \right) \rme^{-\frac{\pi\rmi (t-\beta) n}{g_s}} + \frac{1}{2} \left( \frac{2 \pi (t-\beta)}{n} - \frac{\rmi g_s}{n^2} \right) \rme^{-\frac{2\pi\rmi (t-\beta) n}{g_s}} - \label{Disc multipenner} \\
&
- \left( \frac{\pi (t-2\alpha-\beta)}{n} - \frac{\rmi g_s}{n^2} \right) \rme^{-\frac{\pi\rmi (t-2\alpha-\beta) n}{g_s}} + \frac{1}{2}\left( \frac{2 \pi (t-2\alpha-\beta)}{n} - \frac{\rmi g_s}{n^2} \right) \rme^{-\frac{2\pi\rmi (t-2\alpha-\beta) n}{g_s}} + \nonumber \\
&
\left. + \left( \frac{\pi (2 t-2\alpha-\beta)}{n} - \frac{\rmi g_s}{n^2} \right) \rme^{-\frac{\pi\rmi (2 t-2\alpha-\beta) n}{g_s}} - \frac{1}{2} \left( \frac{2\pi (2 t-2\alpha-\beta)}{n} - \frac{\rmi g_s}{n^2} \right) \rme^{-\frac{2\pi\rmi (2 t-2\alpha-\beta) n}{g_s}} \right\}. \nonumber
\end{align}
\noindent
An alternative approach to obtaining the above result would be to first compute the Borel transform of the free energy (which we can do as we have an exact expression at arbitrary genus which, in particular, tells us that the pre--factor grows as $(2g-3)!$). Then, when performing the inverse Borel transform out of our analytically continued result, one would extract its imaginary part as a sum over residues which would yield precisely the very same result as in \eqref{Disc multipenner} \cite{ps09}.

Overall, \eqref{Disc multipenner} yields twelve different types of nonperturbative effects, all with instanton action associated to an $A$--cycle as in \eqref{t_I}, and all including only one and two--loops contributions around each multi--instanton sector, as further explained in \cite{ps09}. The large--order behavior of perturbation theory will be controled by the instanton whose absolute value of its action is closest to the origin in the complex Borel plane. In this example this is simple to understand without going into Borel analysis, by simply looking at the free energies \eqref{Fg penner}. This expression is essentially a sum of numbers of the form $x^{-2g}$ and, as the genus $g$ grows larger, the dominant term will be the one with the smallest $x$. Having said this, one may now test our nonperturbative discussion of this subsection by matching against large--order results. We shall do this by picking two distinct sets of moduli, namely, $(\alpha,\beta,t) = (-1,-1,1/2)$ and $(\alpha,\beta,t) = (-1,-1/3,2)$. In the first case the leading large--order behavior will be dictated by the dominant contribution in \eqref{Disc multipenner} which is the first one in that expression (of the form $\sim t/2$), while in the second case the leading large--order behavior will dictated by the corresponding dominant contribution in \eqref{Disc multipenner} which will now be the fourth term in that expression (of the form $\sim \beta/2$). The instanton actions are $A = \frac{\rmi\pi}{2}$ and $A = \frac{\rmi\pi}{3}$, respectively. We test the overall nonperturbative structure in figure  \ref{Action multipenner} by plotting, up to genus $g=100$, the behavior of the sequence $1/\sqrt{\alpha_g^{(\CF)}}$ (of its imaginary part, to be precise), alongside with the first three Richardson transforms, and for both sets of moduli. In a straight solid line we plot the two predictions. The error after the third Richardson transform is, in both cases, of the order $10^{-7}\%$, fully validating our nonperturbative analysis.

\FIGURE[ht]{
\label{Action multipenner}
\centering
\includegraphics[width=7cm]{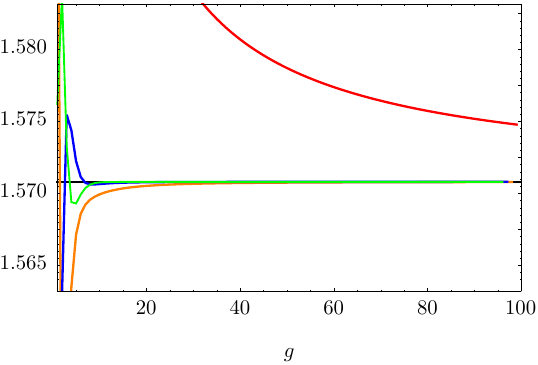}
$\qquad$
\includegraphics[width=7cm]{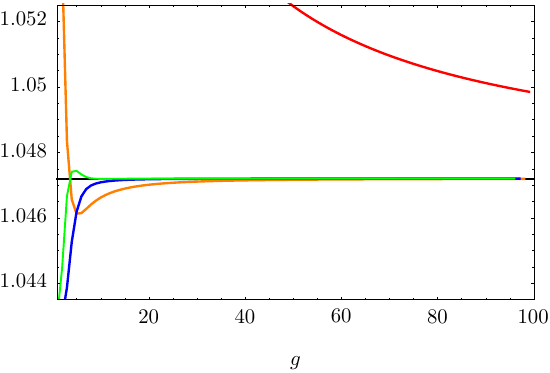}
\caption{We plot the sequence $1/\sqrt{\alpha_g^{(\CF)}}$ associated to the instanton action (see \eqref{seqA}), alongside with its Richardson transforms, for the choices $(\alpha,\beta,t) = (-1,-1,1/2)$ (left image) and $(\alpha,\beta,t) = (-1,-1/3,2)$ (right image). The predictions for the leading asymptotics $A = \pi /2$ and $A = \pi /3$ are given by the horizontal lines. In both cases the errors at genus $g=100$ are of the order $10^{-7}\%$.}
}

\section{Asymptotics of Instantons in the Painlev\'{e} II Equation}\label{sec:P2}

The analysis in the previous section allowed us to check the validity of our one--instanton results, for the Stokes phase of two distinct multi--cut models. In particular, we have checked both the instanton action \eqref{action_sp} and the one--loop one--instanton coefficient \eqref{1inst}, predicted in section \ref{sec:inst}, to very high precision. But our saddle--point analysis also yields multi--instanton results, as for instance in \eqref{Zlresult!}, and the general structure of resurgent transseries solutions further predicts many, new, generalized multi--instanton sectors, as discussed in subsection \ref{sec:ts}. As such, we would now like to check all this multi--instantonic structure, and we shall do so within the context of 2d supergravity, or type 0B string theory, by analyzing the Painlev\'e II equation. This equation arises as a double--scaling limit from the two--cut quartic matrix model we have previously analyzed, but is simpler to analyze from a numerical point--of--view than the full off--critical matrix model.

\subsection{Painlev\'{e} II and Resurgent Transseries}\label{sec:p2&ms}

Recalling the discussions in subsections \ref{sec:backgr} and \ref{sec:quartic}, it should be obvious that the two--cut quartic matrix model has a natural critical point. This is cleary depicted in figure \ref{quartic_numerics}, which shows a critical point for the recursion coefficients at $\lambda t = 3/2$. At this point a phase transition takes place, from the two--cut phase to an unstable one--cut phase. In the double--scaling limit, this critical point is precisely described by the Painlev\'e II equation. At the critical point, and referring to figure \ref{Eff Pot Quartic}, the two cuts collide with each other, having the non--trivial saddle for the eigenvalue instantons at this collision point. In practice what this means is that the smaller endpoints of the cuts will vanish, with $a \to 0$ (and with the non--trivial saddle kept fixed at the origin, $x_0=0$). In terms of $\lambda$, $t$ and $g_s$, the double--scaling limit is defined as
\be\label{dsl1}
g_s \to 0, \qquad \lambda \to \lambda_{\text{c}} = \frac{3}{2}, \qquad t \to 1,
\ee
\noindent
with the variable
\be\label{dsl2}
z = \left( 1 - t \right) g_s^{-\frac{2}{3}}
\ee
\noindent
kept fixed in this limit. As mentioned, it is known that in this limit the matrix model describes 2d supergravity or type $0$B minimal superstrings, see, \textit{e.g.},  \cite{dss90, kms03b, ss04, ss05, kkm05}, and that the physics is encoded in the Painlev\'{e} II equation. This differential equation precisely appears as we take the double--scaling limit in the string equations \eqref{rec eq quartic1} and \eqref{rec eq quartic2}, as discussed in, \textit{e.g.}, \cite{cm91, dss90}. Let us quickly review this point, following \cite{fgz93}, as this will also be important as we connect transseries solutions off and at criticality: start with the string equations \eqref{rec eq quartic1} and \eqref{rec eq quartic2}, and introduce scaling \textit{ans\"{a}tze} for both $P(x)$ and $Q(x)$ \cite{dss90}
\bea
P(x) &\rightarrow& 2 \left( 1 - g_s^{1/3}\, u(z) + g_s^{2/3}\, v(z) \right), 
\label{dslPQ1} \\
Q(x) &\rightarrow& 2 \left( 1 + g_s^{1/3}\, u(z) + g_s^{2/3}\, v(z) \right). \label{dslPQ2}
\eea
\noindent
Plugging these expressions into (an appropriate rewriting of) the string equations, it is simple to obtain in the double--scaling limit
\bea
4 u(z) v(z) - 2 u''(z) &=& 0, \\
2 u^2(z) - 8 v(z) - 2 z &=& 0.
\eea
\noindent
The first equation is readily solved for $v(z)$ which may then be replaced in the second one. As such, one finally obtains a second--order differential equation for $u(z)$,
\be\label{eq2}
2 u''(z) - u^3(z) + z\, u(z) = 0.
\ee
\noindent
The equation above is the Painlev\'{e} II equation in the normalization used in, \textit{e.g.},  \cite{kms03b, kkm05}. In the present paper we are using a slightly different normalization, which follows with a simple rescaling of $u$ and $z$ as $u \to 2^{1/3}\, u$ and $z \to 2^{2/3}\, z$. In this case the Painlev\'{e} II equation becomes
\be\label{P2}
u'' (z) - 2 u^3 (z) + 2z\, u(z) = 0,
\ee
\noindent
which, in particular, also matches the normalization used in \cite{m08}. The perturbative solution corresponds to an expansion around $z \sim + \infty$ where one has $u_{\text{pert}} (z) \sim \sqrt{z}$.

From here on, the procedure to compute the resurgent transseries solution to the Painlev\'e II equation \eqref{P2} follows in parallel, step by step, with what was done in \cite{asv11}. The one--parameter transseries solution to \eqref{P2} was addressed in \cite{m08}, from where we recall the following points. First, the perturbative solution to \eqref{P2} yields
\be
x = z^{-3/2}
\ee
\noindent
as the \textit{open} string coupling. In this case, one may immediately write down an one--parameter transseries solution to the Painlev\'e II equation of the form \cite{m08,  asv11}
\be
u(z) \simeq x^{-1/3} \sum_{n=0}^{+\infty} \sigma^n\, \rme^{-nA/x}\, x^{n \beta} \sum_{g=0}^{+\infty} u_g^{(n)}\, x^g,
\ee
\noindent
where $A$ is the instanton action and $\beta$ a characteristic exponent. Then, plugging this expression back into the Painlev\'e II equation, a solution of this form exists if
\be
A = \pm \frac{4}{3}, \qquad \beta = \frac{1}{2}.
\ee
\noindent
As discussed in \cite{gikm10, asv11} and earlier in this paper, when building nonperturbative solutions with transseries it is important to take into consideration all possible values of the instanton action---in fact, via resurgence, deep in the asymptotics of the solution one will find the need for \textit{both} signs, and thus the need for the two--parameter transseries \textit{ansatz}. As such, we shall now focus on the two--parameter case (but we will also recover some of the results in \cite{m08} along the way). 

Let us begin by writing the Painlev\'e II equation in terms of a different variable
\be
w = x^{1/2} = z^{-3/4}.
\ee
\noindent
This is motivated by having found $\beta = \frac{1}{2}$ above: in the two--parameter case the prefactors will not be of the simple form $x^{n\beta}$ but will depend on two integers, say $n$ and $m$. As we shall see, it will be more convenient to include these contributions inside the perturbative expansions and, as such, to work directly with the variable $x^\beta$. For simplicity of the calculation, it is also convenient to remove the overall factor of $z^{1/2}$ in front of the solution. This motivates us to introduce the new variables (with a slight, but obvious, abuse of notation) 
\be\label{uzsqrtztouw}
u(w) \equiv \left. \frac{u(z)}{\sqrt{z}} \right|_{z = w^{-4/3}}.
\ee
\noindent
It is then a straightforward exercise to rewrite the original equation in terms of this new function
\be\label{new P2}
\frac{9}{16} w^6 \, u''(w) + \frac{9}{16} w^5\, u'(w) - 2 u^3(w) - \left( \frac{w^4}{4} - 2 \right) u(w) = 0.
\ee
\noindent
Our goal is to solve this equation with a two--parameter transseries \textit{ansatz}, along the lines in \cite{asv11}, as (we remind the reader that $w^2=x$ is the open string coupling)
\be\label{tsp2}
u \left( w, \sigma_1, \sigma_2 \right) = \sum_{n=0}^{+\infty} \sum_{m=0}^{+\infty} \sigma_1^n \sigma_2^m\, \rme^{- (n-m) A/w^2}\, \Phi_{(n|m)} (w).
\ee
\noindent
At this stage one might be tempted to assume $\Phi_{(n|m)} (w)$ as a power series in $w$ but, due to resonance effects, this will not work (see \cite{gikm10, asv11}): in order to obtain a solution one further needs to add terms multiplying powers of $\log w$. As such we shall use the following \textit{ansatz} for the asymptotic expansions around generalized multi--instanton sectors
\be\label{tsphi}
\Phi_{(n|m)} (w) = \sum_{k=0}^{\min (n,m)} \log^k (w) \cdot \Phi_{(n|m)}^{[k]} (w) \simeq \sum_{k=0}^{\min (n,m)} \log^k w \cdot \sum_{g=0}^{+\infty} u_g^{(n|m)[k]}\, w^g.
\ee
\noindent
In this case, finding a two--parameter transseries solution to the Painlev\'e II equation now translates to determining the full list of coefficients $u_g^{(n|m)[k]}$. Inserting our \textit{ans\"atze} \eqref{tsphi} and \eqref{tsp2} back into Painlev\'e II  \eqref{new P2}, yields the recursion relation which constructs this transseries:
\bea
&&
2 \sum_{n_1=0}^n \sum_{n_2=0}^{n-n_1} \sum_{m_1=0}^m \sum_{m_2=0}^{m-m_1} \sum_{k_1=0}^k \sum_{k_2=0}^{k-k_1} \sum_{g_1=0}^g \sum_{g_2=0}^{g-g_1}\, u_{g_1}^{(n_1|m_1)[k_1]}\, u_{g_2}^{(n_2|m_2)[k_2]}\, u_{g-g_1-g_2}^{(n-n_1-n_2|m-m_1-m_2)[k-k_1-k_2]} = \nonumber \\
&&
= \left( \frac{9}{4} A^2 (n-m)^2 + 2 \right) u_g^{(n|m)[k]} + \frac{9}{4} A (n-m) (k+1)\, u_{g-2}^{(n|m)[k+1]} + \frac{9}{4} A (n-m)(g-3)\, u_{g-2}^{(n|m)[k]} + \nonumber \\
&&
+ \frac{9}{16} (k+2)(k+1)\, u_{g-4}^{(n|m)[k+2]} + \frac{9}{8} (k+1)(g-4)\, u_{g-4}^{(n|m)[k+1]} + \frac{140 + 9 g (g-8)}{16}\, u_{g-4}^{(n|m)[k]}.
\eea

The above recursion now allows us to see resonance explicitly. Let us consider the case where $|n-m|=1$ and look for the leading terms in the recursion, the $u_g^{(n|m)[k]}$ coefficients. The first term on the second line above is $6 u_g^{(n|m)[k]}$, but the sum in the first line also contains terms with this factor; they are:
\be
2 u_{g}^{(n|m)[k]}\, u_{0}^{(0|0)[0]}\, u_{0}^{(0|0)[0]} + 2 u_{0}^{(0|0)[0]}\, u_{g}^{(n|m)[k]}\, u_{0}^{(0|0)[0]} + 2 u_{0}^{(0|0)[0]}\, u_{0}^{(0|0)[0]}\, u_{g}^{(n|m)[k]}
\ee
\noindent
such that the leading terms in the recursion will cancel\footnote{Recall that $\lim_{z \to + \infty} u_{\text{pert}} (z) \sim \sqrt{z}$, so that one has $u_{0}^{(0|0)[0]} = 1$.}. As explained in greater detail in \cite{asv11} this cancelation describes resonance in the Painlev\'e II equation and thus the need to introduce the ``$[k]$--sectors'', which will still allow us to find a solution for the recursion in spite of the aforementioned cancelation. We refer the reader to \cite{asv11} for further details on this phenomenon.

Another issue which arises when solving the above recursion deals with reparameterization invariance of the transseries \cite{asv11}: the obvious freedom to choose the parameterization of the transseries coefficients $\sigma_1$ and $\sigma_2$ translates to a long list of free coefficients, \textit{i.e.}, coefficients in the transseries which are \textit{not} fixed by the recursion. Do notice that this is not a problem, but rather a requirement from the transseries structure, but we refer the reader to \cite{asv11} for further details on this phenomenon. The punch line is that one needs to choose a prescription to fix these free coefficients. As it turns out, the most natural choice is to set as many free coefficients to zero as possible, as this will also yield the simplest final results. Following \cite{asv11}, we shall fix the reparameterization invariance by setting
\be
u_1^{(m+1|m)[0]} = 0, \qquad \forall m \ge 1, \qquad \text{and} \qquad u_1^{(n|n+1)[0]} = 0, \qquad \forall n \ge 1.
\ee

Having addressed the aforementioned subtleties, all one is left to do is to iterate the recursion in a computer. Results for the lowest sectors follow as
\bea 
\Phi_{(0|0)}^{[0]} (w) &=& 1 - \frac{1}{16} w^4 - \frac{73}{512} w^8 - \frac{10657}{8192} w^{12} - \frac{13912277}{524288} w^{16} - \cdots, \label{00} \\
\Phi_{(1|0)}^{[0]} (w) &=& w - \frac{17}{96} w^3 + \frac{1513}{18432} w^5 - \frac{850193}{5308416} w^7 + \frac{407117521}{2038431744} w^9 - \cdots, \\
\Phi_{(2|0)}^{[0]} (w) &=& \frac{1}{2} w^2 - \frac{41}{96} w^4 + \frac{5461}{9216} w^6 - \frac{1734407}{1327104} w^8 + \frac{925779217}{254803968} w^{10} - \cdots, \\
\Phi_{(1|1)}^{[0]} (w) &=& - 3 w^2 - \frac{291}{128} w^6 - \frac{447441}{32768} w^{10} - \frac{886660431}{4194304} w^{14} - \frac{13316458344441}{2147483648} w^{18} - \cdots. \label{11}
\eea
\noindent
Let us note that, as expected, the first three lines above containing physical multi--instanton sectors precisely agree with the results in \cite{m08} (once we translate from our notation to theirs). Results concerning generalized multi--instanton sectors are new, and we present more details of this explicit transseries solution to the Painlev\'e II equation in appendix \ref{ap3}.

We end this subsection with a few more comments on the (logarithmic) structure of the transseries solution and how it relates---in the double--scaling limit---to the transseries solution of the two--cut quartic matrix model we have discussed in subsection \ref{sec:quartic} and in appendix \ref{ap2}. The first thing to notice is that it is simple to determine the lowest order for which the coefficients $u_{g}^{(n|m)[k]}$ are non--vanishing; let us call this number $2 \beta_{nm}^{[k]}$. The result, which can be immediately checked from the results above and in appendix \ref{ap3}, is the following
\be\label{beta_nm}
2 \beta_{nm}^{[k]} = n + m - 2 \left[\frac{k_{nm}+k}{2} \right]_I,
\ee
\noindent
with $[\star]_I$ denoting the integer part, and 
\be 
k_{nm} = \min(n,m) - m \, \delta_{nm}.
\ee
\noindent
Next, and similarly to what was found for the Painlev\'e I equation in \cite{asv11}, the logarithmic sectors turn out to be related to each other and, in particular, to the non--logarithmic sectors. In fact, we here find a formula very similar to the expression (5.40) in \cite{asv11}, which reads
\be\label{lognonlog}
u_{g}^{(n|m)[k]} = \frac{1}{k!}\, \Big( 8 \left( m-n \right) \Big)^k u_{g}^{(n-k|m-k)[0]}.
\ee
\noindent
This relation will be very useful in reducing the number of independent Stokes constants which enter the game; it provides relations between many of them in the same way as the analogue Painlev\'e I expression was very helpful in \cite{asv11}. As a final point in discussing the structure of the Painlev\'e II transseries solution, let us see how to make the bridge back to the two--cut string equations \eqref{rec eq quartic1} and \eqref{rec eq quartic2}. Its two--parameter transseries solution, \textit{i.e.}, its coefficients $P^{(n|m)}$ and $Q^{(n|m)}$ in \eqref{expansion P and Q}, must agree, in the double--scaling limit, with the coefficients of our present solution $u^{(n|m)}$. That this has to be the case is clear since the Painlev\'{e} II equation itself was derived from the aforementioned string equations via \eqref{dslPQ1} and \eqref{dslPQ2}. But our point here is that this may be made explicit as we find:
\be 
- \left( C\, \sqrt{g_s} \right)^{n+m} g_s^{g-1/3} P^{(n|m)[0]}_g \xrightarrow[\text{DSL}]{} z^{- \frac{ 3 (n+m) + 6 g - 2}{4}} u^{(n|m)[0]}_{2g+n+m}.
\ee
\noindent
In this expression, the ``DSL'' arrow simply means that we have applied the double--scaling limit \eqref{dsl1} and \eqref{dsl2} to the left--hand--side. On the right--hand--side the coefficients which appear are the ones associated to the original variables, \textit{i.e.}, where we have inverted the redefinitions of $u$ and $z$ we did before. There is a similar expression involving the $Q$ coefficients which relates to the one above by a simple change of sign, as can be seen in \eqref{dslPQ2}. Finally, the constant $C$ is given by
\be 
C = \frac{2 \cdot 3^{1/2}}{\sqrt{\lambda}}.
\ee

\subsection{The Resurgence of Multi--Instantons and Stokes Coefficients}\label{sec:m-inst}

We shall now turn to the resurgence of the generalized multi--instanton $(n|m)[k]$ sectors. The resurgence formulae we have discussed in subsection \ref{sec:ts} will verify the validity of these multi--instanton sectors, and they will further allow---upon consistency---to extract many unknown Stokes constants. We shall only focus on effects at exponential order $1^{-g}$ and our analysis will be less detailed than the one in \cite{asv11} where, using more refined techniques, it was possible to ``dig'' deep in the asymptotics and study effects at orders $2^{-g}$, $3^{-g}$, \textit{et cetera}. Nonetheless, our results will fully validate the two--parameter multi--instantonic structure of the Painlev\'e II solutions.

Let us begin by addressing the Stokes constant $S_1^{(0)}$. On what concerns large--order behavior, this constant appears in the perturbative $(0|0)$ sector and we may use the large--order expression \eqref{F00largeorder} to write in the present case
\be 
u_{4g}^{(0|0)[0]} \simeq \frac{S_1^{(0)}}{\rmi \pi}\, \frac{\Gamma \left( 2g-\frac{1}{2} \right)}{A^{2g-\frac{1}{2}}}\, \sum_{h=0}^{+\infty} u_{2h+1}^{(1|0)[0]} A^h \frac{\Gamma \left( 2g-h-\frac{1}{2} \right)}{\Gamma \left( 2g-\frac{1}{2} \right)} + \mathcal{O}(2^{-g}).
\ee
\noindent
Given this expression, it is immediate to construct the sequence
\be 
\frac{\rmi\pi\, A^{2g-\frac{1}{2}}}{\Gamma \left( 2g-\frac{1}{2} \right)}\, u_{4g}^{(0|0)[0]}
\ee
\noindent
which is asymptotic to $S_1^{(0)}$. Taking its Richardson extrapolation, it follows an extremely precise check on the well--known result (see, \textit{e.g.}, \cite{kkm05, m08}), where we found a match of the first $30$ decimal places after $N=20$ Richardson transforms 
\be 
S_1^{(0)} = - \frac{\rmi}{\sqrt{2 \pi}} = - 0.3989422804014327...\, \rmi.
\ee
\noindent
There is a simple relation between the above Stokes constant at criticality, and the corresponding Stokes constant off--criticality, \eqref{2cqmmsc}, which is similar to the relation between the corresponding Stokes constants in \cite{asv11}---\textit{i.e.}, Stokes constant for Painlev\'e I and for the one--cut quartic matrix model. Namely, we find\footnote{For shortness we will avoid the labels referring to either ``Painlev\'{e} II'' or ``Quartic Matrix Model'' throughout the rest of the paper. All constants discussed from here onwards refer to the critical (double--scaled) model.}
\be 
\left. S_1^{(0)} \right|_{\text{PII}} = \frac{\left. S_1^{(0)} \right|_{\text{QMM}}}{C},
\ee
\noindent
where the constant $C$ was defined above. Naturally, this expression is simply encoding the double--scaling limit at the level of Stokes constants (see \cite{asv11} for other Stokes constants).

As we move forward there is one point to have in mind: except for a limited set of empirical relations they satisfy among themselves---which we shall discuss in the following---there are no further analytical predictions for all other Stokes constants. As such, we need to compute them at the same time we test resurgence in an independent fashion. This is done in two steps \cite{gikm10, asv11}: we choose one resurgent formula and validate it via \textit{some} resurgent relations; then we use \textit{different} resurgent relations in this formula to numerically compute new Stokes constants. As one iterates this procedure towards several Stokes constants and several multi--instanton sectors, consistency independently double--checks both the Stokes constants and the resurgence of instantons.

In this spirit, let us move on to the multi--instanton sectors and address the Stokes constant $S_{-1}^{(2)}$ which appears in the $(2|0)$ sector. If we apply our large--order formula for multi--instanton sectors, \eqref{asymn0}, with two physical instantons, $n=2$ and $m=0$, and focus only on the leading contributions to the asymptotics, $k=1$, we arrive at 
\bea 
u_{2g+2}^{(2|0)[0]} &\simeq& \frac{3 S_1^{(0)}}{2\pi\rmi}\, \frac{\Gamma \left( g-\frac{1}{2} \right)}{A^{g-\frac{1}{2}}}\, \sum_{h=0}^{+\infty} \frac{\Gamma \left(g-h-\frac{1}{2} \right)}{\Gamma \left( g-\frac{1}{2} \right)}\, u_{2h+3}^{(3|0)[0]}\, A^h + \label{200} \\
&&
+ \frac{S_{-1}^{(2)}}{2\pi\rmi}\, \frac{\Gamma \left( g+\frac{1}{2} \right)}{(-A)^{g+\frac{1}{2}}}\, \sum_{h=0}^{+\infty} \frac{\Gamma \left( g-h+\frac{1}{2} \right)}{\Gamma \left( g+\frac{1}{2} \right)}\, u_{2h+1}^{(1|0)[0]} \left( -A \right)^h + \nonumber \\
&&
+ \frac{\widetilde{S}_{-1}^{(0)}}{2\pi\rmi}\, \frac{\Gamma \left( g-\frac{1}{2} \right)}{(-A)^{g-\frac{1}{2}}}\, \sum_{h=0}^{+\infty} \frac{\Gamma \left( g-h-\frac{1}{2} \right)}{\Gamma \left( g-\frac{1}{2} \right)}\, u_{2h+3}^{(2|1)[0]} \left( -A \right)^h + \nonumber \\
&&
+ \frac{\widetilde{S}_{-1}^{(0)}}{4\pi\rmi}\, \frac{\Gamma \left( g+\frac{1}{2} \right)}{(-A)^{g+\frac{1}{2}}}\, \sum_{h=0}^{+\infty} \frac{\Gamma \left( g-h+\frac{1}{2} \right)}{\Gamma \left( g+\frac{1}{2} \right)}\, u_{2h+1}^{(2|1)[1]} \left( -A \right)^h \left\{\psi \left( g-h+\frac{1}{2} \right) - \log \left( A \right) - \rmi\pi \right\}. \nonumber
\eea
\noindent
A novel feature of this case is that, adding to the familiar $g!$ large--order growth, the digamma function further produces effects which grow as $g! \log g$ and which will in fact be the \textit{dominant} effects. One procedure to extract and confirm the new Stokes coefficients associated to this expression, via Richardson transforms and when in the presence of $\log g$ factors, was introduced in \cite{gikm10} within the context of the Painlev\'e I equation and further extended in \cite{asv11}. Let us see how to address this issue. We move a factor of
\be
2 \pi \rmi\, \frac{A^{g+\frac{1}{2}}}{\Gamma \left( g+\frac{1}{2} \right)}
\ee
\noindent
to the left--hand--side of the above equation, and expand its right--hand--side in powers of $1/g$. In this way, one obtains a sequence with the following asymptotic behavior:
\be 
A_g \sim B_g\, \log g  + C_g, \qquad \text{where} \qquad B_g \simeq \sum_{k=0}^{+\infty} \frac{b_k}{g^k}, \quad C_g \simeq \sum_{k=0}^{+\infty} \frac{c_k}{g^k}.
\ee
\noindent
To extract the leading coefficient, $b_0$, we may construct a new sequence,
\be\label{log0}
\widetilde{A}_g = g \left( A_{g+1} - A_g \right),
\ee
\noindent
which behaves as
\be 
\widetilde{A}_g \sim \widetilde{B}_g\, \log g + \widetilde{C}_g, \qquad \text{where} \qquad \widetilde{B}_g \simeq \sum_{k=1}^{+\infty} \frac{\widetilde{b}_k}{g^k}, \quad \widetilde{C}_g \simeq b_0 + \sum_{k=1}^{+\infty} \frac{\widetilde{c}_k}{g^k},
\ee
\noindent
thus isolating the coefficient we are looking for. In fact, should we apply a couple of Richardson transform (at least two), we remove the subleading tails in $1/g^k$ and in $\log g/g^k$ and immediately obtain $b_0$ numerically. Similarly, if we want to find $b_1$, we can define 
\be\label{log1}
A^{(1)}_g = g \left( \widetilde{A}_g - b_0 \right),
\ee
\noindent
and now apply the Richardson transforms to the sequence $\widetilde{A}^{(1)}_g = g \left( A^{(1)}_{g+1} - A^{(1)}_g \right)$ in order to extract the coefficient $-b_1$. As we move on to the extraction of the $c_i$ coefficients, the procedure is more or less straightforward. For instance, if we subtract the leading logarithm to the left--hand--side of the original sequence, the new sequence\footnote{Notice that we can subtract further logarithmic terms in order to accelerate the convergence.}
\be\label{order0}
P_g = A_g - b_0\, \log g
\ee
\noindent
will now yield $c_0$. Along the same lines, $P^{(1)}_g = A_g - \left( b_0 + b_1/g \right) \log g$ allows us to extract $-c_1$, and so on. Applying all this in our present context we now have to consider the sequence\footnote{A trivial word on notation: $A$ is the instanton action, $A_g$ the sequence we are addressing.}
\be\label{Ag200}
A_g = 2\pi\rmi\, \frac{A^{g+\frac{1}{2}}}{\Gamma \left( g+\frac{1}{2} \right)}\, u_{2g+2}^{(2|0)[0]},
\ee
\noindent
where we should notice that, due to the factors of $(-1)^g$ in \eqref{200}, we need to look separately at the sequences for $g$ odd and for $g$ even. For simplicity, we shall only discuss the even case, but the odd one is completely analogous. If we now use the sequence \eqref{log0} to compute $\widetilde{A}_{2g}$, we expect it to converge towards the leading coefficient multiplying $\log g$ in the resurgent relation \eqref{200}. What is this number? Using the value of $u_{1}^{(2|1)[1]} = -8$ (simply obtained for instance via \eqref{lognonlog}) and using the fact\footnote{At this precise moment this only adds numerical evidence to the fact that $\widetilde{S}_{-1}^{(0)} = - \rmi\, S_{1}^{(0)}$. But, as we shall see in the following, we can actually show that this relation is true, so we may as well use it already.} that $\widetilde{S}_{-1}^{(0)} = - \rmi\, S_{1}^{(0)}$, if the resurgent formulae hold in the present context then this number should be equal to the analytical value
\be\label{lim-10}
- \frac{\rmi\, \widetilde{S}_{-1}^{(0)}}{2}\, u_{1}^{(2|1)[1]} = 1.59576...\,.
\ee
\noindent
Let us then turn to the sequence and analyze it. This is shown on the first image of figure \ref{fig200}, where we plot the original sequence and some of its Richardson transforms. After $N=20$ Richardson transforms we find the numerical value of $1.59573...$ which differs from the prediction above by less than $0.01 \%$, thus fully validating our resurgent multi--instanton structure.

\FIGURE[ht]{
\label{fig200}
\centering
\includegraphics[width=7cm]{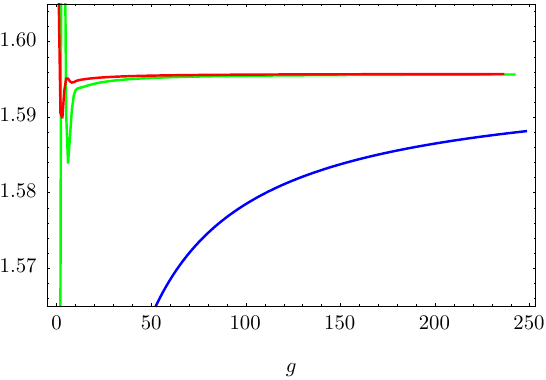}
$\qquad$
\includegraphics[width=7.2cm]{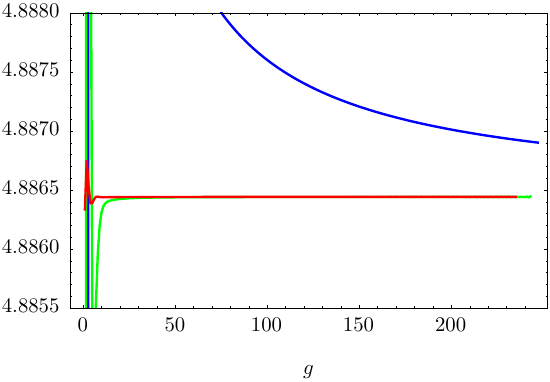}
\caption{The left image shows the sequence $\widetilde{A}_{2g}$ built from \eqref{Ag200} (blue) alongside with its fifth (green) and twentieth (red) Richardson transforms. This can be shown to quickly converge towards our prediction \eqref{lim-10} with errors $\sim 0.01\%$. The right image shows the sequence $P_g$ in \eqref{order0} (blue) alongside its fifth (green) and twentieth (red) Richardson transforms. This quickly converges towards our prediction \eqref{S-12}.}
}

Taking our analysis one step further, we may now extract a new Stokes constant by looking at the leading non--logarithmic term, which may be computed using the sequence \eqref{order0}. According to the same large--order resurgent relation, \eqref{200}, this term should be
\be 
-\frac{\rmi\, \widetilde{S}_{-1}^{(0)}}{2}\, u_{1}^{(2|1)[1]} \left(\log A + \rmi \pi \right) - \rmi\, S_{-1}^{(2)}\, u_{1}^{(1|0)[0]},
\ee
\noindent
where we do not know the value of the Stokes constant $S_{-1}^{(2)}$. However, we may find it by analyzing the sequence \eqref{order0}, as shown in the second image of figure \ref{fig200}: after $N=20$ Richardson iterations we extract the numerical prediction
\be\label{S-12}
S_{-1}^{(2)} = - 5.3455144... - 5.013256493...\, \rmi.
\ee

Before moving on with further Stokes constants, let us make a remark concerning the new Stokes constant we have just computed, \eqref{S-12}: it is a complex number, with both real and imaginary contributions. But, as explained in detail in \cite{asv11}, there are many relations between the Stokes constants and a large number of these depend on each other (although it is unclear how many truly independent Stokes constants exist). Some of these relations may be derived from the general structure of the string genus expansion, and are thus model--independent; while others were found ``experimentally'', and will thus depend upon which equation is under analysis (but see \cite{asv11} for more details on both these points). In particular, all Stokes constants of the form $S_{\ell}^{(n)}$ and $\widetilde{S}_{\ell}^{(n)}$ with $\ell > 0$ are \textit{purely} imaginary. We will thus only list this type of Stokes constants. We shall discuss how these relations arise when we discuss the $(1|1)$ sector below; for the moment let us just mention that, for \eqref{S-12} above, the relation which involves $\widetilde{S}_{1}^{(2)}$ out of $S_{-1}^{(2)}$ is
\be\label{S12}
\widetilde{S}_{1}^{(2)} = - \rmi\, S_{-1}^{(2)} + 4 \pi \rmi\, S_{1}^{(0)} = 5.3455144...\, \rmi.
\ee

Having successfully addressed a two--instanton sector, let us next address a sector involving generalized instantons. In this case, the simplest choice is to study a generalized ``closed string'' sector; the example where we have $n=1=m$ and $k=0$. By ``closed string'' we mean that sectors of the type $(n|n)$ are expected to have an asymptotic expansion in powers of the \textit{closed} string coupling $g_s^2 \sim w^4$ rather than in powers of the \textit{open} string coupling $g_s \sim w^2$, as can be seen in \eqref{11} \cite{asv11}. In this case the $(1|1)$ sector has no logarithmic contributions and the relevant large--order relation is \cite{asv11}
\bea 
 \frac{2\pi\rmi\, A^{g+\frac{1}{2}}}{\Gamma \left( g+\frac{1}{2} \right)}\, u_{2g+2}^{(1|1)[0]} &\simeq& {S}_{1}^{(1)}\, \sum_{h=0}^{+\infty} \frac{\Gamma \left( g-h+\frac{1}{2} \right)}{\Gamma \left( g+\frac{1}{2} \right)}\, u_{2h+1}^{(1|0)[0]}\, A^h + 2 {S}_{1}^{(0)}\, A\, \sum_{h=0}^{+\infty} \frac{\Gamma \left( g-h-\frac{1}{2} \right)}{\Gamma \left (g+\frac{1}{2} \right)}\, u_{2h+3}^{(2|1)[0]}\, A^h - \nonumber \\
&&
- 2 {S}_{1}^{(0)}\, \sum_{h=0}^{+\infty} \frac{\Gamma \left( g-h+\frac{1}{2} \right)}{\Gamma \left( g+\frac{1}{2} \right)}\, u_{2h+1}^{(2|1)[1]}\, A^h\, \widetilde{B}_A \left( g-h-\frac{3}{2} \right) - \nonumber \\
&&
\hspace{-22mm}
-  \frac{\rmi\, \widetilde{S}_{-1}^{(1)}}{(-1)^{g}}\, \sum_{h=0}^{+\infty} \frac{\Gamma \left( g-h+\frac{1}{2} \right)}{\Gamma \left( g+\frac{1}{2} \right)}\, u_{2h+1}^{(0|1)[0]} \left( -A \right)^h + \frac{2 \rmi\, \widetilde{S}_{-1}^{(0)}}{(-1)^{g}}\, A\,  \sum_{h=0}^{+\infty} \frac{\Gamma \left( g-h-\frac{1}{2} \right)}{\Gamma \left( g+\frac{1}{2} \right)}\, u_{2h+3}^{(1|2)[0]} \left( -A \right)^h + \nonumber \\
&&
+ \frac{2\rmi\, \widetilde{S}_{-1}^{(0)}}{(-1)^{g}}\, \sum_{h=0}^{+\infty} \frac{\Gamma \left( g-h+\frac{1}{2} \right)}{\Gamma \left( g+\frac{1}{2} \right)}\, u_{2h+1}^{(1|2)[1]} \left( -A \right)^h B_A \left( g-h-\frac{3}{2} \right).
\label{asym11}
\eea
\noindent
As we have just mentioned, the $(1|1)[0]$ sector will have a standard, topological perturbative expansion, which implies that all terms above with odd $g$ will have to vanish \cite{asv11}. In other words, imposing $u_{2(2g+1)+2}^{(1|1)[0]} = 0$ will result in a tower of relations between the Stokes constants appearing on the right--hand--side of \eqref{asym11}, as this expression needs to vanish order by order in both powers of $1/g^k$ and $\log g/g^k$. For example, expanding the digamma functions we find that imposing that the term proportional to $\log g$ vanishes will imply the condition
\be 
S_{1}^{(0)} - \rmi\, \widetilde{S}_{-1}^{(0)} = 0 \quad \Rightarrow \quad \widetilde{S}_{-1}^{(0)} = - \rmi\, S_{1}^{(0)},
\ee
\noindent
which we had already put forward and checked numerically---now being ``theoretically'' justified. On the other hand, the term at order $\mathcal{O}(1)$ yields a relation involving two unknown constants,
\be\label{Stokes11}
S_{1}^{(1)} + \rmi\, \widetilde{S}_{-1}^{(1)} + 8 \pi \rmi\, S_{1}^{(0)} = 0,
\ee
\noindent
where we have used \eqref{lognonlog} to relate $u_{1}^{(1|2)[1]} = 8 u_{1}^{(0|1)[0]}$. Continuing along these lines and looking at further required cancelations, one may use this procedure in order to extract similar relations between further Stokes constants, such as \eqref{S12} which we have discussed above. Our goal now is to apply the same reasoning as used within the $(2|0)$ sector in order to compute this new Stokes constant, $S_{1}^{(1)}$ (and, along the way, $\widetilde{S}_{-1}^{(1)}$ as well). This is very similar to what we have done before, with the slight difference that now only the sequences for even $g$ are relevant. Once again the term proportional to $\log g$ offers just a consistency check on the resurgent structure of the transseries solution and on (already) known Stokes constants, and we show in figure \ref{fig110} that this is indeed working perfectly: the relevant sequence, after Richardson extrapolation, converges towards the correct number, $-S_{1}^{(0)}\, u_{1}^{(2|1)[1]} + \rmi\, \widetilde{S}_{-1}^{(2)}\, u_{1}^{(1|2)[1]}$, with an error smaller than $0.001 \%$. 

\FIGURE[ht]{
\label{fig110}
\centering
\includegraphics[width=7cm]{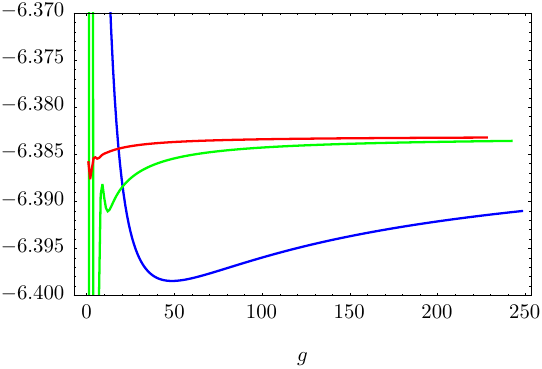}
$\qquad$
\includegraphics[width=7cm]{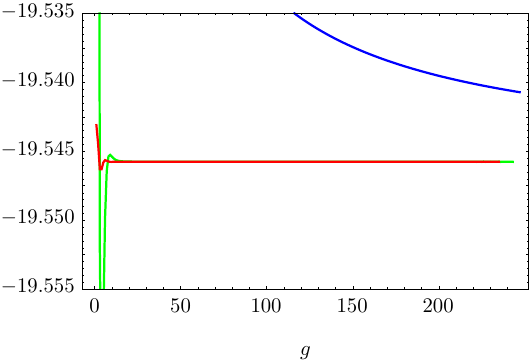}
\caption{The left image shows the sequence which tests the leading, $\log g$, coefficient of the large--order relation \eqref{asym11} (blue), alongside with its fifth (green) and twentieth (red) Richardson transforms. This sequence quickly converges towards the expected limit $- S_{1}^{(0)}\, u_{1}^{(2|1)[1]} + \rmi\, \widetilde{S}_{-1}^{(2)}\, u_{1}^{(1|2)[1]}$ with an error smaller than $0.001 \%$. The right image shows the sequence which tests the leading, order $\mathcal{O}(1)$, term in the large--order relation \eqref{asym11} (blue), alongside its fifth (green) and twentieth (red) Richardson transforms. Again, this quickly leads to our prediction \eqref{predS11}.}
}

The new constants we are after appear at order $\mathcal{O}(1)$, without logarithmic contributions. After using the relevant sequence, \eqref{order0}, we find, as we show in figure \ref{fig110}, a fast convergence towards the number $\xi = -19.54576...\, \rmi$ that resurgence sets to
\be\label{predS11}
\xi = S_{1}^{(0)}\, u_{1}^{(2|1)[1]}\, \log A + S_{1}^{(1)}\, u_{1}^{(1|0)[0]} - \rmi\, \widetilde{S}_{-1}^{(1)}\, u_{1}^{(0|1)[0]} - \rmi\, \widetilde{S}_{-1}^{(0)}\, u_{1}^{(1|2)[1]} \left( \log A + \rmi \pi \right).
\ee
\noindent
Using this result together with the previous relation, \eqref{Stokes11}, we find $S_{1}^{(1)}$ (which is, as expected, a purely imaginary number)
\be 
S_{1}^{(1)} = -10.6910288...\, \rmi.
\ee

\begin{table}[htb]
\begin{center}
\begin{tabular}{|l|r|r|l|l|}
\hline
& & Precision & From \\
\hline
$S_{1}^{(0)}$ & $- 0.39894228... \; \rmi$ & $\infty$ & $\Phi_{(0|0)}^{[0]}$ \\
\hline
$S_{1}^{(1)}$ & $- 10.6910288... \; \rmi$ & 7 & $\Phi_{(1|1)}^{[0]}$ \\
\hline 
$\widetilde{S}_{1}^{(2)}$ & $5.3455144... \; \rmi$ & 7 & $\Phi_{(2|0)}^{[0]}$ \\
\hline
\end{tabular}
\end{center}
\caption{The independent Stokes constants we have calculated. The third column gives the number of decimal places to which the answer is explicitly computed, while the fourth column shows the instanton sector where each constant appears for the first time. All constants we address first appear at order $1^{-g}$.}
\label{table:P2stokescoeff}
\end{table}

The (independent) Stokes constants we have computed are summarized in table \ref{table:P2stokescoeff}. It is interesting to notice that a further ``experimental'' relation $\widetilde{S}_{1}^{(2)} = -\frac{1}{2}\, S_{1}^{(1)}$ is (apparently) true in this case. The exact same relation was also found in \cite{asv11}, in the context of the Painlev\'e I equation, alongside with some other extra relations, all of them emerging from purely numerical relations. We expect that by examining further data in the present Painlev\'e II context also many similar relations will be found. However, at this stage, we have no first principles explanation for these extra relations: determining the minimal set of independent Stokes constants is a very interesting open problem for future research.

\subsection{The Nonperturbative Free Energy of Type 0B String Theory}\label{sec:dslF}

The final point we wish to address is the construction of the nonperturbative free energy for 2d supergravity or 1d type 0B string theory. In fact, using the results of our transseries analysis of the Painlev\'e II equation, we may now build its associated double--scaled free energy. This free energy is obtained from the solution of the Painlev\'e II equation via \cite{dss90, kms03b, ss04, ss05}
\be\label{P2toF}
F_{\text{ds}}'' (z) = - \frac{1}{4}\, u(z)^2.
\ee
\noindent
For convenience, from this point on we shall drop the double--scaled label, but we will always be talking about the free energy at the critical point. The first thing to notice is that there is now a fundamental difference with respect to the Painlev\'{e} I case, studied in \cite{asv11}: the relation between the (twice derived) free energy and the solution of the differential equation is no longer linear. Nonetheless, the right--hand--side of \eqref{P2toF} still has a transseries expansion
\be\label{usq_ts}
-\frac{1}{4}\, u(z)^2 \equiv \sum_{n=0}^{+\infty} \sum_{m=0}^{+\infty} \sigma_1^n \sigma_2^m\, \rme^{-(n-m)\, A\, z^{3/2}}\, \varphi_{(n|m)} (z),
\ee
\noindent
but where now one has
\be 
\varphi_{(n|m)} (z) = \sum_{n'=0}^{n} \sum_{m'=0}^{m} \Phi_{(n'|m')} (z)\, \Phi_{(n-n'|m-m')} (z) \simeq \sum_{g=0}^{+\infty} \frac{{\widetilde{u}}_g^{(n|m)}}{z^{\frac{3g}{4}}}.
\ee
\noindent
Relating this expression to the free energy now just requires a double--integration, as follows from \eqref{P2toF}. Let us begin by looking at the perturbative sector, where we bring back the $\sqrt{z}$ factor we had pulled out in \eqref{uzsqrtztouw}. In this case, the integration leads to
\be\label{F00dsl}
F^{(0|0)} (z) = -\frac{1}{4} \iint \rmd z\, z \left( \Phi_{(0|0)} (z) \right)^2 = - \frac{z^3}{24} - \frac{\log \left( z \right)}{32} + \frac{3}{512 z^3} +\frac{63}{4096 z^6} + \cdots.
\ee
\noindent
As a check on this result, notice that if we apply the double--scaling limit, \eqref{dsl1} and \eqref{dsl2}, to the quartic matrix model free energies, $F_g (t)$, which we have computed via the Euler--Maclaurin formula in \eqref{F_EM} (the first few of which are presented in appendix \ref{ap1}), and if we further implement the rescalings $u \rightarrow 2^{1/3}\, u$ and $z \rightarrow 2^{2/3}\, z$ associated to our choice of normalization, then the answer one obtains precisely matches the above result.

Having understood how to construct the free energy in the perturbative sector, one may move on towards multi--instanton sectors. Beginning with the one--instanton sector arising from the product $\Phi_{(0|0)} (z)\, \Phi_{(1|0)} (z)$, the first coefficient to compute is simply given by
\be\label{Fts_dsl_first}
-\frac{1}{2} \sigma_1\, u_1^{(1|0)[0]} \iint \rmd z\, z^{1/4}\, \rme^{-A z^{3/2}} = - \frac{1}{8}\, \sigma_1\, u_1^{(1|0)[0]}\, z^{-3/4}\, \rme^{-A z^{3/2}} + \cdots.
\ee
\noindent
In the expression above we have kept only the leading term and we have explicitly displayed the coefficient $u_1^{(1|0)[0]}$. Recall that when solving the Painlev\'{e} II equation we chose to set $u_1^{(1|0)[0]} = 1$, and recall that this freedom in choosing the normalization was a consequence of a reparameterization invariance of the double--transseries solution \cite{asv11}. One now needs to readdress this point in order to properly fix the free energy transseries. As shown in \cite{asv11}, rescaling the transseries parameters as $\sigma_1 = c_1\, \widehat{\sigma}_1$ and $\sigma_2 = c_2\, \widehat{\sigma}_2$ makes the following quantities scale accordingly
\bea
\Phi_{(n|m)} &=& c_1^{-n}\, c_2^{-m}\, \widehat{\Phi}_{(n|m)}, \label{rescale0phi} \\
S_\ell^{(k)} &=& c_1^{1-k}\, c_2^{1-k-\ell}\, \widehat{S}_\ell^{(k)}, \label{rescale1} \\
\widetilde{S}_\ell^{(k)} &=& c_1^{1+\ell-k}\, c_2^{1-k}\, \widehat{\widetilde{S}}_\ell^{(k)}. \label{rescale2}
\eea
\noindent
The convenient scaling to do, when dealing with the free energy, is
\be 
\sigma_{1,2} = S_1^{(0)}\, \sigma_{1,2}^F.
\ee
\noindent
In fact, this immediately implies that the leading coefficient of the one--instanton free energy is
\be 
F_0^{(1|0)[0]} = -\frac{1}{8}\, S_1^{(0)}\, u_1^{(1|0)[0]} = \frac{\rmi}{8 \sqrt{2 \pi}},
\ee
\noindent
and thus the free energy Stokes constant is very simply
\be\label{S10F}
S_1^{(0)F} = 1.
\ee
\noindent
This convenient normalization may further be double--checked by using the large--order connection \eqref{Fg large order}
\be 
F_g^{(0|0)[0]} \sim \frac{S_1^{(0)F}}{\rmi\pi}\, \frac{\Gamma \left( 2g-\frac{5}{2} \right)}{A^{2g-\frac{5}{2}}}\, F_0^{(1|0)[0]}.
\ee
\noindent
After just a few Richardson transforms we find that \eqref{S10F} is indeed consistent. It is important to remark that \textit{physical} quantities should not depend on normalization choices, so that only combinations which are left invariant by the above rescalings are physical. In this particular case, the physical quantity is
\be 
S_1^{(0)F} \cdot F_0^{(1|0)[0]}.
\ee
\noindent
Had we chosen to have $\sigma_1^F = \sigma_1$, then we would have found $S_1^{(0)F} = S_1^{(0)}$, but the combination above would not have changed. A longer discussion on  normalizations may be found in \cite{asv11}.

We are now ready to proceed and explicitly compute generalized $(n|m)$ multi--instanton sectors in the free energy of 2d supergravity or 1d type 0B string theory. From the point--of--view of the double--integration, the only complicated sectors are the ones with logarithms. In fact, when $n=m$ the procedure is immediate and a straightforward generalization of what we did for the perturbative $(0|0)$ sector in \eqref{F00dsl}. As such, and always having in mind that we are now dealing with the function $u(z)^2$, in \eqref{usq_ts}, we have for general $n \neq m$, 
\be
\sigma_1^n \sigma_2^m\, \rme^{-(n-m)A/w^2}\, \varphi_{(n|m)}^{[0]} (w) \simeq \sigma_1^n \sigma_2^m\, \rme^{-(n-m)A/w^2}\, \sum_{g=0}^{+\infty} {\widetilde{u}}^{(n|m)[0]}_{2g+2\beta_{nm}^{[0]}}\, w^{2g+2\beta_{nm}^{[0]}}.
\ee
\noindent
It can be shown---and easily checked---that the relation \eqref{lognonlog} connecting  logarithmic $(n+k|m+k)[k]$ to non--logarithmic $(n|m)[0]$ sectors still holds in the precise same form for $\varphi_{(n|m)}^{[k]}$ and its components. In this case, it is convenient to assemble together all sectors which are related to the $(n|m)[0]$ sector; due to the aforementioned relation each of these is of the form (transseries parameters and logarithmic factor included)
\be
\frac{1}{k!} \left( 8 \left( m-n \right) \sigma_1 \sigma_2\, \log w \right)^k \sigma_1^n \sigma_2^m\, \rme^{-(n-m)A/w^2}\, \varphi_{(n|m)}^{[0]} (w).
\ee
\noindent
Finally summing over $k$, one finds
\be
\varphi_{(n|m)}^{\text{[sum]}} (w) = \rme^{8 \left( m-n \right) \sigma_1 \sigma_2\, \log w}\, \varphi_{(n|m)}^{[0]} (w) = w^{8 \left( m-n \right) \sigma_1 \sigma_2}\, \varphi_{(n|m)}^{[0]} (w). 
\ee
\noindent
In order to do the double--integration, let us first move back to the $z$ variable so that the $(n|m)$ contribution becomes
\be\label{nm_contribution}
\sigma_1^n \sigma_2^m\, \rme^{- (n-m) A z^{3/2}}\, \sum_{g=0}^{+\infty} {\widetilde{u}}^{(n|m)[0]}_{2g+2\beta_{nm}^{[0]}} z^{-\frac{3g+3\beta_{nm}^{[0]}-2}{2} + \frac{8(n-m) \sigma_1 \sigma_2}{A}}. 
\ee
\noindent
The double--integration may now be carried through using that
\be\label{int_dsl}
\iint \rmd z\, z^q\, \rme^{-\ell A z^{3/2}} = \frac{2}{3 \ell A}\, z^{q+1/2}\, \rme^{-\ell A z^{3/2}}\, \sum_{m=1}^{+\infty} a_m (q) \cdot \left(-\ell A z^{3/2} \right)^{-m},
\ee
\noindent
where the coefficients $a_m (q)$ are given by
\be 
a_m(q) = \frac{\Gamma \left( m - \frac{2q+1}{3} \right)}{\Gamma \left( - \frac{2q+1}{3} \right)} - \frac{\Gamma \left( m - \frac{2q-1}{3} \right)}{\Gamma \left( - \frac{2q-1}{3} \right)}.
\ee
\noindent
Notice that the $a_m(q)$ coefficients are polynomials in $q$ of degree $m-1$. Further, given the integrand in \eqref{nm_contribution}, the variable $q$ is actually linear in $\sigma_1 \sigma_2$ and, as such, the $a_m(q)$ coefficients will be polynomials of degree $m-1$ in $\sigma_1 \sigma_2$. This effectively means that the double--integration of the $(n|m)$ sector of $u(z)^2$ contributes not only to the $(n|m)$ sector of the free energy, but to all other $(n+r|m+r)$ sectors as well (with $r > 0$).

We are now essentially done. Using a computer, we can apply the integral \eqref{int_dsl} systematically and find that the free energy has the structure
\be 
F(z, \sigma_1^F, \sigma_2^F) = \sum_{n=0}^{+\infty} \sum_{m=0}^{+\infty} \left( S_1^{(0)} \right)^{n+m} \left( \sigma_1^F \right)^n \left( \sigma_2^F \right)^m  \rme^{- (n-m) A z^{3/2}}\, z^{\frac{3}{\pi} (m-n) \sigma_1^F \sigma_2^F}\, F^{(n|m)}(z),
\ee
\noindent
where the ``coefficients'' $F^{(n|m)}(z)$ will be asymptotic expansions in powers of $z^{-3/2}$ (both integer and half--integer, and also containing the occasional logarithm). The first few sectors of the critical free energy are the following
\bea 
F^{(0|0)}(z) &=& - \frac{1}{24} z^3 - \frac{1}{32} \log z + \frac{3}{512} z^{-3} + \frac{63}{4096} z^{-6} + \cdots, \\
F^{(1|0)}(z) &=& - \frac{1}{8} z^{-\frac{3}{4}} + \frac{65}{768} z^{-\frac{9}{4}} - \frac{19273}{147456} z^{-\frac{15}{4}} + \frac{13647905}{42467328} z^{-\frac{21}{4}} - \cdots, \\
F^{(1|1)}(z) &=& \frac{4}{3} z^{\frac{3}{2}} +\frac{25}{96} z^{-\frac{3}{2}} + \frac{6323}{24576} z^{-\frac{9}{2}} + \frac{5015413}{3145728} z^{-\frac{15}{2}} + \cdots, \\
F^{(2|0)}(z) &=& - \frac{1}{32} z^{-\frac{3}{2}} + \frac{59}{1536} z^{-3} - \frac{9745}{147456} z^{-\frac{9}{2}} + \frac{3335669}{21233664} z^{-6} - \cdots, \\
F^{(2|1)}(z) &=& - \frac{9}{16} z^{-\frac{9}{4}} + \frac{737}{512} z^{-\frac{15}{4}} - \frac{398375}{98304} z^{-\frac{21}{4}} + \frac{142017823}{9437184} z^{-\frac{27}{4}} - \cdots, \\
F^{(2|2)}(z) &=& - 3 \log z + \frac{111}{64} z^{-3} + \frac{54507}{8192} z^{-6} + \frac{15245711}{196608} z^{-9} + \cdots, \\
F^{(3|0)}(z) &=& - \frac{1}{96} z^{-\frac{9}{4}} + \frac{59}{3072} z^{-\frac{15}{4}} - \frac{7645}{196608} z^{-\frac{21}{4}} + \frac{1836031}{18874368} z^{-\frac{27}{4}} - \cdots, \\
F^{(3|1)}(z) &=& - \frac{17}{64} z^{-3} + \frac{1211}{1536} z^{-\frac{9}{2}} - \frac{655883}{294912} z^{-6} + \frac{161783969}{21233664} z^{-\frac{15}{2}} - \cdots, \\
F^{(3|2)}(z) &=& \frac{17}{8} z^{-\frac{9}{4}} - \frac{2267}{384} z^{-\frac{15}{4}} + \frac{3488915}{147456} z^{-\frac{21}{4}} - \frac{251878099}{2654208} z^{-\frac{27}{4}} + \cdots, \\
F^{(3|3)}(z) &=& \frac{17}{3} z^{-\frac{3}{2}} + \frac{35675}{2304} z^{-\frac{9}{2}} + \frac{11452163}{81920} z^{-\frac{15}{2}} + \frac{157674856009}{58720256} z^{-\frac{21}{2}} + \cdots.
\eea
\noindent
In the list above we presented the sectors $(n|m)$ with $n \geq m$. The coefficients with $n < m$ differ at most by signs, obeying the rule
\be 
F_g^{(m|n)} = (-1)^{g+\left[n/2\right]_I} F_g^{(n|m)}, \qquad n > m.
\ee
\noindent
The starting powers in the free energy coefficients $F^{(n|m)}$ can be easily related to the starting powers $\beta_{nm}^{[0]}$ of the Painlev\'e II coefficients $u^{(n|m)}$, for instance by looking at \eqref{int_dsl}. At the end of the day we find
\bea 
F^{(n|n)} &\sim& z^{-\frac{3}{2} \beta_{nn}^{[0]}+3}, \\
F^{(n|m)} &\sim& z^{-\frac{3}{2} \beta_{nm}^{[0]}}, \qquad n \neq m,
\eea
\noindent
where $\beta_{nm}^{[0]}$ was defined above in \eqref{beta_nm}. For completeness, we also recall that the logarithmic sectors are ``hidden'' inside the term
\be 
z^{\frac{3}{\pi} (m-n) \sigma_1^F \sigma_2^F} = \exp \left(\frac{3}{\pi} \left( m-n \right) \sigma_1^F \sigma_2^F \, \log z \right).
\ee
\noindent
As a final point, we should also comment on the Stokes constants for the free energy. Since the Stokes constants for $u(z)$ and $u(z)^2$ are the same, the Stokes constants for the free energy are related to those of Painlev\'{e} II via the rescalings described above, \eqref{rescale1} and \eqref{rescale2}, and so
\bea 
S_{\ell}^{(k)F} &=& \ell^2 \left( S_1^{(0)} \right)^{2k + \ell -2} S_{\ell}^{(k)}, \label{stokesF1} \\
\widetilde{S}_{\ell}^{(k)F} &=& \ell^2 \left( S_1^{(0)} \right)^{2k - \ell -2} \widetilde{S}_{\ell}^{(k)}. \label{stokesF2}
\eea
\noindent
The extra $\ell^2$ appearing above comes from taking two derivatives on the factor $\exp \left(\pm \ell A z^{3/2} \right)$. On what concerns the independent Stokes constants we computed in section \ref{sec:m-inst}, the respective values for the independent free energy Stokes constants are presented in table \ref{table:Fstokescoeff}.
\begin{table}[htb]
\begin{center}
\begin{tabular}{|l|r|r|l|}
\hline
& & Precision\\
\hline
$S_{1}^{(0)F}$ & $1.0000000000... $ & $\infty$ \\
\hline
$S_{1}^{(1)F}$ & $- 4.26510341... \; \rmi$ & 8 \\
\hline 
$\widetilde{S}_{1}^{(2)F}$ & $2.13255170... \; \rmi$ & 8  \\
\hline
\end{tabular}
\end{center}
\caption{The independent Stokes constants for the free energy of 2d supergravity or 1d type 0B string theory. They are related to the Stokes constants of the Painlev\'{e} II equation via \protect\eqref{stokesF1} and \protect\eqref{stokesF2}.}
\label{table:Fstokescoeff}
\end{table}

\section{Conclusions and Outlook}\label{sec:concl}

In this paper we have continued our analysis of the nonperturbative structure of the large $N$ limit, and of string theory, along the lines in \cite{asv11}. We have generalized many results \cite{msw07, m08, msw08, ps09, gikm10, kmr10} to the multi--cut realm, with emphasis on proceeding with the nonperturbative study of the quartic matrix model initiated in \cite{msw07, m08, asv11}, this time around in its two--cut phase with its Painlev\'e II double--scaling limit. Our results support the need for resurgent analysis and transseries, in particular the need for two--parameter transseries solutions including new nonperturbative sectors. As in previous work, the question remains to explain, semi--classically, what these new sectors are: for example, in the Painlev\'e II context, while the physical multi--instanton sectors correspond to ZZ--branes \cite{akk03, mmss04, m08} there is no similar understanding of the generalized sectors. Partial discussions may be found in \cite{gikm10, kmr10, asv11, ciy12} but no conclusive answer has yet been reached. This question is also related to finding a first--principles calculation of the many ``experimental'' Stokes constants we have found: for one of these constants, $S_1^{(0)}$, in the Painlev\'e II framework, there are many analytical methods which determine it, see, \textit{e.g.}, \cite{dz95, bi99, kkm05, m08, ciy10}, but for all others finding one such analytical method is still an open problem. This is probably related to first determining how many truly independent Stokes constants there are for each problem, and further explaining the empirical relations we have found among them.

As one looks towards future research, some natural generalizations of our present work quickly come to mind. For example, one natural extension would be to ``dig'' deeper into the asymptotics of our examples, putting the full resurgent formulae on even stronger grounds. Recall that in \cite{asv11}, both for the one--cut quartic model and for its Painlev\'e I double--scaling limit, techniques of Borel--Pad\'e resummation were used in order to analyze contributions to the large--order behavior arising at exponentially suppressed orders of $2^{-g}$, $3^{-g}$, and so on. It would be very interesting to extend those results within the present examples of the two--cut quartic model and its Painlev\'e II double--scaling limit. Another interesting line of work would be to further explore the connection to the AGT framework, along the lines in \cite{dv09, sw09}. For example, one could address the calculation of (symmetric) higher--point correlation functions, or, in a different line, compare our present nonperturbative construction with other nonperturbative completions suggested within the AGT set--up as in \cite{cdv10}. Yet another interesting line of work would be to address extensions of our Painlev\'e II results towards its deformations which arise within the type 0B minimal superstring context, when turning on RR flux or in the presence of charged ZZ--branes \cite{kms03, ss05}. This flux is controled by a parameter, $q$, and the equation which describes the minimal string set--up is now a deformation of Painlev\'e II, namely
\be
u'' (z) - 2 u^3 (z) + 2z\, u(z) = - \frac{q^2}{u^3 (z)}.
\ee
\noindent
This equation is certainly addressable within our framework and it would be very interesting to fully carry out its resurgent transseries analysis, extending our Painlev\'e II results.

In order to be fully explicit when addressing multi--cut Stokes phases, we have focused on the two--cut case where the Stokes phase is essentially related to the $\BZ_2$ symmetry of the spectral curve configuration. But one may extend this calculation for an arbitrary number of cuts, $k$, as long as one keeps the corresponding spectral geometry configuration having a $\BZ_k$ symmetry on its cuts---this symmetry will ensure that, although generically dealing with hyperelliptic configurations, at the end of the day all calculations reduce to elliptic integrals (very much along the same lines as it occurred for us in subsection \ref{sec:oneinst}). Afterwards, and still following our own guidelines from the $\BZ_2$ case, a proper treatment of the sum over instantons will further ensure that these elliptic functions will cancel in the end, thus producing adequate results for a Stokes phase. Setting up such $\BZ_k$ symmetric spectral configurations is very simple, as it is to compute their corresponding instanton actions. The multi--instanton analysis should then follow with some extra work. Another interesting point of this example is its own double--scaling limit \cite{cm91, cisy09} which seems to lead to new integrable hierarchies. For a $\BZ_k$--symmetric configuration the string equations get more complicated, but are certainly solvable within our framework. In this way, it should be possible to say a lot about the nonperturbative structure of their corresponding solutions and, thus, about the general structure of these new integrable hierarchies.

The two--parameter transseries solution we have obtained for the Painlev\'e II equation is, in principle, its full nonperturbative solution. How may we understand the information encoded in this solution? When addressing 2d supergravity, or type 0B string theory, we are looking for a real solution to this equation, \eqref{P2}, for all $z \in \BR$. Recall from, \textit{e.g.}, \cite{m08} that this is naturally associated to the two phases of the Painlev\'e II solutions: the weak--coupling phase, when $z \to + \infty$, and the strong--coupling phase, when $z \to - \infty$, where one finds the asymptotic behaviors \cite{hm80, dz95}
\bea
u (z) &\sim& \sqrt{z}, \qquad\qquad\qquad\qquad\qquad\qquad\qquad\,\,\, z \to + \infty,\\
u (z) &\sim& \frac{1}{2^{1/12}\, \sqrt{2\pi}} \left( -z \right)^{-1/4} \rme^{- \frac{2 \sqrt{2}}{3} \left( -z \right)^{3/2}}, \qquad z \to - \infty.
\label{hmsolutionminus}
\eea
\noindent
There is in fact one such real solution, interpolating between the above weak and strong couplings, the Hastings--McLeod solution \cite{hm80}. Notice that there are many  solutions to the Painlev\'e II equation: for example, in \cite{dz95} one finds a large class of global purely imaginary solutions, alongside another large class of global real solutions of which the Hastings--McLeod solution is a particular (singular limit) case---and the asymptotics of all these solutions are well known \cite{dz95}. In particular, all these solutions should be encoded in our transseries solution, but in here we shall only discuss the Hastings--McLeod solution which was also addressed in \cite{m08}. This in itself is already a non--trivial problem: clearly, the ``instanton action'' in \eqref{hmsolutionminus} is different from the Painlev\'e II instanton actions appearing in its two--parameter transseries solution. The natural question that follows is: how is the Hastings--McLeod solution encoded in our two--parameter transseries solution, and how can it provide for both types of weak and strong coupling behaviors, displayed above? In particular, how may $A$ and $-A$ of Painlev\'e II ``conspire'' to yield the extra $\sqrt{2}$ factor? This question was partially addressed in \cite{m08}, in the context of an one--parameter transseries solution. In there, it was shown that---upon Borel resummation---one may perform a median resummation of the transseries along the Stokes line in the positive real axis to yield a real solution of the Painlev\'e II equation (see the final discussion in \cite{asv11} as well), \textit{i.e.},
\be
u_{\BR} (z,\sigma) \equiv \CS_{+} u \left( z, \sigma - \frac{1}{2} S_1 \right),
\ee
\noindent
where $\CS_{+}$ denotes a left Borel resummation along the positive real axis (see, \textit{e.g.}, \cite{asv11}). Once this is done, the Hastings--McLeod solution is that particular real solution which has $\sigma=0$ in the expression above \cite{m08}. In particular, this median resummation of the one--parameter transseries reproduces the Hastings--McLeod content for $z \in \BR^+$. But one question remained open: what happens along the \textit{negative} real axis instead? To answer this question one needs the full two--parameter transseries solution we have constructed in this paper, but yet this is not the full story: constructing a median resummation along the negative real axis, where one now finds an infinite number of highly non--trivial Stokes constants, is a much harder problem, and moving from positive $z$ to negative $z$ will also entail crossing Stokes lines. These crossings will make Stokes constants jump, not only as overall factors but also inside exponential terms due to the logarithmic sectors as we discussed in subsection \ref{sec:dslF}. As such, it would be a very interesting project to make this strong/weak coupling interpolation completely explicit, within the resurgent transseries framework. We hope to return to some of these ideas in the future.

\acknowledgments
We would like to thank Hirotaka Irie, Marcos Mari\~no, Marcel Vonk and Niclas Wyllard for useful discussions and/or comments. The authors would further like to thank CERN TH--Division for hospitality, where a part of this work was conducted. RS would further like to thank the Department of Mathematics of the University of Geneva and CERN--TH Division for extended hospitality. The research of RS was supported by the Swiss National Science Foundation Grant IZKOZ2--140198, the European Science Foundation Grant ITGP--3685, and the FCT--Portugal Grant PTDC/MAT/119689/2010. The research of RV was partially supported by the FCT--Portugal Grant SFRH/BD/70613/2010.

\newpage

\appendix

\section{The Two--Cut Quartic Matrix Model: Structural Data}\label{ap2}

In this appendix we present some explicit results concerning the two--parameter transseries solution to the two--cut quartic matrix model. Let us recall that in subsection \ref{sec:quartic} we have solved the string equations of this model, \eqref{eq(n|m)1} and \eqref{eq(n|m)2}, by introducing the \textit{ansatz}
\be 
\CP (x) = \sum_{n=0}^{+\infty} \sum_{m=0}^{+\infty} \sigma_1^{n} \sigma_2^{m}\, P^{(n|m)} (x),
\ee
\noindent
with
\be\label{(n|m)ap}
P^{(n|m)} (x) \simeq \rme^{- (n-m) A(x)/g_s} \sum_{g=\beta_{nm}}^{+\infty} g_s^g\, P^{(n|m)}_g (x), 
\ee
\noindent
and similarly for $\CQ (x)$. In the table below we show the maximum order (in $g$) in the string coupling to which we have recursively computed the above nonperturbative coefficients:
\begin{table}[ht]
\centering
\begin{tabular}{c|ccccc}
\begin{picture}(20,20)(0,0)
\put(17,8){$n$}
\put(3.6,18){\line(1,-1){22.3}}
\put(3,0){$m$}
\end{picture}
& 0 & 1 & 2 & 3 & 4 \\
\hline
0 & 60 & 10 & 10 & 10 & 5 \\
1 &    & 10 &  5 &  5 & 5 
\end{tabular}
\caption{Values for the highest $g$ for which we have calculated $P^{(n|m)}_g$ and $Q^{(n|m)}_g$.}
\label{tablequartic}
\end{table}

\noindent
Do note that, since the sums in \eqref{(n|m)ap} have a ``starting genus'' which is (in general) $\beta_{nm} = -\min (m,n) \leq 0$, the actual number of coefficients that we have computed is \textit{bigger} than the numbers displayed in table \ref{tablequartic}. It is also worth pointing out that in the cases where $n=m$ the asymptotic expansions contain only even powers of $g_s$, which implies half of the indicated coefficients vanish. Finally, the sectors $(m|n)$ and $(n|m)$ are trivially related via (similar for $Q (x)$)
\be 
P^{(n|m)}_g (x) = (-1)^g\, P^{(m|n)}_g (x).
\ee

Let us begin by presenting explicit results for the first few coefficients in the perturbative sector\footnote{Recall from the main body of the text that we are using the variable $p = \sqrt{9 - 6 \lambda x}$.}
\bea
P^{(0|0)}_0 &=& \frac{1}{\lambda} \left( 3 - p \right), \qquad \qquad \quad \;\;\, Q^{(0|0)}_0 = \frac{1}{\lambda} \left( 3 + p \right), \\
P^{(0|0)}_1 &=& \lambda\, \frac{162 - 27 p - 9 p^2}{2 p^5}, \qquad Q^{(0|0)}_1 = \lambda\,\frac{- 162 - 27 p + 9 p^2}{2 p^5}, \\
P^{(0|0)}_2 &=& \lambda^3\, \frac{1915812 - 314928 p - 181521 p^2 + 18711 p^3 + 1944 p^4}{8 p^{11}}, \\
Q^{(0|0)}_2 &=& \lambda^3\, \frac{- 1915812 - 314928 p + 181521 p^2 + 18711 p^3 - 1944 p^4}{8 p^{11}}.  
\eea
\noindent
Proceeding with the multi--instanton sectors (and just explicitly showing results for $\CP (x)$ from now on), the first few coefficients in the $(1|0)$ sector are
\bea 
P^{(1|0)}_0 &=& - \sqrt{\frac{3-p}{p}}, \\
P^{(1|0)}_1 &=& \lambda\, \frac{459 - 45 p^2 + 6 p^3}{8\, p^{7/2} \left( 3 + p \right) \left( 3 - p \right)^{1/2}}, \\
P^{(1|0)}_2 &=& \lambda ^2\, \frac{9}{128\, p^{13/2} \left( 3 + p \right)^2 \left( 3 - p \right)^{3/2}} \times \\
&&
\times \left( - 122553 - 15552 p + 27270 p^2 + 2844 p^3 - 1593 p^4 - 132 p^5 + 4 p^6 \right), \nonumber
\eea
\noindent
while in the $(2|0)$ sector we find
\bea 
P^{(2|0)}_0 &=& - \lambda\, \frac{3-p}{2 p^2}, \\
P^{(2|0)}_1 &=& \lambda^2\, \frac{1107 + 108 p - 117 p^2 - 6 p^3}{8\, p^5 \left( 3 + p \right)}, \\
P^{(2|0)}_2 &=& \lambda^3\, \frac{9}{64\, p^8 \left( 3 + p \right)^2 \left( 3 - p \right)} \times \nonumber \\
&&
\times \left( - 442341 - 65448 p + 102330 p^2 + 12924 p^3 - 6669 p^4 - 636 p^5 + 80 p^6 \right).
\eea
\noindent
One of the main features of using multi--parameters transseries is the appearance of generalized multi--instanton sectors, which may have different signs of the instanton action within the nonperturbative exponential contribution. This may lead, sometimes, to the cancelation of all terms in this exponential contribution---for example, in the present setting this happens when $n=m$---and we will be left with a (perturbative) expansion in the closed string coupling. The first sector with this feature is the $(1|1)$ sector, where the first few coefficients are
\bea 
P^{(1|1)}_0 &=& \lambda\, \frac{9-p}{p^2}, \\
P^{(1|1)}_2 &=& \lambda^3\, \frac{70713 - 10125 p - 4617 p^2 + 261 p^3}{8\, p^8}, \\
P^{(1|1)}_4 &=& \lambda^5\, \frac{1}{128\, p^{14}} \left( 8806981203 - 1369011699 p - 959100102 p^2 + 103563198 p^3 + \right. \nonumber \\
&&
\left. + 18833715 p^4 - 787563 p^5 \right).
\eea
\noindent
The general case $n \neq m$ is more complicated. Generically, asymptotic expansions will be in powers of the string coupling, $g_s$, and the ``starting genus'' may start taking negative values. Furthermore, logarithmic contributions begin to appear \cite{asv11}. For example, one may compute the following coefficient in the $(2|1)$ sector (this is the second non--vanishing coefficient in this sector):
\bea
P^{(2|1)}_0 &=& \lambda^2\, \frac{1}{16\, p^{7/2} \left( 3 + p \right) \left( 3 - p \right)^{1/2}}\, \bigg\{ \left( - 432 - 180 p + 24 p^2 + 12 p^3 \right) + \nonumber \\
&&
\left. + \left( 153 - 15 p^2 + 2 p^3 \right) \log \left( \frac{p^6}{9-p^2}\right) \right\}.
\eea

Even though we have not produced as much data as in the one--cut solution discussed in \cite{asv11}, we are still able to conjecture the general form of all these coefficients. Our data, together with the experience gathered in \cite{asv11}, indicate that the nonperturbative coefficients take the form
\be 
P^{(n|m)}_g (x) = \sum_{k=0}^{\min (n,m)} \log^k \left( f(x) \right) \cdot P^{(n|m)[k]}_g (x),
\ee
\noindent
where
\be
P^{(n|m)[k]}_g (x) = \frac{\lambda^{c_1}}{p^{c_2} \left( 3 - p \right)^{c_3} \left( 3 + p \right)^{c_4}}\, {\mathfrak{P}}^{(n|m)[k]}_g (x),
\ee
\noindent
and where the function $f(x)$, written in terms of the variable $p$, is
\be 
f(p) = \frac{p^6}{9-p^2}.
\ee
\noindent
Above, the coefficients $c_i$ are given by
\bea 
c_1 &=& n + m + g - 1, \\
c_2 &=& \frac{3}{2} \left( n + m \right) + 3g - 1, \\
c_3 &=& \left( 1 - \delta_{nm} \right) \frac{1}{2} \left( 3m - n + 2g \right), \\
c_4 &=& \left( 1 - \delta_{nm} \right) \left( m + g \right),
\eea
\noindent
and they are valid whenever $n \geq m$. The ${\mathfrak{P}}^{(n|m)[k]}_g (x)$ are polynomials in $p$ of degree $3 \left( m+g \right)$. When $n=m$, these polynomials get reduced and have degree $n+g$. Concerning the pattern for the $Q^{(n|m)}_g (x)$ coefficients, we find a similar result, but with the roles of $c_3$ and $c_4$ interchanged,
\be
Q^{(n|m)[k]}_g (x) = \frac{\lambda^{c_1}}{p^{c_2} \left( 3 - p \right)^{c_4} \left( 3 + p \right)^{c_3}}\, {\mathfrak{Q}}^{(n|m)[k]}_g (x),
\ee
\noindent
and with the extra condition
\be
{\mathfrak{Q}}^{(n|m)[k]}_g (p) = - {\mathfrak{P}}^{(n|m)[k]}_g (-p).
\ee

Finally, upon further analyzing our data, a relation emerges between the coefficients in the $(n|m)[k]$ and the $(n-k|m-k)[0]$ sectors (this is very similar to the relation \eqref{lognonlog} which we have found for the nonperturbative Painlev\'e II coefficients in the main body of the text). We find
\be\label{lognonlogP}
P^{(n|m)[k]}_g = \frac{1}{k!} \left( \frac{\lambda \left( n-m \right)}{6} \right)^k P^{(n-k|m-k)[0]}_{g+k} \,. 
\ee

For completeness, let us be fully specific on a few of the polynomials ${\mathfrak{P}}^{(n|m) [k]}_g$, which we have explicitly computed. These polynomials take the form $c \sum a_i p^i$, where $p$ is their variable and $c$ and $a_i$ their coefficients. We list these coefficients in the tables that follow.

\begin{table}[t]
\centering
\begin{tabular}{c|rrr}
$g$     & 0 & 1 & 2 \\
\hline
$c$     & $-\frac{3}{4}$ & $\frac{1}{32}$ & $-\frac{3}{512}$ \\
\hline
$p^0$ &   36 &    670680 &    811753164 \\
$p^1$ &   15 &      8991 &    163196127 \\
$p^2$ & $-$2 & $-$159732 & $-$277986654 \\
$p^3$ & $-$1 &       486 &  $-$53318331 \\
$p^4$ &      &     11340 &     33149088 \\
$p^5$ &      &    $-$261 &      6037173 \\
$p^6$ &      &    $-$208 &   $-$1519542 \\
$p^7$ &      &           &    $-$264789 \\
$p^8$ &      &           &        16680 \\
$p^9$ &      &           &         3068
\end{tabular} 
\hspace{2em}
\begin{tabular}{c|rrr}
$g$     & $-$1 & 0 & 1 \\
\hline
$c$     & $-\frac{1}{6}$ & $\frac{1}{16}$ & $-\frac{3}{256}$ \\
\hline
$p^0$ & 1 &   153 &   122553 \\
$p^1$ &   &     0 &    15552 \\
$p^2$ &   & $-$15 & $-$27270 \\
$p^3$ &   &     2 &  $-$2844 \\
$p^4$ &   &       &     1593 \\
$p^5$ &   &       &      132 \\
$p^6$ &   &       &     $-$4
\end{tabular}
\caption{Prefactor $c$ and coefficients of the polynomials ${\mathfrak{P}}^{(2|1)[0]}_g$ (left) and ${\mathfrak{P}}^{(2|1)[1]}_g$ (right).}
\end{table}

\begin{table}[t]
\centering
\begin{tabular}{c|rrr}
$g$     & 0 & 1 & 2 \\
\hline
$c$     & $-\frac{1}{4}$ & $\frac{1}{8}$ & $-\frac{3}{128}$ \\
\hline
$p^0$ &   297 &    528525 &   2176342749 \\
$p^1$ &    54 &     69255 &    368793810 \\
$p^2$ & $-$27 & $-$118584 & $-$766103913 \\
$p^3$ &  $-$4 &  $-$12744 & $-$118696752 \\
$p^4$ &       &      7533 &     96370155 \\
$p^5$ &       &       549 &     13149378 \\
$p^6$ &       &    $-$94  &   $-$4972095 \\
$p^7$ &       &           &    $-$549324 \\
$p^8$ &       &           &        81696 \\
$p^9$ &       &           &         5288
\end{tabular} 
\hspace{2em}
\begin{tabular}{c|rrr}
$g$     & $-$1 & 0 & 1 \\
\hline
$c$     & $-\frac{1}{6}$ & $\frac{1}{8}$ & $-\frac{3}{64}$ \\
\hline
$p^0$ & 1 &   369 &    442341 \\
$p^1$ &   &    36 &     65448 \\
$p^2$ &   & $-$39 & $-$102330 \\
$p^3$ &   &  $-$2 &  $-$12924 \\
$p^4$ &   &       &      6669 \\
$p^5$ &   &       &       636 \\
$p^6$ &   &       &     $-$80
\end{tabular}
\caption{Prefactor $c$ and coefficients of the polynomials ${\mathfrak{P}}^{(3|1)[0]}_g$ (left) and ${\mathfrak{P}}^{(3|1)[1]}_g$ (right).}
\end{table}

\begin{table}[t]
\centering
\begin{tabular}{c|rrr}
$g$     & 0 & 1 & 2 \\
\hline
$c$     & $-\frac{3}{16}$ & $\frac{3}{128}$ & $-\frac{9}{2048}$ \\
\hline
$p^0$ &   351 &   2998377 &   14430217473 \\
$p^1$ &    54 &    451980 &    2674565406 \\
$p^2$ & $-$33 & $-$654156 & $-$4961854665 \\
$p^3$ &  $-$4 &  $-$82620 &  $-$841785048 \\
$p^4$ &       &     39393 &     605366703 \\
$p^5$ &       &      3600 &      90717246 \\
$p^6$ &       &   $-$382  &   $-$29788263 \\
$p^7$ &       &           &    $-$3633804 \\
$p^8$ &       &           &        442368 \\
$p^9$ &       &           &         30856
\end{tabular} 
\hspace{2em}
\begin{tabular}{c|rrr}
$g$     & $-$1 & 0 & 1 \\
\hline
$c$     & $-\frac{1}{8}$ & $\frac{3}{64}$ & $-\frac{9}{1024}$ \\
\hline
$p^0$ & 1 &  1269  &   3852765 \\
$p^1$ &   &   126  &    593892 \\
$p^2$ &   & $-$135 & $-$876582 \\
$p^3$ &   &  $-$8  & $-$115236 \\
$p^4$ &   &        &     55341 \\
$p^5$ &   &        &      5544 \\
$p^6$ &   &        &    $-$572
\end{tabular} 
\caption{Prefactor $c$ and coefficients of the polynomials ${\mathfrak{P}}^{(4|1)[0]}_g$ (left) and ${\mathfrak{P}}^{(4|1)[1]}_g$ (right).}
\end{table}

\section{Perturbative Free Energy in the Quartic Matrix Model}\label{ap1}

In the main text we have discussed how the Euler--Maclaurin formula (suitably adapted to the period--two case) provides for a recipe in order to extract the genus $g$ perturbative free energies, $\CF^{(0|0)}_g$, out of the recursion coefficients in the orthogonal polynomial framework. However, it is important to notice that this method is computationally very time consuming (even more so that in the one--cut case addressed in \cite{asv11}) and thus ends up providing for less data in the resurgence tests than directly using the coefficients $P^{(0|0)}_g$ or $Q^{(0|0)}_g$. Nonetheless, we explicitly need to know these coefficients as they are used to determine the Stokes coefficient out of the large--order sequence \eqref{oneseq}. In here, we shall explicitly list a few of these results for the $\CF^{(0|0)}_g$ (as usual, written in terms of the variable $p$). We find
\bea 
\CF^{(0|0)}_0 &=& \frac{\left( 9 - p^2 \right)^2}{576 \lambda^2}\, \log \left( \frac{1296}{\lambda^4} \right), \\
\CF^{(0|0)}_1 &=& \frac{1}{4} \log \left( \frac{3+p}{2 p} \right), \\
\CF^{(0|0)}_2 &=& \frac{\lambda^2}{320\, p^6 \left( 9 - p^2 \right)^2} \times \\
&&
\times \left( 787320 - 174960 p - 215055 p^2 + 43740 p^3 + 18630 p^4 - 3780 p^5 - 471 p^6 \right), \nonumber \\
\CF^{(0|0)}_3 &=& \frac{\lambda^4}{1792\, p^{12} \left( 9 - p^2 \right)^4} \times \\
&&
\times \left( 1214950653504 - 234633327264 p - 653277037896 p^2 + 119905844184 p^3 + \right. \nonumber \\
&&
+ 141553030437 p^4 - 24374010024 p^5 - 15592951332 p^6 + 2467933272 p^7 + \nonumber \\
&&
\left. + 895852062 p^8 - 125778744 p^9 - 23861628 p^{10} + 2857680 p^{11} + 181989 p^{12} \right). \nonumber
\eea
\noindent
As the genus increases, the expressions become exponentially longer and we shall not show any more explicit formulae. However, our results do indicate a clear pattern for the perturbative genus $g$ free energies: for genus $g \geq 2$ they have the form
\be 
\CF^{(0|0)}_g (\lambda, p) = \frac{\lambda^{2(g-1)}}{p^{6(g-1)} \left( 9-p^2 \right)^{2(g-1)}}\, \mathfrak{F}_g (p),
\ee
\noindent
where $\mathfrak{F}_g (p)$ is a polynomial in $p$ of degree $6(g-1)$. Finally, as we have discussed in subsection \ref{sec:dslF}, applying the double--scaling limit to these results, and taking two derivatives, yields a precise match with (the square of) the perturbative data arising within the Painlev\'e II equation.

\section{The Painlev\'e II Equation: Structural Data}\label{ap3}

In this appendix we present some explicit results concerning the two--parameter transseries solution to the Painlev\'e II equation. Let us recall that in subsection \ref{sec:p2&ms} we have solved this equation, \eqref{P2}, by introducing the \textit{ansatz}
\be\label{TSp2}
u \left( w, \sigma_1, \sigma_2 \right) = \sum_{n=0}^{+\infty} \sum_{m=0}^{+\infty} \sigma_1^n \sigma_2^m\, \rme^{-(n-m) A/w^2}\, \Phi_{(n|m)} (w),
\ee
\noindent
with
\be\label{TSphi}
\Phi_{(n|m)} (w) \simeq \sum_{k=0}^{\min (n,m)} \log^k w \cdot \sum_{g=0}^{+\infty} u_g^{(n|m)[k]}\, w^g.
\ee
\noindent
For shortness we introduce 
\be 
\Phi_{(n|m)}^{[k]} (w) \simeq \sum_{g=0}^{+\infty} u_g^{(n|m)[k]}\, w^g.
\ee
\noindent
As we discussed in the main text, this \textit{ansatz} turns the original differential equation into a recursive equation for the coefficients $u_g^{(n|m)[k]}$. In table \ref{tablep2} we show the maximum order in $w$ to which we have calculated these coefficients. We have only listed the $n \geq m$ cases, but we shall see below how the $(m|n)$ and $(n|m)$ sectors are trivially related. We shall also see that there is a relation between the $[k]$th and the $[0]$th logarithmic sectors. In the following we will reproduce a few examples concerning all this data.
\begin{table}[ht]
\centering
\begin{tabular}{c|ccccc}
\begin{picture}(20,20)(0,0)
\put(17,8){$n$}
\put(3.6,18){\line(1,-1){22.3}}
\put(3,0){$m$}
\end{picture}
& 0 & 1 & 2 & 3 & 4 \\
\hline
0  & 1000 & 1000 & 1000 & 500 & 500 \\
1  & & 1000 & 100 & 100 & 100 \\
2  & & & 100 & 100 & 100 \\
3  & & & & 100 & 100 \\
4  & & & & & 100
\end{tabular}
\caption{Maximum order in $w$ for which we have calculated $u_g^{(n|m)[k]}$.}
\label{tablep2}
\end{table}

The first few $(n|0)$ sectors we found are:
\bea 
\Phi_{(0|0)}^{[0]} &=& 1 - \frac{1}{16} w^4 - \frac{73}{512} w^8 - \frac{10657}{8192} w^{12} - \frac{13912277}{524288} w^{16} - \cdots, \label{ts00} \\
\Phi_{(1|0)}^{[0]} &=& w - \frac{17}{96} w^3 + \frac{1513}{18432} w^5 - \frac{850193}{5308416} w^7 + \frac{407117521}{2038431744} w^9 - \cdots, \\
\Phi_{(2|0)}^{[0]} &=& \frac{1}{2} w^2 - \frac{41}{96} w^4 + \frac{5461}{9216} w^6 - \frac{1734407}{1327104} w^8 + \frac{925779217}{254803968} w^{10} - \cdots.
\eea
\noindent
The lowest $\Phi_{(n|1)}^{[0]}$ are
\bea 
\Phi_{(1|1)}^{[0]} &=& - 3 w^2 - \frac{291}{128} w^6 - \frac{447441}{32768} w^{10} - \frac{886660431}{4194304} w^{14} - \cdots, \label{110} \\
\Phi_{(2|1)}^{[0]} &=& w^3 - \frac{115}{48} w^5 + \frac{30931}{18432} w^7 - \frac{4879063}{663552} w^9 + \cdots.
\eea
\noindent
The first time we encounter logarithmic terms is for $n=2$, $m=1$, where we have
\be\label{211}
\Phi_{(2|1)}^{[1]} = -8 w + \frac{17}{12} w^4 - \frac{1513}{2304} w^6 + \frac{850193}{663552} w^8 + \cdots.
\ee
\noindent
From the full list of data we computed, one finds a relation between the coefficients in sectors $(n|m)[k]$ (logarithmic) and $(n-k|m-k)[0]$ (non--logarithmic), which is the following
\be\label{logklog0}
u_{g}^{(n|m)[k]} = \frac{1}{k!}\, \big( 8 \left( m-n \right) \big)^k\, u_{g}^{(n-k|m-k)[0]}.
\ee
\noindent
Finally, the sectors with $n < m$ are very closely related to the ones with $n > m$ as\footnote{Similarly to what was found in \cite{asv11} for the Painlev\'e I equation, we may suspect that this is just an ``apparent'' relation, only to be falsified at some high $n$, $m$ and $g$ (in \cite{asv11} one had to go to $n=3$, $m=4$ and genus $g=11$ to falsify it). However, all the data we have produced is consistent with this relation.}
\be 
u_{g}^{(m|n)[k]} = \left| u_{g}^{(n|m)[k]} \right|, \qquad \text{for} \,\, n > m.
\ee
\noindent
As a final note, we add that all our off--criticality transseries results, \textit{i.e.}, the results for the two--cut quartic matrix model partially presented in appendix \ref{ap2}, match the present transseries solution of Painlev\'{e} II, when in the double--scaling limit. A \textit{Mathematica} notebook with the complete explicit results we have obtained is available upon request.

\newpage


\bibliographystyle{plain}

\end{document}